\documentclass{xlinear}  

\usepackage{makeidx}
\usepackage{graphicx}
\usepackage{axodraw}
\usepackage{cite}
\makeindex
\begin{document}
%





\newenvironment{Eqnarray}%
   {\arraycolsep 0.14em\begin{eqnarray}}{\end{eqnarray}}
\def\beqa{\begin{Eqnarray}}
\def\eeqa{\end{Eqnarray}}
\def\CR{\nonumber \\ }


\def\leqn#1{(\ref{#1})}






\def\st{\scriptstyle}
\def\sst{\scriptscriptstyle}

\def\overbar#1{\overline{#1}}
\let\littlebar=\bar
\let\bar=\overbar



\def\etal{{\it et al.}}
\def\ie{{\it i.e.}}
\def\eg{{\it e.g.}}
\def\opcit{{\it op.~cit.}}



\def\VEV#1{\left\langle{ #1} \right\rangle}
\def\bra#1{\left\langle{ #1} \right|}
\def\ket#1{\left| {#1} \right\rangle}
\def\vev#1{\langle #1 \rangle}



\def\lsim{\mathrel{\raise.3ex\hbox{$<$\kern-.75em\lower1ex\hbox{$\sim$}}}}
\def\gsim{\mathrel{\raise.3ex\hbox{$>$\kern-.75em\lower1ex\hbox{$\sim$}}}}


\def\D{\mathcal{D}}
\def\L{\mathcal{L}}
\def\M{\mathcal{M}}
\def\W{\mathcal{W}}



\def\One{{\bf 1}}
\def\hc{{\mbox{\rm h.c.}}}
\def\tr{{\mbox{\rm tr}}}
\def\half{\frac{1}{2}}
\def\thalf{\frac{3}{2}}
\def\third{\frac{1}{3}}
\def\tthird{\frac{2}{3}}

\def\Dslash{\not{\hbox{\kern-4pt $D$}}}
\def\dslash{\not{\hbox{\kern-2pt $\del$}}}



\def\Pl{{\mbox{\scriptsize Pl}}}
\def\eff{{\mbox{\scriptsize eff}}}
\def\CM{{\mbox{\scriptsize CM}}}
\def\BR{\mbox{\rm BR}}
\def\ee{e^+e^-}
\def\sstw{\sin^2\theta_w}
\def\cstw{\cos^2\theta_w}
\def\mz{m_Z}
\def\gz{\Gamma_Z}
\def\mw{m_W}
\def\mt{m_t}
\def\mb{m_b}
\def\gt{\Gamma_t}
\def\mh{m_h}
\def\gmu{G_\mu}
\def\GF{G_F}
\def\alphas{\alpha_s}
\def\msb{{\bar{\ssstyle M \kern -1pt S}}}
\def\lmsb{\Lambda_{\msb}}
\def\ELER{e^-_Le^+_R}
\def\EREL{e^-_Re^+_L}
\def\ELEL{e^-_Le^+_L}
\def\ERER{e^-_Re^+_R}
\def\eps{\epsilon}



\def\cnone{\widetilde \chi_1^0}
\def\ch#1{\widetilde\chi^+_{#1}}
\def\chm#1{\widetilde\chi^-_{#1}}
\def\neu#1{\widetilde\chi^0_{#1}}
\def\s#1{\widetilde{#1}}
\def\mssm{MSSM}


\interfootnotelinepenalty=10000

\makeatletter
\def\section{\@startsection{section}{0}{\z@}{5.5ex plus .5ex minus
 1.5ex}{2.3ex plus .2ex}{\large\bf}}
\def\subsection{\@startsection{subsection}{1}{\z@}{3.5ex plus .5ex minus
 1.5ex}{1.3ex plus .2ex}{\normalsize\bf}}
\def\subsubsection{\@startsection{subsubsection}{2}{\z@}{-3.5ex plus
-1ex minus  -.2ex}{2.3ex plus .2ex}{\normalsize\sl}}

\renewcommand{\@makecaption}[2]{%
   \vskip 10pt
   \setbox\@tempboxa\hbox{\small #1: #2}
   \ifdim \wd\@tempboxa >\hsize     
       \small #1: #2\par          
     \else                        
       \hbox to\hsize{\hfil\box\@tempboxa\hfil}
   \fi}

\makeatother


\def\BR{{\rm BR}}
\def\what{\widehat}
\def\anti{\overline}
\def\tbtb{t\anti b \, \anti t b}
\def\bbbb{b\anti b b\anti b}
\def\bb{b\anti{b}}
\def\anti{\overline}
\def\gam{\gamma}
\def\gev{~\mbox{GeV}}
\def\mev{~\mbox{MeV}}
\def\tev{~\mbox{TeV}}
\def\fbi{~\mbox{fb$^{-1}$}}
\def\abi{~\mbox{ab$^{-1}$}}
\def\epem{e^+e^-}
\def\rts{\sqrt s}
\def\lam{\lambda}
\def\sweff{\sin^2\theta_{\mathrm{eff}}}

\def\mpl{M_{\rm PL}}
\def\crr{\crcr\noalign{\vskip .1in}}
\def\fig#1{fig.~\ref{#1}}
\def\Fig#1{Fig.~\ref{#1}}
\def\figs#1#2{figs.~\ref{#1} and \ref{#2}}
\def\Figs#1#2{Figs.~\ref{#1} and \ref{#2}}
\def\Ref#1{\cite{#1}}
\def\Refs#1#2{\cite{#1} and \cite{#2}}
\def\Rref#1{Ref.~\cite{#1}}
\def\Rrefs#1#2{Refs.~\cite{#1} and \cite{#2}}
\def\sect#1{Section~\ref{#1}}
\def\Sect#1{Section~\ref{#1}}
\def\eq#1{eq.~(\ref{#1})}
\def\eqs#1#2{eqs.~(\ref{#1})--(\ref{#2})}
\def\Eq#1{Eq.~(\ref{#1})}
\def\Eqs#1#2{Eqs.~(\ref{#1})--(\ref{#2})}
\def\eqns#1#2{eqs.~(\ref{#1}) and (\ref{#2})}
\def\tanb{\tan\beta}
\def\sinb{\sin\beta}
\def\cosb{\cos\beta}
\def\sina{\sin\alpha}
\def\cosa{\cos\alpha}
\def\sinbma{\sin(\beta-\alpha)}
\def\cosbma{\cos(\beta-\alpha)}
\def\hsm{h_{\rm SM}}
\def\mhsm{m_{h_{\rm SM}}}
\def\h{h}
\def\mh{m_{\h}}
\def\hl{h^0}
\def\ha{A^0}
\def\hh{H^0}
\def\hpm{H^\pm}
\def\hp{H^+}
\def\hm{H^-}
\def\mha{m_{\ha}}
\def\mhl{m_{\hl}}
\def\mhh{m_{\hh}}
\def\mhpm{m_{\hpm}}
\def\mhmax{m_h^{\rm max}}
\def\mstopa{M_{\widetilde t_1}}
\def\mstopb{M_{\widetilde t_2}}
\def\msusy{M_{\rm SUSY}}
\def\mgl{m_{\tilde g}}
\def\SM{Standard Model}
\def\sm{SM}
\def\phm{\phantom{-}}
\def\ifmath#1{\relax\ifmmode #1\else $#1$\fi}
\def\ls#1{\ifmath{_{\lower1.5pt\hbox{$\scriptstyle #1$}}}}

\renewcommand{\thefootnote}{\arabic{footnote}}

\def\inondecoupling{\index{radiative corrections!non-decoupling effects}}
\def\ihnondecoup{\index{Higgs bosons!general two-Higgs-doublet model (2HDM)!special non-decoupling scenarios}}
\def\imajorana{\index{Higgs bosons!Majorana couplings}}
\def\ihiggsbounds{\index{Higgs bosons!mass bounds}}
\def\ilrmodels{\index{left-right symmetric models}}
\def\ihdmm{\index{Higgs bosons!triplets!doubly-charged member}}
\def\ihtheoryc{\index{Higgs bosons!theory contraints on masses and parameters}}
\def\icollparam{\index{collider/detector parameters, luminosities or features}}
\def\igigazhmssm{\index{Giga-$Z$!indirect constraints on MSSM particles}}
\def\ipngb{\index{pseudo-Nambu-Goldstone bosons}}
\def\ihnolose{\index{Higgs bosons!discovery parameter regmess}}
\def\ihddiff{\index{Higgs bosons!difficult parameter regimes}}
\def\ihdirect{\index{Higgs bosons!experimental limits}}
\def\ihmeas{\index{Higgs bosons!precision measurements}}
\def\ihextended{\index{extended or exotic Higgs scenarios}}
\def\ihsinglets{\index{Higgs bosons!singlets}}
\def\iradiative{\index{radiative corrections}}
\def\iradiativesusy{\index{radiative corrections!supersymmetric particle loops}}
\def\irge{\index{renormalization group equations}}
\def\iyuk{\index{Yuakwa couplings}}
\def\iperturbativity{\index{perturbativity constraints}}
\def\iunitarity{\index{unitarity constraints}}
\def\ihdiscovery{\index{Higgs bosons!discovery channels}}
\def\ihwidths{\index{Higgs bosons!widths}}
\def\ihprod{\index{Higgs bosons!production mechanisms}}
\def\ihcpviol{\index{CP-violating Higgs sectors}}
\def\ihmass{\index{Higgs bosons!mass(es)}}
\def\ihbrs{\index{Higgs bosons!branching ratios}}
\def\ihbrssusymod{\index{Higgs bosons!branching ratios!MSSM results}}
\def\ihcoups{\index{Higgs bosons!couplings}}
\def\ihself{\index{Higgs bosons!self-couplings}}
\def\itriviality{\index{triviality}}
\def\inat{\index{naturalness}}
\def\impl{\index{Planck mass ($\mpl$)}}
\def\ieft{\index{effective field theory}}
\def\ihcp{\index{Higgs bosons!CP properties}}
\def\ihspinparity{\index{Higgs bosons!spin and parity}}
\def\iewsb{\index{electroweak symmetry breaking (EWSB)}}
\def\ihtriplets{\index{Higgs bosons!triplets}}
\def\icoupu{\index{coupling constant unification}}
\def\itwohdm{\index{Higgs bosons!general two-Higgs-doublet model (2HDM)}}
\def\idecoup{\index{decoupling limit}}
\def\idecoupdelayed{\index{decoupling limit!delayed}}
\def\idecouppremature{\index{decoupling limit!premature}}
\def\igigaz{\index{Giga-$Z$}}
\def\ipew{\index{precision electroweak constraints}}
\def\igamc{\index{photon-photon collider}}


\begin{flushright}
UCD-02-18 \\
SCIPP 02/37 \\
IUHEX-202 \\
December, 2002 \\
hep-ph/0301023 \\
\end{flushright}
\vskip1cm
\begin{center}
{\LARGE\bf
Higgs Physics at the Linear Collider}\\[1cm]

{\large John F. Gunion$^*$, Howard E. Haber$^\dagger$
and Rick Van Kooten$^\ddagger$}\\[5pt]
{\it $^*$ Davis Institute for High Energy Physics \\
University of California, Davis, CA 95616, U.S.A.}\\[.3cm]
{\it $^\dagger$ Santa Cruz Institute for Particle Physics  \\
University of California, Santa Cruz, CA 95064, U.S.A.} \\[.3cm]
{\it $^\ddagger$ Department of Physics \\
Indiana University, Bloomington, IN 47405, U.S.A.}\\[1.5cm]

\thispagestyle{empty}

{\bf Abstract}\\[1pc]

\begin{minipage}{11cm}
We review the theory of Higgs bosons, with emphasis
on the Higgs scalars of the
Standard Model and its non-supersymmetric and supersymmetric extensions.
After surveying the expected knowledge of Higgs boson physics after 
the Tevatron and LHC experimental programs, we examine in detail
expectations for precision Higgs measurements at a future $e^+e^-$
linear collider (LC).  A comprehensive phenomenological profile can be
assembled from LC Higgs studies (both in $e^+e^-$ and $\gamma\gamma$
collisions).  The Giga-$Z$ option can provide important constraints and
consistency checks for the theory of electroweak symmetry breaking.

\end{minipage}  \\
\vskip1.5cm

{\large To appear in {\it Linear Collider Physics in the 
New Millennium}, edited by K.~Fujii, D.~Miller and A.~Soni
(World Scientific)}
\end{center}
\clearpage


\setcounter{page}{1}

\setcounter{chapter}{0}

\chapter{HIGGS PHYSICS AT THE LINEAR COLLIDER}

\markboth{J.F.~Gunion, H.E.~Haber and R.~Van Kooten}
{Higgs Physics at the Linear Collider}

\author{John F.~Gunion}

\address{Davis Institute for High Energy Physics \\
University of California, Davis, CA 95616, U.S.A. \\
E-mail: jfgucd@higgs.ucdavis.edu
}

\author{Howard E.~Haber}

\address{Santa Cruz Institute for Particle Physics  \\
University of California, Santa Cruz, CA 95064, U.S.A. \\
E-mail: haber@scipp.ucsc.edu
}

\author{Rick Van Kooten}

\address{Department of Physics \\
Indiana University, Bloomington, IN 47405, U.S.A. \\
E-mail: rickv@paoli.physics.indiana.edu}

\begin{abstract}
We review the theory of Higgs bosons, with emphasis
on the Higgs scalars of the
Standard Model and its non-supersymmetric and supersymmetric extensions.
After surveying the expected knowledge of Higgs boson physics after 
the Tevatron and LHC experimental programs, we examine in detail
expectations for precision Higgs measurements at a future $e^+e^-$
linear collider (LC).  A comprehensive phenomenological profile can be
assembled from LC Higgs studies (both in $e^+e^-$ and $\gamma\gamma$
collisions).  The Giga-$Z$ option can provide important constraints and
consistency checks for the theory of electroweak symmetry breaking.

\end{abstract}

\clearpage
\tableofcontents  

\section{Introduction}
\label{seca}

This chapter shows how an $e^+e^-$ linear collider (LC) 
can contribute to our
understanding of the Higgs sector through detailed studies
of the physical Higgs boson state(s).
Although this subject has been reviewed several times in the 
past~\cite{hhg,Gunion:1996cn,snow96,Murayama:1996ec,Accomando:1998wt,higgsreview},
there are at least two reasons to revisit the subject.
First, the completion of the LEP Higgs search, together with 
precise measurements from SLC, LEP, and the Tevatron, 
provides a clearer idea of what to expect.
The simplest interpretation of these results point to a light Higgs boson
with (nearly) standard couplings to $W$ and~$Z$.
The key properties of such a particle can be investigated
with a LC with a center-of-mass energy of $\sqrt{s}=500$~GeV.\icollparam\
Second, the luminosity expected from the LC is now higher:
200--300~fb$^{-1}$yr$^{-1}$ at $\sqrt{s}=500$~GeV, and
300--500~fb$^{-1}$yr$^{-1}$ at $\sqrt{s}=800$~GeV.
In particular, detailed designs have been developed for the LC
by the American/Asian
collaboration (the NLC/JLC design~\cite{nlc,jlc}) and by the 
DESY-based European
collaboration (the TESLA design~\cite{tesla}). 
At $\sqrt s=500$~GeV,
the nominal $10^7$ sec year integrated luminosities
are 220~fb$^{-1}$yr$^{-1}$
and 340~fb$^{-1}$yr$^{-1}$ for the NLC/JLC and TESLA designs, 
respectively~\cite{resourcebook}.\icollparam\
Consequently, several tens of thousands of Higgs bosons should be
produced in each year of operation.  With such samples, 
numerous precision Higgs measurements become feasible, and 
will provide fundamental insights into the properties of the Higgs
boson(s) and the underlying dynamics of electroweak symmetry breaking.
 
The $e^+e^-$ LC with center-of-mass energy $\sqrt{s}$ can also be
designed to operate as a photon-photon collider.\igamc\ This is
achieved by using Compton backscattered photons in the scattering of
intense laser photons on the initial polarized $e^+e^-$
beams~\cite{Ginzburg:1983vm,Ginzburg:1984yr}.  The resulting
$\gamma\gamma$ center of mass energy is peaked at about $0.8\sqrt{s}$
for the appropriate choices of machine parameters.  The luminosity achievable
as a function of the photon beam energy depends strongly on the
machine parameters (in particular, the choice 
of laser polarizations).\icollparam\
The photon collider provides additional opportunities for Higgs
physics~\cite{hggpheno,Gunion:1993ce,hjikia,mmelles,gunasner,velasco,Asner:2002aa}.

Finally,  we note that substantial improvements are possible 
for precision measurements of 
$m_W$, $m_t$ and electroweak mixing angle measurements 
at the LC~\cite{futureprecision}.\ipew\
But, the most significant improvements can be achieved at
the Giga-$Z$~\cite{gigaz},
where the LC operates at $\sqrt{s}=m_Z$ and
$\sqrt{s}\simeq 2m_W$.\igigaz\  With an
integrated luminosity of 50~fb$^{-1}$, one can collect
$1.5\times 10^9$ $Z$ events and about $10^6$ $W^+W^-$ pairs in the
threshold region.\icollparam\ Employing a global fit to the precision 
electroweak data in the
Standard Model, the anticipated fractional Higgs mass uncertainty
achievable would be about $8\%$.
This would provide a stringent test for the theory of the Higgs boson,
as well as very strong
constraints on any new physics beyond the Standard Model that couples
to the $W$ and $Z$ gauge bosons.

There is an enormous literature on the Higgs boson and, more generally,
on possible mechanisms of electroweak symmetry breaking.
It is impossible to review all theoretical approaches here.
To provide a manageable, but nevertheless illustrative, survey of the
LC capabilities, we focus mostly on the Higgs boson of the Standard
Model~(SM), and on the Higgs bosons of the minimal supersymmetric
extension of the Standard Model (MSSM).
Although this choice is partly motivated by simplicity, a stronger
impetus comes from the precision data collected over the past few years,
and some other related considerations.

The SM, which adds to the observed particles a single complex doublet 
of scalar fields, is economical.
It provides a good fit to the precision electroweak data.\ipew\
Many extended models 
of electroweak symmetry breaking possess a limit,
called the decoupling limit~\cite{decoupling},
that is experimentally almost
indistinguishable from the SM at low energies.\ihextended\idecoup\
These models agree with the data equally well, and even away from the
decoupling limit they usually predict a weakly coupled Higgs boson
whose mass is at most several hundred~GeV.
Thus, the SM serves as a basis for discussing the Higgs phenomenology of
a wide range of models, all of which are compatible with present experimental
constraints.

However, the SM suffers from several theoretical problems, which are either
absent or less severe with weak-scale supersymmetry.
The Higgs sector of the minimal supersymmetric extension of the Standard
Model (MSSM) is a constrained 
two-Higgs-doublet model (2HDM),
consisting of two CP-even Higgs bosons, $\hl$ and $\hh$, a CP-odd
Higgs boson, $\ha$, and a charged Higgs pair, $\hpm$.\itwohdm\ 
The MSSM is especially attractive because the superpartners modify
the running of the strong, weak, and electromagnetic gauge couplings in
just the right way as to yield unification  at about $10^{16}$~GeV
\cite{susyguts}.\icoupu\
For this reason, the MSSM is arguably the most compelling extension of
the SM.
This is directly relevant to Higgs phenomenology, because 
a theoretical bound requires the mass of the lightest CP-even Higgs
boson $\hl$ of the MSSM to be less than about 135~GeV 
(in most non-minimal supersymmetric models, $\mhl\lsim 200$~GeV),
as discussed in Section~\ref{seceb}.\ihtheoryc\
Furthermore, the MSSM offers, in some regions of parameter space,
very non-standard Higgs phenomenology, so the full range of 
possibilities in the MSSM
can be used to indicate how well the LC performs in non-standard
scenarios.
Thus, we use the SM to show how the LC fares when there is only one
observable Higgs boson, and the MSSM to illustrate how extra
fields can complicate the phenomenology.\ihddiff\
We also use various other models to illustrate important exceptions
to conclusions that would be drawn from the SM and MSSM alone.

The rest of this chapter is organized as follows.
Section~\ref{secb} presents, in some detail, the argument that one should
expect a weakly coupled Higgs boson with a mass that is probably
below about 200~GeV.
In Section~\ref{secc}, we summarize the theory of the SM Higgs boson.
In Section~\ref{secd}, we review the expectations for Higgs discovery and the
determination of Higgs boson properties at the Tevatron and LHC.
In Section~\ref{sece}, we introduce the Higgs sector of the MSSM
and discuss its theoretical properties.
The present direct search limits are reviewed, and expectations for
discovery at the Tevatron and LHC 
are described in Section~\ref{secf}.
In Section~\ref{secg}, we treat the theory of the non-minimal Higgs sector
more generally.
In particular, we focus on the decoupling limit, in which the properties
of the lightest Higgs scalar are nearly identical to those of the
Standard Model Higgs boson, and discuss how to distinguish the two.\idecoup\
We also discuss some non-decoupling exceptions to the usual decoupling
scenario.  In Section~\ref{seck}, we briefly discuss the 
case of a Higgs sector containing triplet Higgs representations
and also consider the Higgs-like particles that can arise if the
underlying assumption of a weakly coupled elementary Higgs sector is
{\it not} realized in Nature.\ihtriplets\

Finally, we turn to the program of Higgs measurements that can be
carried out at the LC,  focusing on $e^+e^-$ collisions at
higher energy, but also addressing
the impact of the Giga-$Z$ operation 
and $\gamma\gamma$ collisions.\igigaz\igamc\
This material summarizes and updates 
the results of the American Linear Collider
Working Group that were presented in \cite{resourcebook}.
The measurement of Higgs boson properties in $e^+e^-$
collisions is outlined in Section~\ref{sech}, and
includes a survey of the measurements that can be made for a
SM-like Higgs boson for all masses up to 500~GeV.
We also discuss measurements of the extra Higgs bosons that appear
in the MSSM.
Because the phenomenology of the decoupling limit mimics that of
the SM Higgs boson, we emphasize how the precision that stems from high
luminosity helps to diagnose the underlying dynamics.\idecoup\
In Section~\ref{seci}, we outline the impact of the Giga-$Z$ operation
on constraining and exploring various scenarios.\igigaz\
In Section~\ref{secj}, the most important gains from a 
photon-photon collider
are reviewed.\igamc\  We end this chapter with some brief
conclusions in Section~\ref{secl}.

\section{Expectations for electroweak symmetry breaking}
\label{secb}

With the recent completion of experimentation at the LEP collider,
the Standard Model of particle physics appears close to final
experimental verification.  After more than ten
years of precision measurements of
electroweak observables at LEP, SLC and the Tevatron, no 
definitive departures from Standard Model predictions have been found
\cite{precision}.\ipew\  In some cases, theoretical predictions have been
checked with an accuracy of one part in a thousand or better.
However, the dynamics responsible for electroweak symmetry breaking 
have not yet been directly identified.  Nevertheless, these dynamics  
affect predictions for currently observed electroweak processes at
the one-loop quantum level.  Consequently, the analysis of precision
electroweak data can already provide some useful constraints on the nature of
electroweak symmetry breaking dynamics.

In the Standard Model, electroweak symmetry
breaking dynamics arise via a
self-interacting complex doublet of scalar fields, which consists of four
real degrees of freedom.\iewsb\  Renormalizable interactions are
arranged in such a way that the neutral component of the scalar doublet
acquires a vacuum expectation value, $v=246$~GeV, which sets the scale
of electroweak symmetry breaking.
Hence, three massless Goldstone bosons 
are generated that are absorbed by the $W^\pm$ and $Z$, 
thereby providing the resulting massive gauge
bosons with longitudinal components.  
The fourth scalar degree of freedom remains in the physical spectrum, 
and it is the CP-even neutral Higgs boson 
of the Standard Model.
It is further assumed in the Standard Model that the scalar doublet also 
couples to fermions through Yukawa interactions.\iyuk\
After electroweak symmetry breaking, these interactions
are responsible for the generation of quark and charged lepton masses.

The global analysis of electroweak
observables provides a good fit to the Standard Model predictions.
Such analyses take the Higgs mass as a free parameter.  The 
electroweak observables depend logarithmically
on the Higgs mass through its one-loop effects.\iradiative\
The accuracy of the current data (and the reliability of the
corresponding theoretical computations) already provides a significant
constraint on the value of the Higgs mass. 
In \cite{precision},
the constraints of the global precision electroweak
analysis yield a one-sided 95\% CL upper limit of
$\mhsm\leq 193$~GeV at 95\% CL.\ipew\
Meanwhile, direct searches
for the Higgs mass at LEP achieved a 95\% CL limit of
$m_{h_{\rm SM}}>114.4$~GeV~\cite{LEPHiggs}.

One can question the significance of these results.  After all,
the self-interacting scalar field is only one  model of
electroweak symmetry breaking; other approaches,
based on different dynamics, are also possible.
For example, one can introduce new fermions and new forces, in
which the Goldstone bosons are a consequence of the strong
binding of the new fermion fields \cite{techni}.\iewsb\  Present
experimental data does not sufficiently constrain the
nature of the dynamics responsible for electroweak symmetry breaking.
Nevertheless, one can attempt to classify alternative scenarios and
study the constraints of the global precision electroweak fits and the
implications for phenomenology at future colliders.\ipew\  Since electroweak
symmetry dynamics must affect the one-loop corrections to
electroweak observables, constraints on alternative approaches can
be obtained by generalizing the global precision electroweak fits to
allow for new contributions at one-loop.  These enter primarily
through corrections to the self-energies of the gauge bosons (the
so-called ``oblique'' corrections).  Under the assumption that new
physics is characterized by a mass scale $M\gg m_Z$, one can
parameterize the leading oblique corrections by three constants, 
$S$, $T$, and $U$, first introduced by Peskin and 
Takeuchi~\cite{takeuchi}.\ipew\ In
almost all theories of electroweak symmetry breaking dynamics, $U\ll
S$, $T$, so it is sufficient to consider a global electroweak fit in
which $m_{h_{\rm SM}}$, $S$ and $T$ are free parameters.  (The zero of the
$S$--$T$ plane must be defined relative to some fixed value of the
Higgs mass, usually taken to be 100~GeV.)   New electroweak symmetry
breaking dynamics could generate non-zero values of $S$ and $T$, while
allowing for a much heavier Higgs mass (or equivalent).  Various
possibilities have been classified by 
Peskin and Wells \cite{peskinwells}, who
found that all models in the literature allowing a significantly heavier Higgs
boson also generate experimental signatures of new physics at the TeV scale.
At the LC, such new physics could be studied either directly by producing new
particles or indirectly by improving precision measurements of electroweak
observables.\ipew\

In this chapter, we mainly consider the case of a light weakly coupled
Higgs boson, corresponding to the simplest 
interpretation of the precision electroweak data.
Nevertheless, this still
does not fix the theory of electroweak symmetry breaking.  It is easy
to construct extensions of the scalar boson dynamics and generate
non-minimal Higgs sectors.  Such theories can contain charged Higgs
bosons and neutral Higgs bosons of opposite (or indefinite) CP-quantum
numbers.\ihcp\ Although some theoretical constraints exist,
there is still considerable freedom in constructing models which
satisfy all known experimental constraints.  Moreover, in most
extensions of the Standard Model, there exists a large range of
parameter space in which the properties of the lightest Higgs scalar 
are virtually indistinguishable from those of the Standard Model Higgs
boson.\idecoup\ Once
the Higgs boson is discovered, one of the challenges for experiments 
at future colliders is to detect deviations from the 
properties expected for the Standard Model Higgs boson, in order
to better constrain the underlying scalar dynamics.

Although the Standard Model provides a remarkably successful description of the
properties of the quarks, leptons and spin-1 gauge bosons at energy scales
of $\mathcal{O}(100)$~GeV and below, the Standard Model is not the
ultimate theory of the fundamental particles and their interactions.
At an energy scale above the Planck scale, $\mpl\simeq 10^{19}$~GeV, 
quantum gravitational effects become significant
and the Standard Model must
be replaced by a more fundamental theory that incorporates gravity.\impl\
Furthermore, on theoretical grounds it is very likely that the 
Standard Model breaks down at some energy scale, $\Lambda$.
In this case, the Standard Model degrees of freedom are no longer
adequate for describing the physics above $\Lambda$ and new physics
must enter. 
Thus, the Standard Model is not a {\it fundamental} theory;
at best, it is an {\it effective field theory}~\cite{EFT}.\ieft\
At energies below the scale~$\Lambda$,  the Standard Model (with
higher-dimension operators to parameterize the new physics at the scale
$\Lambda$) provides an extremely good description of all observable
phenomena.

To assess the potential of
future experiments, it is essential to ask how large 
$\Lambda$ can be.\ieft\
After all, if $\Lambda$ is as large as $\mpl$, it does not matter much
whether the Standard Model is a fundamental 
or an effective field theory.\impl\ 
The energy scale $\Lambda$ arises when analyzing the scale dependence 
of the Higgs self-coupling and the Higgs-top quark Yukawa coupling, so the 
value of $\Lambda$ depends on the Higgs mass $m_{h_{\rm SM}}$ and 
the top quark mass~$m_t$.\iyuk\ (Recall that
$m_{h_{\rm SM}}^2=\half\lambda v^2$ and $m_t=h_t v/\sqrt{2}$, 
where $\lambda$ is the Higgs quartic self-coupling and $h_t$ is the
Higgs-top quark Yukawa coupling.)\iyuk\
If $m_{h_{\rm SM}}$ is too large, then the perturbatively evolved
Higgs self-coupling $\lambda$ blows up at some scale
$\Lambda$ \cite{Dashen,hambye}.\ieft\
If $m_{h_{\rm SM}}$ is too small (compared to $m_t$), then the 
(perturbative) Higgs potential develops
a second (global) minimum at a large value of the scalar field of order
$\Lambda$ \cite{quiros}.  
Thus, new physics must enter at or below $\Lambda$,
in order that the true minimum of the theory correspond to the
observed SU(2)$\times$U(1) broken vacuum with $v=246$~GeV for scales
above~$\Lambda$.\ihtheoryc\

The arguments just given are based on perturbation theory so, taken on 
their own, are not completely persuasive.
Once the couplings become strong, it is \emph{a priori} conceivable that
a novel, non-perturbative scaling behavior would rescue the theory.
This possibility has been checked, mostly in an approximation 
neglecting the gauge couplings~\cite{Dashen}.
Let us consider first the case of large Higgs mass (large $\lambda$), 
so the Yukawa coupling can be neglected.
By a number of methods, particularly numerical lattice
methods~\cite{npHiggs,finalNeuberger} and strong coupling
expansions~\cite{LuscherWeisz}, one finds that non-perturbative
renormalization effects rapidly bring the self-coupling back into the
perturbative regime.\itriviality\
This is the so-called triviality phenomenon: if one starts with 
finite $\Lambda$ and tries to take the limit $\Lambda\to\infty$,
one ends up with a renormalized self-coupling $\lambda\to 0$.
A similar phenomenon occurs for large Yukawa couplings.\iyuk\
In these studies \cite{npYukawa}, the vacuum is never unstable 
(non-perturbatively).
Instead, strong renormalization effects drive the Yukawa coupling 
back into the perturbative regime.\ihtheoryc\

The non-perturbative studies very much favor a finite $\Lambda$, 
despite exhaustive attempts~\cite{finalNeuberger} to find non-trivial
behavior for $\Lambda\to\infty$.\ieft\
They also justify the use of perturbation theory, because the 
renormalized self- and Yukawa couplings end up being small enough at 
scales below $\Lambda$.\iperturbativity\
Thus, with the perturbative analysis and a given value of
$\Lambda$, one can compute the minimum and maximum Higgs mass allowed.
The results of such an analysis (with shaded bands indicating the
theoretical uncertainty of the result) are illustrated in
\fig{trivial} \cite{hambye,Riesselmann}.\ihtheoryc\
\begin{figure}
\begin{center}
\includegraphics*[width=9cm]{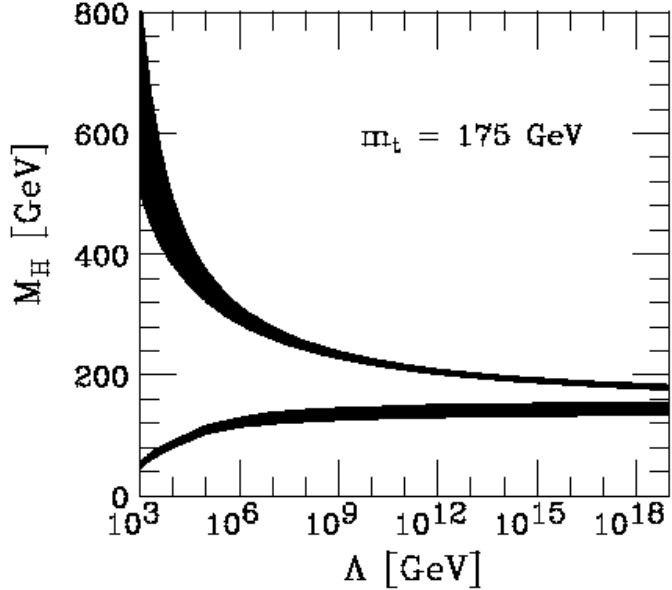}
\end{center}
  \caption[0]{\label{trivial} \small
The upper \protect\cite{Dashen,hambye} and the
lower \protect\cite{quiros} Higgs mass bounds
as a function of the
energy scale $\Lambda$ at which the Standard Model breaks down,
assuming $m_t=175$~GeV and $\alpha_s(m_Z)=0.118$.\ihiggsbounds\  The shaded
areas above reflect the theoretical uncertainties in the
calculations of the Higgs mass bounds
From~\protect\cite{Riesselmann}.\ieft\itriviality\ihtheoryc}
\end{figure}
From the upper bound one sees that if $\Lambda\sim 10^3~{\rm GeV}=1$~TeV,
then $m_{h_{\rm SM}}\lsim 800$~GeV.
If $\Lambda$ is more remote, say $\Lambda\gsim 10^6$~GeV, then 
$m_{h_{\rm SM}}\lsim 300$~GeV.\ieft\ihtheoryc\

In the Higgs mass range 130~GeV~$\lsim m_{h_{\rm SM}}\lsim 180$~GeV,
it seems that an effective Standard Model
could survive all the way to the Planck scale.\footnote{The
constraint on $\Lambda$ due to vacuum stability in \fig{trivial}
can be relaxed somewhat if
one allows for the electroweak vacuum
to be metastable, with a lifetime greater than the age of the universe.
The analysis of \cite{Isidori:2001bm} finds that for a
sufficiently long-lived electroweak vacuum, the Higgs mass
lower limit of 130~GeV just quoted is reduced to about 115~GeV.}\impl\ieft\
However, such a possibility seems unlikely, based on the following
``naturalness'' argument \cite{naturally}.\inat\
In an effective field theory, 
masses and dimensionless couplings are calculable in terms of parameters
of a more fundamental theory that
describes physics at the energy scale $\Lambda$.  All
low-energy couplings and fermion masses are logarithmically sensitive to
$\Lambda$.  In contrast, scalar squared-masses are {\it quadratically}
sensitive to $\Lambda$.  The Higgs
mass (at one-loop) has the following heuristic form:\inat\
\begin{equation} \label{natural}
m_h^2= (m_h^2)_0+\frac{cg^2}{16\pi^2}\Lambda^2\,,
\end{equation}
where $(m_h^2)_0$ is a parameter of the fundamental theory and $c$ is a
constant, presumably of $\mathcal{O}(1)$, that depends on the physics near
the scale~$\Lambda$.  The ``natural'' value
for the scalar squared-mass is $g^2\Lambda^2/16\pi^2$.  Thus, the expectation
for $\Lambda$ is
\begin{equation} \label{tevscale}
\Lambda\simeq \frac{4\pi m_h}{g}\sim \mathcal{O}(1~{\rm TeV})\,.
\end{equation}
If $\Lambda$ is significantly larger than 1~TeV 
then the only way
for the Higgs mass to be of order the scale of electroweak symmetry
breaking is to have an ``unnatural'' cancellation between the two terms
of \eq{natural}.
This seems highly unlikely given that the two terms of
\eq{natural} have completely different origins.  

An attractive theoretical framework that incorporates weakly coupled Higgs
bosons and satisfies the constraint of \eq{tevscale} is that of
``low-energy'' or ``weak-scale'' 
supersymmetry \cite{Nilles84,Haber85}.\inat\
In this framework, supersymmetry is
used to relate fermion and boson masses and interaction strengths.  
Since fermion masses are only
logarithmically sensitive to $\Lambda$, boson masses will exhibit the
same logarithmic sensitivity if supersymmetry is exact.  Since no
supersymmetric partners of Standard Model particles have yet been
found, supersymmetry cannot be an exact symmetry of nature.
Thus, $\Lambda$ should be identified
with the supersymmetry breaking scale.\ieft\  The
naturalness constraint of \eq{tevscale} is still relevant.   It
implies that the scale of supersymmetry
breaking should not be much larger than 1~TeV, to preserve the
naturalness of scalar masses.\ieft\inat\
The supersymmetric extension
of the Standard Model would then replace the Standard Model as the
effective field theory of the TeV scale.  

One advantage of the supersymmetric
approach is that the effective low-energy supersymmetric theory {\it
can} be valid all the way up to the Planck scale, while still being
natural.  The unification of the three gauge couplings at an energy
scale close to the Planck scale, which does not occur in the Standard
Model, is seen to occur in the minimal supersymmetric extension of the
Standard Model~\cite{susyguts}, 
and provides an additional motivation for seriously 
considering the low-energy supersymmetric framework.\icoupu\
However, the fundamental origin of supersymmetry breaking
is not known at present.  Without a fundamental theory of
supersymmetry breaking, one ends up with an effective low-energy
theory characterized by over 100 unknown parameters that in principle
would have to be measured by experiment.\ieft\
This remains one of the main stumbling
blocks for creating a truly predictive 
supersymmetric model of fundamental particles
and their interactions.  Nevertheless, the Higgs sectors of the
simplest supersymmetric models are quite strongly constrained and
exhibit very specific phenomenological profiles.

\section{The Standard Model Higgs boson---theory}
\label{secc}

In the Standard Model, the Higgs mass,
$m_{h_{\rm SM}}^2=\half\lambda v^2$, depends on the Higgs self-coupling
$\lambda$.\ihself\ Since $\lambda$ is unknown at present, the value of
the SM Higgs mass is not predicted, although other
theoretical considerations (discussed in Section~\ref{secb}) place
constraints on the Higgs mass, as exhibited 
in \fig{trivial}.\itriviality\ The
Higgs couplings to fermions [gauge bosons] are proportional to
the corresponding particle masses [squared-masses].\ihcoups\ 
As a result, Higgs
phenomenology is governed primarily by the couplings of the Higgs
boson to the $W^\pm$ and  $Z$ and the third-generation quarks and
leptons. Two-loop-induced Higgs couplings are also 
phenomenologically important:
the $h_{\rm SM} gg$  coupling ($g$ is the gluon), which 
is induced by the one-loop graph in which the Higgs boson
couples to a virtual $t\bar t$ pair, and the
$h_{\rm SM}\gamma\gamma$ coupling, for which
the $W^+W^-$ loop contribution is dominant. Further details of the
SM Higgs boson properties are given in \cite{hhg}.

\subsection{Standard Model Higgs boson decay modes}
\label{secca}

\begin{figure}[b!]
\begin{center}
\includegraphics*[width=0.75\textwidth]{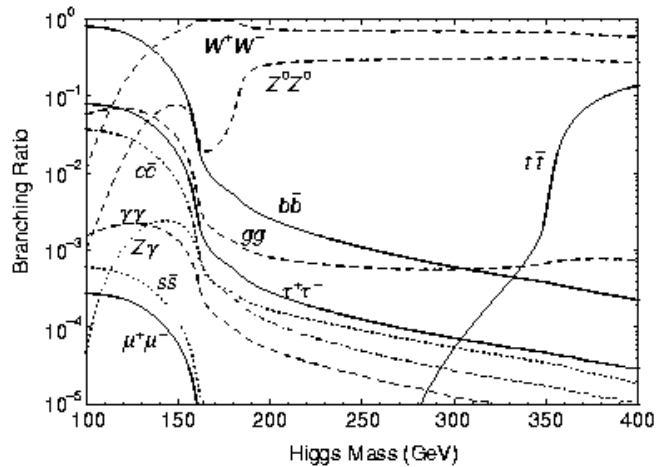}
\end{center}
\vskip -0.3cm
\caption[0]{\label{fg:1}  \small Branching ratios of the dominant decay
modes of the \SM\ Higgs boson.  These results
have been obtained with the program HDECAY \protect\cite{hdecay}, and include
QCD corrections beyond the leading order.\ihbrs\
}
\end{figure}

The Higgs boson mass is the only unknown parameter in the Standard
Model. Thus, one can compute Higgs boson branching ratios and
production cross sections as a function of $m_{h_{\rm SM}}$.\ihmass\ 
The branching
ratios for the dominant decay modes of a Standard Model Higgs
boson are shown as a function of Higgs boson mass in \fig{fg:1}.\ihbrs\
Note that subdominant channels are important to establish a complete
phenomenological profile of the Higgs boson, and to check consistency
(or look for departures from) Standard Model predictions.
For $115~{\rm GeV}\sim m_{h_{\rm SM}}\lsim 2\mw$ many decays modes are large
enough to measure, as discussed in Section~\ref{sech}.

For $m_{h_{\rm SM}}\lsim 135$~GeV, the main Higgs decay mode is
$h_{\rm SM}\to b\bar b$, while the decays $h_{\rm SM}\to \tau^+\tau^-$ and
$c\bar c$ can also be phenomenologically relevant. In addition,
although one--loop suppressed, the decay $h_{\rm SM}\to gg$ is
competitive with other decays for $m_{h_{\rm SM}}\lsim 2\mw$ because of the
large top Yukawa coupling and the color factor.
As the Higgs mass increases above 135~GeV, the branching ratio to
vector boson pairs becomes dominant. In particular, the main Higgs
decay mode is $h_{\rm SM}\to WW^{(*)}$, where one of the $W$'s must be
off-shell (indicated by the star superscript) if $m_{h_{\rm SM}}<2\mw$. For
Higgs bosons with $m_{h_{\rm SM}}\gsim 2m_t$, the decay $h_{\rm SM}\to t\bar t$
begins to increase until it reaches its maximal value of about 20\%.\ihbrs\

Rare Higgs decay modes can also play an important role.\ihbrs\ The
one-loop decay $h_{\rm SM}\to\gamma\gamma$ is a  suppressed
mode.   For
$\mw\lsim m_{h_{\rm SM}}\lsim 2\mw$, ${\rm BR}(h_{\rm SM}\to\gamma\gamma)$ is
above $10^{-3}$.  This decay channel provides an important Higgs
discovery mode at the LHC for $100~{\rm GeV}\lsim m_{h_{\rm SM}}\lsim
150$~GeV.  At the LC, the direct observation of
$h_{\rm SM}\to\gamma\gamma$ is also possible in $Zh_{\rm SM}$
production despite its suppressed branching
ratio.  The partial width $\Gamma(\hl\to\gamma\gamma)$ is also
significant in that it
governs the Higgs production rate at a $\gamma\gamma$ collider.\igamc\

\subsection{Standard Model Higgs boson production at the LC}
\label{seccb}

In the Standard Model there are two main processes to produce
the Higgs boson in $e^+e^-$ annihilation.\ihprod\
These processes are also relevant in many extensions of the Standard
Model, particularly near the decoupling limit, in which the
lightest CP-even Higgs boson possesses properties nearly identical to
those of the SM Higgs boson.  
In the  ``Higgsstrahlung'' process, a virtual $Z$~boson
decays to an on-shell~$Z$ and the $h_{\rm SM}$, as
depicted in \fig{higgs:fig:HZdiagram}(a).
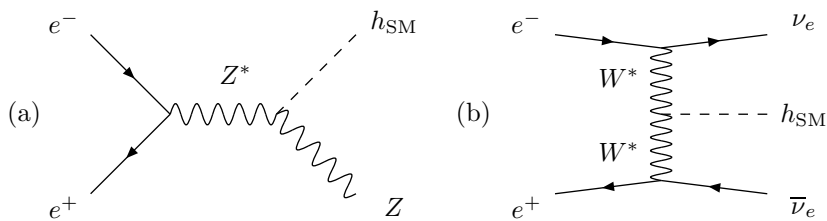
\begin{figure}[h!]
\begin{center}
\begin{picture}(375,90)(0,0)
\SetOffset(-10,0)
\ArrowLine(45,75)(75,45)
\ArrowLine(75,45)(45,15)
\Photon(75,45)(115,45){4}{5}
\Photon(115,45)(145,15){4}{5}
\DashLine(115,45)(145,75){4}
\ArrowLine(220,75)(260,70)
\ArrowLine(260,70)(300,75)
\ArrowLine(260,20)(220,15)
\ArrowLine(300,15)(260,20)
\Photon(260,70)(260,45){4}{5}
\Photon(260,20)(260,45){4}{5}
\DashLine(260,45)(300,45){4}
\Text(20,45)[]{(a)}
\Text(35,80)[]{$e^-$}
\Text(35,10)[]{$e^+$}
\Text(100,60)[]{$Z^*$}
\Text(160,80)[]{$h_{\rm SM}$}
\Text(160,10)[]{$Z$}
\Text(190,45)[]{(b)}
\Text(210,80)[]{$e^-$}
\Text(210,10)[]{$e^+$}
\Text(245,59)[]{$W^*$}
\Text(245,31)[]{$W^*$}
\Text(315,80)[]{$\nu_e$}
\Text(315,10)[]{$\overline\nu_e$}
\Text(315,45)[]{$h_{\rm SM}$}
\end{picture}
\end{center}
\caption[0]{\label{higgs:fig:HZdiagram}\small
Main production processes for Higgs
production in $e^+e^-$ annihilation. (a)~Higgsstrahlung.
(b)~$WW$ fusion.\ihprod}
\end{figure}

\begin{figure}[t!]
\begin{center}
\includegraphics*[width=0.65\textwidth,angle=-90]{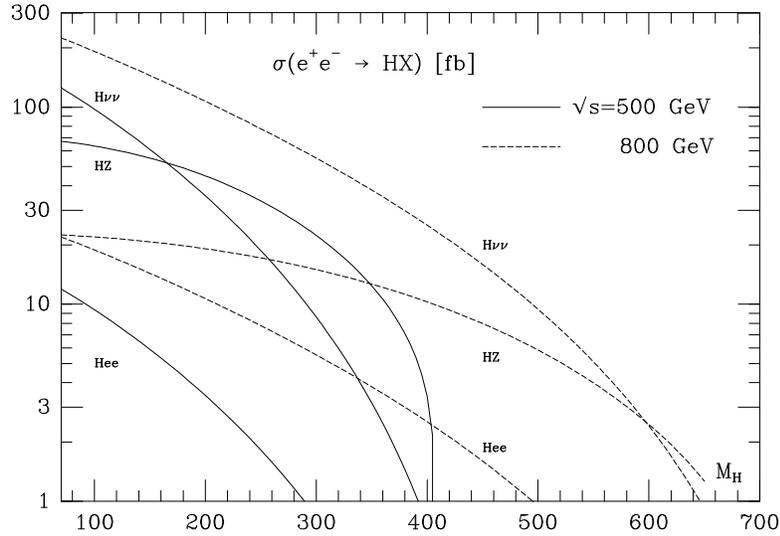}
\end{center}
\caption[0]{\label{smhiggsxsec} Cross sections for Higgsstrahlung 
($e^+e^-\to Zh_{\rm SM}$)
and Higgs production via $W^+W^-$ fusion ($e^+e^-\to \nu\bar\nu h_{\rm
SM}$) and $ZZ$ fusion ($e^+e^-\to e^+e^-h_{\rm SM}$) as a function of
$m_{h_{\rm SM}}$ for two center-of-mass energies, $\sqrt{s}=500$ and
800~GeV \protect\cite{Accomando:1998wt}.\ihprod}
\end{figure}

The cross section for Higgsstrahlung rises sharply at threshold to a
maximum a few tens of GeV above $m_h+m_Z$, and then falls off as $s^{-1}$,
as shown in \fig{smhiggsxsec}.\ihprod\
The associated production of the $Z$ provides an important trigger 
for Higgsstrahlung events.  In particular, in some theories beyond the
Standard Model, the Higgs boson decays into invisible modes, in which case
the ability to reconstruct the Higgs boson mass peak in 
the spectrum of the missing mass recoiling against the~$Z$ 
will be crucial.
The other production process is called ``vector boson fusion'', where
the incoming $e^+$ and $e^-$ each emit a virtual vector boson,
followed by vector boson fusion to the $h_{\rm SM}$. 
\Fig{higgs:fig:HZdiagram}(b) depicts the $W^+W^-$ fusion process.
Similarly, the $ZZ$ fusion process corresponds to $e^+e^-\to e^+e^-h_{\rm SM}$.
In contrast to Higgsstrahlung, the vector boson fusion cross section
grows as $\ln s$, and thus
is the dominant Higgs production mechanism for 
$\sqrt{s}\gg m_{h_{\rm SM}}$.
The cross section for $WW$ fusion is about ten times larger than that
for $ZZ$ fusion.  Nevertheless, the latter provides complementary
information on the $ZZh_{\rm SM}$ vertex.   Note that at an $e^-e^-$
collider, the Higgsstrahlung and $W^+W^-$ fusion processes are absent,
so that $ZZ$ fusion is the dominant Higgs production process~\cite{ememhan}.

\begin{figure}[thb]
\begin{center}
\includegraphics*[width=0.75\textwidth]{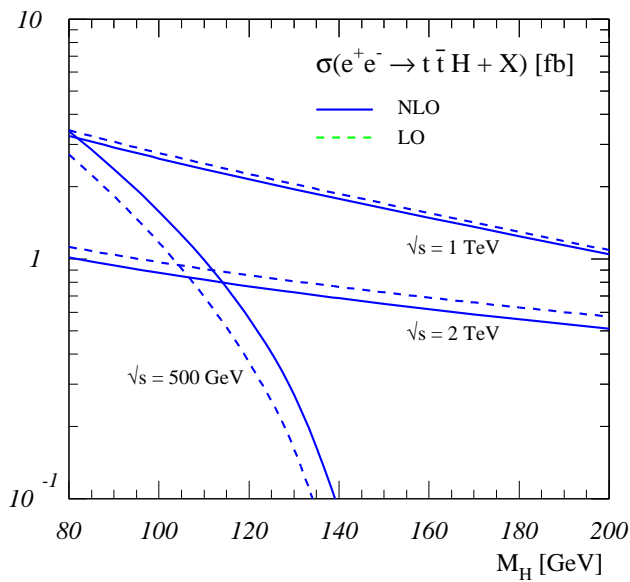}
\end{center}
\caption[0]{\label{ttbarhiggs} \small
Cross-sections for $e^+e^-\to t\bar th_{\rm SM}$ in fb for three
choices of center-of-mass energy.  The dashed lines correspond to the
tree-level result \protect\cite{Djouadi:1992tk}, and the solid lines include
the next-to-leading order QCD corrections \protect\cite{Dittmaier:1998dz}.}
\end{figure}

Other relevant processes for producing Higgs bosons are associated
production with a fermion-antifermion pair,
and multi-Higgs production.\ihprod\  Among production mechanisms of
the former class, only $e^+e^-\to t\bar{t}h_{\rm SM}$ has a
significant cross section, around the femtobarn level in the Standard
Model, as depicted in \fig{ttbarhiggs}.
As a result, if $m_{h_{\rm SM}}$ is small enough (or $\sqrt{s}$ is large
enough), this process can be used for determining
the Higgs--top quark Yukawa coupling.\iyuk\  
The cross section for double Higgs production 
($e^+e^-\to Zh_{\rm SM}h_{\rm SM}$) is even smaller, of
order 0.1~fb for $100~{\rm GeV}\lsim m_{h_{\rm SM}}\lsim 150$~GeV and 
$\sqrt{s}$ ranging between 500~GeV and 1~TeV.
With sufficient luminosity, the latter can be used
for extracting the triple Higgs self-coupling.

At the $\gamma\gamma$ collider, a Higgs boson is produced 
as an $s$-channel resonance via the one-loop 
triangle diagram.\igamc\ihprod\
Every charged particle whose mass is generated  by
the Higgs boson contributes to this process.
In the Standard Model, the main contributors are the $W^\pm$ 
and the $t$-quark loops.  See Section~\ref{secj} for further discussion.

\section{SM Higgs searches before the linear collider}
\label{secd}

\subsection{Direct search limits from LEP}
\label{secda}

The LEP collider completed its final run in 2000, and presented
tantalizing hints for the possible observation of the Higgs boson.  
The combined data from all four LEP collaborations
exhibits
a slight preference for the signal plus background hypothesis
for a Standard Model Higgs boson mass in the vicinity of 116~GeV, as
compared to the background only hypothesis~\cite{LEPHiggs}.
However, the excess over the expected background was
not sufficiently significant to support a claim of discovery
or even an ``observation'' of evidence for the Higgs boson.  
A more conservative interpretation of the
data yields a 95\%~CL lower limit of $m_{h_{\rm SM}}>114.4$~GeV.\ihdirect\

\subsection{Implications of precision electroweak measurements}
\label{decdb}

Indirect constraints on the Higgs boson mass within the SM can be
obtained from confronting the SM predictions with results of electroweak
precision measurements.\ipew\ In the case of the top quark mass,
the indirect determination turned out to be in remarkable agreement
with the actual experimental value~\cite{precision}. In comparison,  
to obtain constraints on $m_{h_{\rm SM}}$ of similar precision,
much higher accuracy is required for both the 
experimental results and the theory predictions.
This is due to the fact that
the leading dependence of the precision observables on $m_{h_{\rm SM}}$ is only
logarithmic, while the dominant effects of the top-quark mass enter
quadratically. 

\begin{figure}[t!]
\begin{center}
\includegraphics*[width=0.7\textwidth]{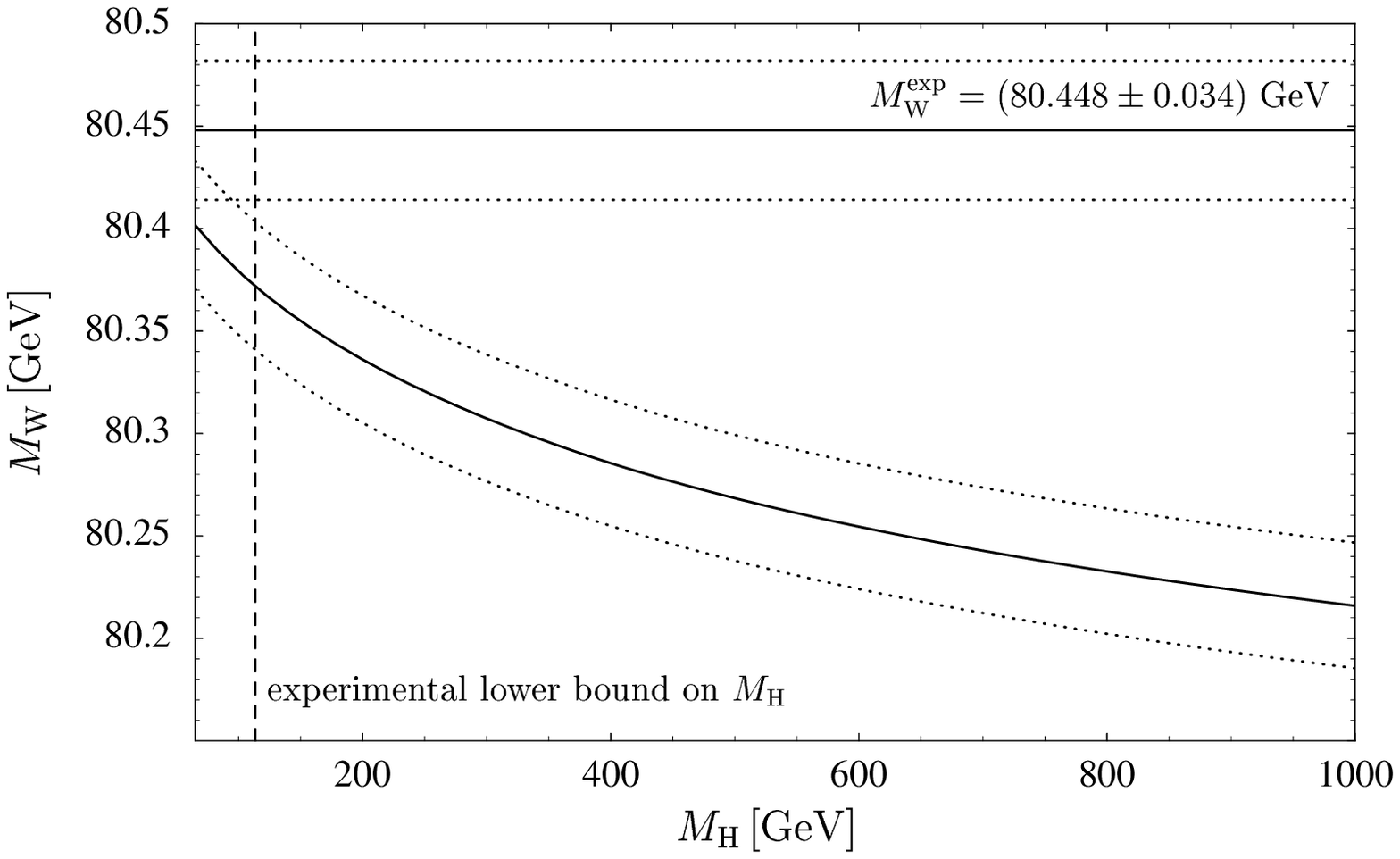}
\includegraphics*[width=0.7\textwidth]{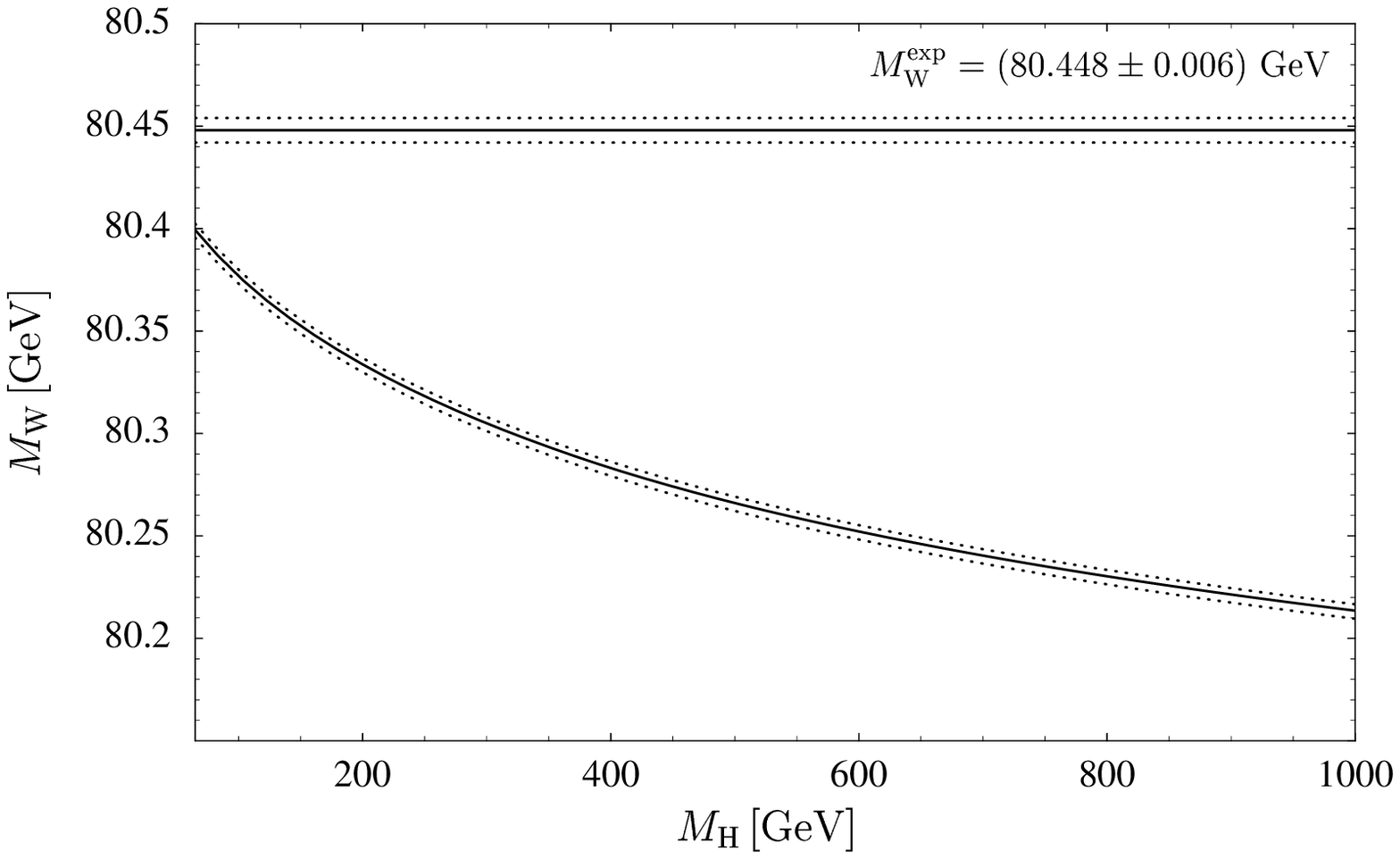}
\end{center}
\caption[]{\label{fig:mw2l} \small
The prediction for $m_W$ as a function of $m_{h_{\rm SM}}$ is compared
with the experimental value of $m_W$ (a) for the current experimental
accuracies of $m_W$ and $\mt$, and (b) for the prospective
future accuracies at the LC with the Giga-$Z$ option and a precise
$W$ threshold scan, assuming the present
experimental central values~\cite{deltaral2}.
The present experimental 95\% CL\ lower bound on the
Higgs-boson mass, $m_{h_{\rm SM}} = 114.4$~GeV, is indicated in (a) by
the vertical dashed line.\ipew\
}
\end{figure}

\Fig{fig:mw2l}(a) shows the currently most precise
result for $m_W$ as function of $m_{h_{\rm SM}}$ in the SM, 
and compares it with the present experimental value of $m_W$.  The
calculation incorporates 
the complete electroweak fermion-loop contributions at 
$\mathcal{O}(\alpha^2)$~\cite{deltaral2}. 
Based on this result, the remaining theoretical
uncertainty from unknown higher-order corrections has been estimated to
be about $6 \mev$~\cite{deltaral2}. This is about a factor five smaller than
the uncertainty induced by the current experimental error on the
top-quark mass, $\Delta\mt^{\rm exp} = \pm 5.1 \gev$, 
which presently dominates 
the theoretical uncertainty. \Fig{fig:mw2l}(b) shows the
prospective situation at a future $e^+e^-$ linear 
collider after the Giga-$Z$ operation and a threshold measurement
of the $W$ mass (keeping the
present experimental central values for simplicity), which are expected to
reduce the experimental errors to $\Delta m_W^{\rm exp} = 6 \mev$
and $\Delta\mt^{\rm exp} = 200 \mev$. 
The plot clearly shows a considerable 
improvement in the sensitivity to $m_{h_{\rm SM}}$ 
achievable at the LC via very precise
measurements of $m_W$ and $\mt$. Since furthermore the experimental
error of $\sweff$ is expected to be reduced by almost a factor of 20 at
Giga-$Z$, the accuracy in the indirect determination of the Higgs-boson
mass from all data will improve by about a factor of 10 compared to the
present situation~\cite{gigaz}.\igigaz\ipew\

\subsection{Expectations for Tevatron searches}
\label{secdc}

The upgraded Tevatron began taking data in the spring of 2001.  This
is the only collider at which the Higgs boson can be produced 
until the LHC begins operation in 2007.
The Tevatron Higgs Working Group~\cite{tevreport}
presented a detailed analysis of the
Higgs discovery reach at the upgraded Tevatron. 
Here, we summarize the main results.  Two Higgs mass ranges were
considered separately: (i)~100~GeV$\lsim m_{h_{\rm SM}}\lsim 135$~GeV 
and (ii)~135~GeV$\lsim m_{h_{\rm SM}}\lsim 190$~GeV, 
corresponding to the two different
dominant Higgs decay modes: $h_{\rm SM}\to b\bar b$ for the lighter mass
range and $h_{\rm SM}\to WW^{(*)}$ for the heavier mass range.

In mass range (i), the relevant production mechanisms are $q_i\bar
q_j\to Vh_{\rm SM}$, where $V=W$ or $Z$. In all cases, the dominant
$h_{\rm SM}\to b\bar b$ decay was employed.\ihprod\ihdiscovery\ 
The most relevant final-state signatures
correspond to events in which the vector boson decays leptonically
($W\to\ell\nu$, $Z\to\ell^+\ell^-$ and $Z\to\nu\bar\nu$, where
$\ell=e$ or $\mu$), resulting in $\ell\nu b\bar b$, $\nu\bar\nu b\bar
b$ and $\ell^+\ell^- b\bar b$ final states.  In mass range (ii), the
relevant production mechanisms include 
$gg\to h_{\rm SM}$, $V^* V^*\to h_{\rm SM}$ 
and $q_i\bar q_j\to Vh_{\rm SM}$, with decays $h_{\rm SM}\to WW^{(*)}$,
$ZZ^{(*)}$.\ihbrs\
The most relevant phenomenological signals are those in which two of
the final-state vector bosons decay leptonically, resulting in
$\ell^+\ell^-\nu\bar\nu$ or $\ell^\pm\ell^\pm jjX$, where $j$ is a
hadronic jet and $X$ consists of two additional leptons (either
charged or neutral).  For example, the latter can arise from $Wh_{\rm SM}$
production followed by $h_{\rm SM}\to WW^{(*)}$, where the two like-sign $W$
bosons decay leptonically, and the third $W$ decays into hadronic
jets.  In this case $X$ is a pair of neutrinos.

\begin{figure}[t!]
  \begin{center}
\includegraphics*[width=\textwidth]{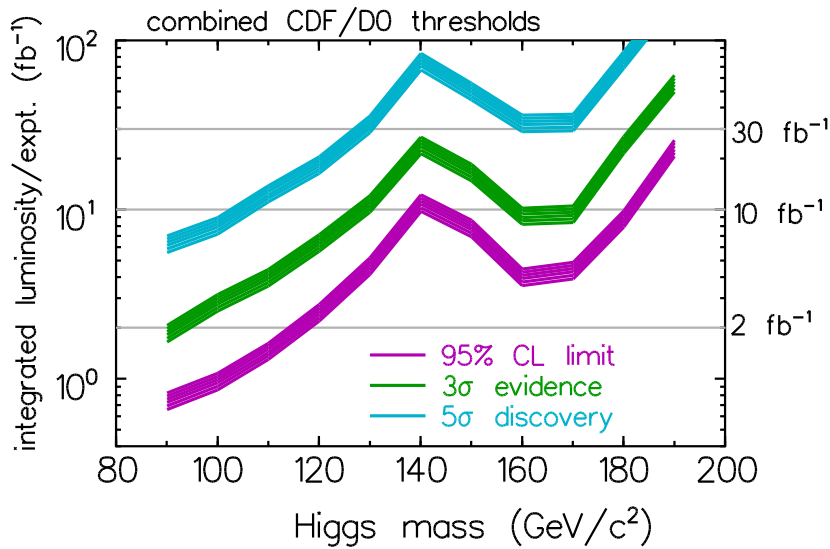}
  \end{center}
\caption[0]{  \label{f:combined-final} \small
The integrated luminosity required per experiment, to
            either exclude a SM Higgs boson at 95\% CL or discover it at the
            $3\sigma$ or $5\sigma$ level, as a function of the Higgs mass.  
            These results are based on the combined
            statistical power of both experiments.  The curves shown  
            are obtained by combining the $\ell\nu b\bar b$,
            $\nu\bar\nu b\bar b$ and $\ell^+\ell^-b\bar b$ channels
            using the neural network selection in the low-mass Higgs region
            ($90~{\rm GeV}~\lsim m_{h_{\rm SM}}\lsim 135$~GeV), and the
            $\ell^\pm\ell^\pm jjX$ and $\ell^+\ell^-\nu\bar\nu$ 
            channels in the high-mass Higgs region
            ($135~{\rm GeV}~\lsim m_{h_{\rm SM}}\lsim 190$~GeV).  
            The lower edge of 
            the bands is the calculated threshold; the bands extend upward 
            from these nominal thresholds by 30\% as an indication of the
            uncertainties in $b$-tagging efficiency, background rate,
            mass resolution, and other effects.  
            Taken from \protect\cite{tevreport}.}
\end{figure}

\Fig{f:combined-final} summarizes 
the Higgs discovery reach versus the total integrated luminosity
delivered to the Tevatron (and by assumption, delivered to each
detector).
As the plot shows, the required integrated luminosity increases
rapidly with Higgs mass to 140 GeV, beyond which the high-mass
channels play the dominant role.  With 2~fb$^{-1}$ per detector,
the 95\% CL limits will barely extend the
expected LEP limits, but with 10 fb$^{-1}$, the SM Higgs boson can be
excluded up to 180 GeV if the Higgs boson
does not exist in that mass range.\ihdiscovery\

With further machine improvements, the Tevatron may provide a total
integrated luminosity as high as $15~{\rm fb}^{-1}$ before the LHC
starts to yield significant results.  If such an integrated 
luminosity could be reached and if $m_{h_{\rm SM}}\simeq 115$~GeV, 
then the Tevatron experiments will be able to achieve a $5\sigma$
discovery of the Higgs boson.  If no Higgs events are detected, the LEP
limits will be significantly extended, with a 95\%~CL exclusion
possible up to about $m_{h_{\rm SM}}\simeq 185$~GeV.\footnote{The 
Higgs mass region around
140 GeV might require more luminosity, depending on the
magnitude of systematic errors due to uncertainties in $b$-tagging
efficiency, background rate, the $b\bar b$ mass resolution, {\it etc.}}  
Moreover, evidence for a
Higgs boson at the $3\sigma$ level could be achieved up to about
$m_{h_{\rm SM}}\simeq 175$~GeV.\ihdiscovery\

Evidence for or discovery of a Higgs boson at
the Tevatron would be a landmark in high energy physics.
However, even if a Higgs boson is seen, the Tevatron 
data would only provide a very rough phenomenological profile of the
Higgs boson.
In contrast, the LHC and the LC (the latter with greater precision)
could measure enough of its properties to 
verify that the interactions of the Higgs boson 
provide the dynamics responsible for the generation of mass for the
vector bosons, quarks and charged leptons.

\subsection{Expectations for LHC searches}
\label{secdd}

\begin{figure}[t!]
\begin{center}
\includegraphics*[width=0.84\textwidth]{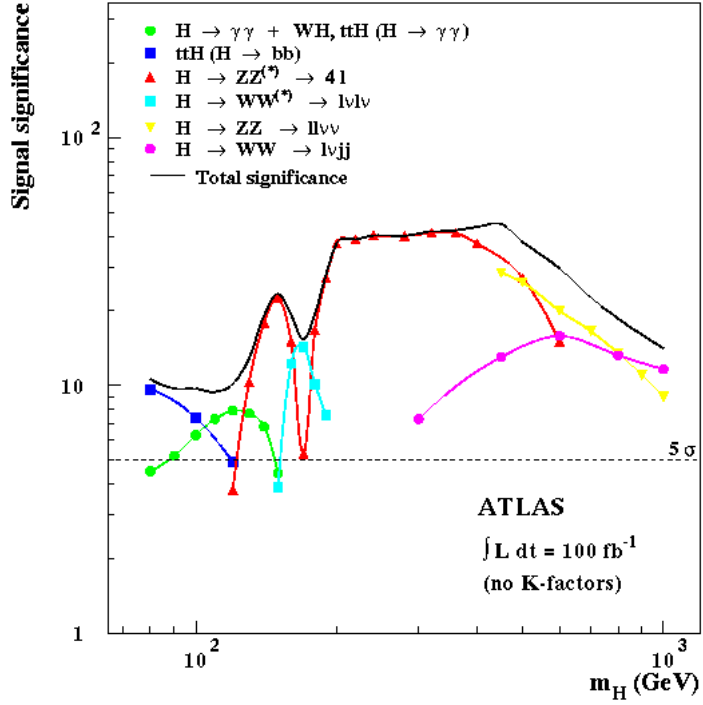}
\end{center}
\caption[0]{\label{f:atlashsm} \small
Statistical significance levels as a function of the
Higgs mass for the ATLAS experiment at the LHC,
assuming an integrated luminosity of
100~fb$^{-1}$.  Taken from \protect\cite{atlashsmref}.\ihdiscovery}
\end{figure}

At the LHC, the ATLAS and CMS detectors have been specifically
designed to guarantee observation of a SM Higgs boson, regardless
of its mass (below about 1~TeV).
The most important production processes for the $h_{\rm SM}$
are the gluon fusion process, $gg\to h_{\rm SM}$, and the vector boson
fusion process, $WW\to h_{\rm SM}$.\ihprod\ In particular, for $m_{h_{\rm
SM}}\lsim 130\gev$ the important discovery modes are $gg,WW\to h_{\rm
SM}\to\gam\gam$, $\tau^+\tau^-$.\ihdiscovery\  
At high luminosity, $q_i\anti q_j\to
W^\pm h_{\rm SM}$ and $gg\to t\anti th_{\rm SM}$ with $h_{\rm
SM}\to\gam\gam$ and $h_{\rm SM}\to b\anti b$ should also be visible.
For $m_{h_{\rm SM}}>130\gev$, $gg\to h_{\rm SM} \to ZZ^{(*)}\to
4\ell$ is extremely robust except for the small mass region with
$m_{h_{\rm SM}}$ just above $2\mw$ in which $h_{\rm SM}\to WW$ is
allowed and ${\rm BR}(h_{\rm SM}\to ZZ^*)$ drops sharply.  In this region,
$gg,WW\to h_{\rm SM}\to WW\to \ell\nu\ell\nu$ provides a strong Higgs
signal. For $m_{h_{\rm SM}}> 300$--400~GeV, the final states
$h_{\rm SM}\to WW\to \ell\nu jj$ and $h_{\rm SM}\to ZZ\to
\ell\ell\nu\nu$, where the $h_{\rm SM}$ is produced by a combination
of $gg$ and $WW$ fusion, provide excellent discovery channels for 
Higgs masses up to about 1~TeV ({\it
i.e.}, well beyond the $m_{h_{\rm SM}}\sim 800\gev$ limit of viability
for the $h_{\rm SM}\to 4\ell$ mode).  These results are summarized in
\fig{f:atlashsm}, from which we observe that the net statistical
significance for the $h_{\rm SM}$, after combining channels, exceeds
$10\sigma$ for all $m_{h_{\rm SM}}>80\gev$, assuming accumulated
luminosity of $L=100\fbi$ at the ATLAS detector~\cite{atlashsmref}.
Similar results are obtained by the 
CMS collaboration~\cite{cmshsmref,higgs-CMS_Higgs_lumi}, with the
$\hsm \gam\gam$ discovery mode even more prominent for 
$\mhsm\lsim 150$~GeV.\ihdiscovery\

\begin{figure}[t!]
\begin{center}
\resizebox{\textwidth}{!}{
\includegraphics*[angle=90]{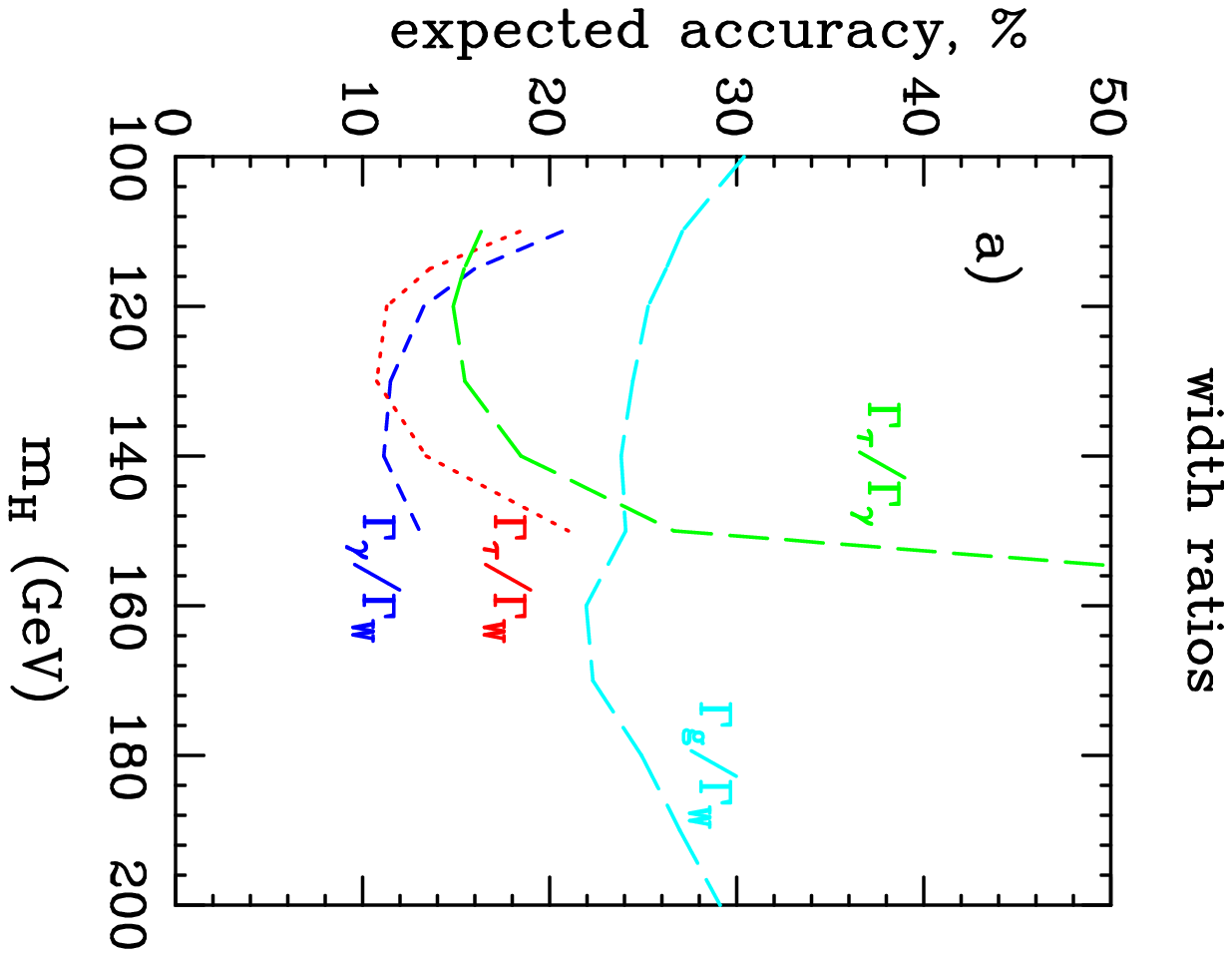}
\hspace*{3mm}
\includegraphics*[angle=90]{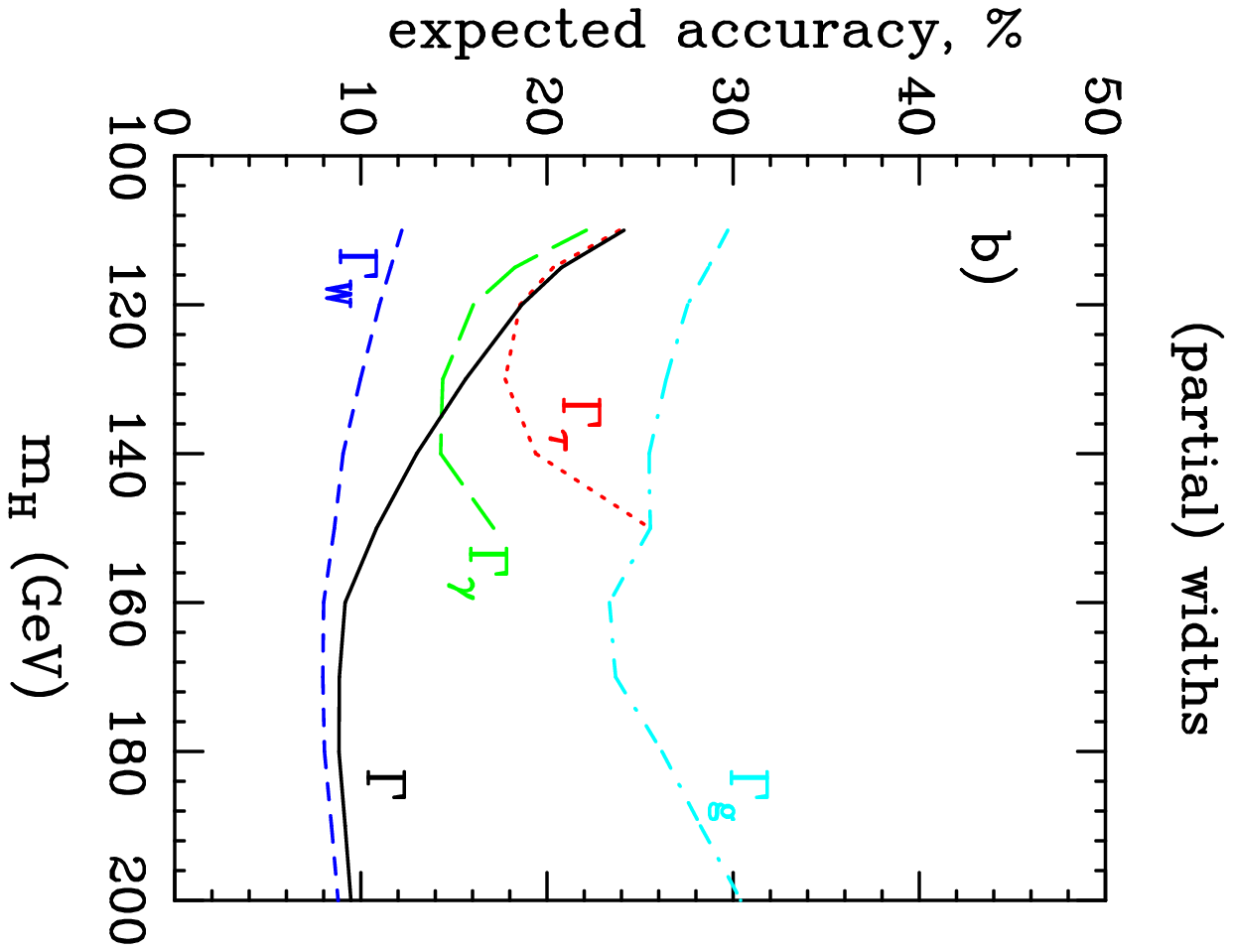}
}
\end{center}
\caption[0]{\label{f:zeppenfeld} \small 
Relative accuracy expected at the LHC with
200~fb${^{-1}}$ of data for (a)~various ratios of Higgs boson partial
widths and (b)~the indirect determination of partial and total
widths.  Expectations for width
ratios assume $W$, $Z$ universality;
indirect width measurements also assume $b$, $\tau$ universality and
a small branching ratio for unobserved modes.  Taken from the parton-level
analysis of \protect\cite{zeppenfeld}.\ihcoups\ihwidths\ihmeas}
\label{fig:LHCwidths}
\end{figure}

Precision measurements for a certain number of quantities will be
possible, depending upon the exact value of $m_{h_{\rm SM}}$.\ihmeas\ 
  For example, in
\cite{atlashsmref2} it is estimated that $m_{h_{\rm SM}}$ can be measured to
$<0.1\%$ for $m_{h_{\rm SM}}<400\gev$ and to $0.1$--$1\%$ for
$400<m_{h_{\rm SM}}<700\gev$. 
Using the $4\ell$ final state, $\Gamma^T_{h_{\rm SM}}$ can
be determined for $m_{h_{\rm SM}}\gsim 250\gev$ 
from the shape of the $4\ell$ mass
peak.  Various ratios of branching ratios and a selection of cross
sections times branching ratios can be measured in any given mass
region.\ihbrs\ihmeas\  Some early estimates of possibilities and achievable
accuracies appear in \cite{Gunion:1996cn}.  A more recent, but
rather optimistic parton-level theoretical study
\cite{Zeppenfeld:2000de,zeppenfeld} (see \fig{f:zeppenfeld}) finds 
that if $m_{h_{\rm SM}}\lsim 200\gev$ then good
accuracies can be achieved for many absolute partial widths and for
the total width provided: (a)~$WW$ fusion production can be reliably
separated from $gg$ fusion; (b)~the $WW/ZZ$ coupling ratio is as
expected in the SM from the SU(2)$\times$U(1) symmetry; (c)~the $WW^*$
final state can be observed in both $gg$ and $WW$ fusion; and
(d)~there are no unexpected decays of the $h_{\rm SM}$.
Errors estimated in this way for $L=200\fbi$ of accumulated data are given in
\fig{f:zeppenfeld}. However, errors found in recent ATLAS studies of 
a few of these channels are substantially larger~\cite{deRoeck}.\ihmeas\
For example, for $L=300\fbi$ ATLAS finds that the smallest
error on $\Gamma_\tau/\Gamma_W$ is achieved for $\mhsm\sim 130\gev$
and is of order $\pm27\%$ 
(as compared to the $\pm 11\%$ of \fig{f:zeppenfeld}) and that the
error rises rapidly for $\mhsm$ away from $130\gev$, 
reaching $\pm60\%$ for $\mhsm=150\gev$.
Invisible Higgs decays have also been addressed 
in the theoretical work~\cite{higgs-Eboli}; 
CMS simulations show some promise for this channel.\ihbrs\ihcoups\ihdiscovery\

\section{Higgs bosons in low-energy supersymmetry}
\label{sece}

The Higgs sector of the MSSM, the
simplest realistic model of low-energy supersymmetry,
consists of the two-Higgs-doublet extension of the 
Standard Model plus the corresponding superpartners. 
Two Higgs doublets, one  
with $Y=+1$ and one with $Y=-1$, are needed in order
that gauge anomalies due to the higgsino superpartners 
are exactly canceled.  In particular, the Higgs
sector contains eight scalar degrees of freedom: one complex
$Y=-1$ doublet, {\boldmath $\Phi_d$}$=(\Phi_d^0,\Phi_d^-)$ and one
complex $Y=+1$ doublet, {\boldmath
$\Phi_u$}$=(\Phi_u^+,\Phi_u^0)$.  This notation reflects the fact
that in the MSSM, the interaction Lagrangian that describes
the Higgs couplings to fermions 
is constrained by supersymmetry and obeys the
following property: $\Phi_d^0$ couples exclusively to down-type
fermion pairs and $\Phi_u^0$ couples exclusively to up-type
fermion pairs.\ihcoups\ This pattern of Higgs-fermion couplings defines the
Type-II two-Higgs-doublet model \cite{wise,hhgchap4}.\itwohdm\

When the Higgs potential is minimized, the neutral
components of the Higgs fields acquire vacuum expectation 
values\footnote{The phases of the
Higgs fields can be chosen such that the vacuum expectation values
are real and positive. That is, the tree-level MSSM Higgs sector
conserves CP, which implies that the neutral Higgs mass
eigenstates possess definite CP quantum numbers.}\ihcp\
$\langle\Phi_d^0\rangle=v_d/\sqrt{2}$ and
$\langle\Phi_u^0\rangle=v_u/\sqrt{2}$,
where the normalization has been chosen such that $v^2\equiv
v_d^2+v_u^2=(246~{\rm GeV})^2$.  The ratio 
of vacuum expectation values is denoted by 
\begin{equation}
\tan\beta\equiv v_u/v_d\,.
\end{equation}
The physical Higgs spectrum consists of a charged Higgs pair
\begin{equation} \label{hpmstate}
\hpm=\Phi_d^\pm\sinb+ \Phi_u^\pm\cosb\,,
\end{equation}
one CP-odd scalar
\begin{equation} \label{hastate}
\ha= \sqrt{2}\left({\rm Im\,}\Phi_d^0\sinb+{\rm Im\,}\Phi_u^0\cosb
\right)\,,
\end{equation}
and two CP-even scalars:
\beqa
\hl &=& -(\sqrt{2}\,{\rm Re\,}\Phi_d^0-v_d)\sin\alpha+
(\sqrt{2}\,{\rm Re\,}\Phi_u^0-v_u)\cos\alpha\,,\nonumber\\
\hh &=& (\sqrt{2}\,{\rm Re\,}\Phi_d^0-v_d)\cos\alpha+
(\sqrt{2}\,{\rm Re\,}\Phi_u^0-v_u)\sin\alpha\,,
\label{scalareigenstates}
\eeqa
(with $\mhl\leq \mhh$).
The angle $\alpha$ arises when the CP-even Higgs
squared-mass matrix (in the $\Phi_d^0$---$\Phi_u^0$ basis) is
diagonalized to obtain the physical CP-even Higgs states.
The Goldstone bosons, $G^\pm$ and $G^0$, which are orthogonal to
$H^\pm$ and $\ha$ respectively, provide the 
longitudinal components of the massive $W^\pm$ and $Z$ via the Higgs
mechanism.

In two-Higgs-doublet models with Type-II Higgs-fermion couplings,
the $Y=-1$ Higgs doublet generates mass for ``up''-type quarks and 
the $Y=+1$ Higgs doublet generates mass for ``down''-type
quarks (and charged leptons) \cite{Inoue82,Gunion86}.\itwohdm\
The tree-level
relations between the quark masses and the Higgs-fermion Yukawa 
couplings (using 3rd family notation) are given by:\iyuk\ihcoups\
\beqa h_b &=&
\frac{\sqrt{2}\,m_b}{v_d}=\frac{\sqrt{2}\,m_b}{v\cos\beta}
\,,\label{hbdef} \\
h_t &=& \frac{\sqrt{2}\,m_t}{v_u}=\frac{\sqrt{2}\,m_t}{v\sin\beta}
\,. \label{htdef}
\eeqa
Radiative corrections to these relations will be examined in
Section~\ref{seceb}.\iradiative\

At this stage, $\tanb$ is a free parameter of the model.  However,
theoretical considerations suggest that $\tanb$ cannot be too small or
too large.\ihtheoryc\ The
crudest bounds arise from unitarity constraints.\iunitarity\  If $\tanb$
becomes too small, then the Higgs coupling to top quarks becomes
strong and the tree-unitarity of processes
involving the Higgs-top quark Yukawa coupling is violated.\iunitarity\
A rough lower bound advocated by \cite{hewett}, $\tanb\gsim 0.3$,
corresponds to a value of $h_t$ in the perturbative
region.\iperturbativity\  A similar argument involving $h_b$
would yield $\tanb\lsim 120$. A more restrictive theoretical
constraint is based on the requirement that Higgs--fermion Yukawa 
couplings, $h_t$ and $h_b$,
remain finite when running from the electroweak scale to
some large energy scale $\Lambda$, above which
new physics enters.\ihtheoryc\iyuk\ The limits on $\tanb$ depend on 
the choice of $\Lambda$.\ieft\ Integrating the
Yukawa coupling renormalization group equations 
from the electroweak scale to $\Lambda$ (allowing for
the possible existence of a supersymmetry-breaking scale,
$\mz\leq\msusy\leq \Lambda$), one can determine the range of
$\tanb$ for which $h_t$ and $h_b$
remain finite.\irge\ This exercise has been carried
out at two-loops in \cite{schrempp}.  Suppose that the low-energy
theory at the electroweak scale is the MSSM, and that there is no
additional new physics below the grand unification scale of
$\Lambda=2\times 10^{16}$~GeV.  Then, for $\mt=175$~GeV, the
Higgs-fermion Yukawa couplings remain finite
at all energy scales below $\Lambda$ if 
$1.5\lsim\tanb\lsim 65$.\iperturbativity\ihtheoryc\ Note that this result
is consistent with the scenario of radiative electroweak symmetry
breaking in low-energy supersymmetry based on supergravity, which
requires that $1\lsim\tanb\lsim \mt/\mb$.\footnote{Here, the quark
masses are evaluated at $m_Z$.  We take $m_t(m_Z)\simeq 165$~GeV and
$m_b(m_Z)\simeq 3$~GeV.}  Thus,  we expect $\tanb$ to lie in the range
$1\lsim\tanb\lsim 55$.\footnote{The lower bound on $\tanb$ can be taken
to be $\tanb\gsim 2.4$, based on the LEP MSSM Higgs search discussed in 
Section~\ref{secfa}.}

\subsection{MSSM Higgs sector at tree level}
\label{secea}

The supersymmetric structure of the theory imposes constraints on
the Higgs sector of the model. As a result, all Higgs sector
parameters at tree level are determined by two free parameters:
$\tanb$ and one Higgs mass, conveniently chosen to be $\mha$. In
particular,\ihmass\ihtheoryc\
\begin{equation} \mhpm^2 =\mha^2+\mw^2\,, \label{susymhpm} 
\end{equation}
and the CP-even Higgs bosons $\hl$ and $\hh$ are eigenstates of
the following squared-mass matrix (with respect
to the ${\rm Re}~\Phi_d^0$--${\rm Re}~\Phi_u^0$ basis):
\begin{equation} \mathcal{M}_0^2 =    \left(
\begin{array}{ll}
\phantom{-}\mha^2 \sin^2\beta + m^2_Z \cos^2\beta&\quad
           -(\mha^2+m^2_Z)\sin\beta\cos\beta \\[4pt]
  -(\mha^2+m^2_Z)\sin\beta\cos\beta&\quad
  \phantom{-}\mha^2\cos^2\beta+ m^2_Z \sin^2\beta 
\end{array}
   \right)\,.\label{kv}
\end{equation}
The eigenvalues of $\mathcal{M}_0^2$ are
the tree-level squared-masses of the two CP-even Higgs scalars
\begin{equation}
  m^2_{\hh,\hl} = \half \left( \mha^2 + m^2_Z \pm
                  \sqrt{(\mha^2+m^2_Z)^2 - 4m^2_Z \mha^2 \cos^2 2\beta}
                  \; \right)\,,\label{kviii}
\end{equation}
Note that \eq{kviii} yields an upper bound
to the tree-level mass of the light CP-even Higgs boson:
$\mhl^2\leq\mz^2|\cos 2\beta|\leq\mz^2$.\ihtheoryc\ihiggsbounds\  
Radiative corrections
can significantly increase this upper bound by as much as 50\%
as described in Section~\ref{seceb}.\iradiative\
Nevertheless, it is already apparent that the
MSSM favors a CP-even
Higgs boson whose mass is not much larger than $m_Z$,
a result that is consistent with the inferred Higgs mass limit  
from precision electroweak measurements discussed in 
Section~\ref{secb}.\ihmass\ihtheoryc\

From the above results, one also obtains:
\begin{equation}
\cos^2(\beta-\alpha)=\frac{\mhl^2(\mz^2-\mhl^2)}{\mha^2(\mhh^2-\mhl^2)}\,,
\label{cbmasq}
\end{equation}
where $\alpha$ is the angle that diagonalizes the CP-even Higgs
squared-mass matrix [see \eq{scalareigenstates}].  
In the convention where $\tanb$ is positive ({\it i.e.},
$0\leq\beta\leq\pi/2$), the angle $\alpha$ lies in the range
$-\pi/2\leq\alpha\leq 0$.\ihtheoryc\

The limit of $\mha\gg\mz$ is of particular interest.\idecoup\
The expressions for the
Higgs masses and mixing angle simplify in this limit and one finds
\beqa
\mhl^2 &\simeq&\ \mz^2\cos^2 2\beta\,, \label{largema1}\\[3pt]
\mhh^2 &\simeq&\ \mha^2+\mz^2\sin^2 2\beta\,,\label{largema2}\\[3pt]
\mhpm^2& =& \ \mha^2+\mw^2\,,\label{largema3} \\[3pt]
\cos^2(\beta-\alpha)&\simeq&\ {\mz^4\sin^2 4\beta\over 4\mha^4}\,.
\label{largema4} 
\eeqa 
Two consequences are immediately apparent.
First, $\mha\simeq\mhh \simeq\mhpm$, up to corrections of $\mathcal{O}
(\mz^2/\mha)$.\ihtheoryc\  Second, $\cos(\beta-\alpha)=0$ up to corrections
of $\mathcal{O}(\mz^2/\mha^2)$. This is the {\it
decoupling} limit \cite{decoupling,gunhabdecoup} because when $\mha$ is large,
the effective low-energy theory below the scale of $\mha$ contains
a single CP-even Higgs boson, $\hl$, whose properties are nearly
identical to those of the Standard Model Higgs boson, $\hsm$.\idecoup\

The phenomenology of the Higgs sector is determined by the various
couplings of the Higgs bosons to gauge bosons, Higgs bosons and
fermions.\ihcoups\ The couplings of the two CP-even Higgs bosons to $W$ and
$Z$ pairs are given in terms of the angles $\alpha$ and $\beta$ 
by\ihcoups\
\begin{equation} 
g\ls{\hl VV}= g\ls{V} m\ls{V}\sinbma \,,\qquad\quad
           g\ls{\hh VV}= g\ls{V} m\ls{V}\cosbma\,,\label{vvcoup}
\end{equation}
where
$g\ls V\equiv 2m_V/v$ for $V=W$ or $Z$.  There are
no tree-level couplings of $\ha$ or $\hpm$ to $VV$.  The couplings
of one gauge boson to two neutral Higgs bosons are given by:
\begin{equation}
g_{\hl\ha Z}={g\cosbma\over 2\cos\theta_W} \,,\qquad\quad
           g_{\hh\ha Z}={-g\sinbma\over 2\cos\theta_W}\,.
           \label{hvcoup}
\end{equation}

As noted earlier, the pattern of Higgs-fermion couplings in the MSSM 
are those of the Type-II two-Higgs-doublet model \cite{wise,hhgchap4}.\itwohdm\
The couplings of the neutral Higgs bosons to $f\bar
f$ relative to the Standard Model value, $gm_f/2\mw$, are given by\ihcoups\
\beqa
 \label{qqcouplings}
\hl b\bar b \;\;\; ({\rm or}~ \hl \tau^+ \tau^-):&&~~ -
{\sin\alpha\over\cos\beta}=\sin(\beta-\alpha)
-\tan\beta\cos(\beta-\alpha)\,,\nonumber\\[3pt]
\hl t\bar t:&&~~~ \phm{\cos\alpha\over\sin\beta}=\sin(\beta-\alpha)
+\cot\beta\cos(\beta-\alpha)\,,\nonumber\\[3pt]
\hh b\bar b \;\;\; ({\rm or}~ \hh \tau^+ \tau^-):&&~~~
\phm{\cos\alpha\over\cos\beta}=
\cos(\beta-\alpha)
+\tan\beta\sin(\beta-\alpha)\,,\nonumber\\[3pt]
\hh t\bar t:&&~~~ \phm{\sin\alpha\over\sin\beta}=\cos(\beta-\alpha)
-\cot\beta\sin(\beta-\alpha)\,,\nonumber\\[3pt]
\ha b \bar b \;\;\; ({\rm or}~ \ha \tau^+
\tau^-):&&~~~\phm\gamma_5\,{\tan\beta}\,,
\nonumber\\[3pt]
\ha t \bar t:&&~~~\phm\gamma_5\,{\cot\beta}\,,
\eeqa
(the $\gamma_5$ indicates a pseudoscalar coupling), and the
charged Higgs boson couplings to fermion pairs
(with all particles pointing into the vertex) are given by
\beqa
\label{hpmqq}
g_{H^- t\bar b}&= &{g\over{\sqrt{2}\mw}}\
\Bigl[m_t\cot\beta\,P_R+m_b\tan\beta\,P_L\Bigr]\,,
\nonumber\\[3pt]
g_{H^- \tau^+ \nu}&= & {g\over{\sqrt{2}\mw}}\
\Bigl[m_{\tau}\tan\beta\,P_L\Bigr]\,,
\eeqa
where $P_{R,L}\equiv\half(1\pm\gamma_5)$ are the right and left-handed
projection operators, respectively.

The neutral Higgs couplings to fermion pairs
[\eq{qqcouplings}] have been written in such a way that their
behavior can be immediately ascertained in the decoupling limit
($\mha\gg\mz$) by setting $\cosbma=0$.\idecoup\ In
particular, in the decoupling limit, the couplings of $\hl$ to
vector bosons and fermion pairs are equal to the corresponding
couplings of the SM Higgs boson.

\begin{figure}[t!]
\begin{center}
\includegraphics[width=3.2in]{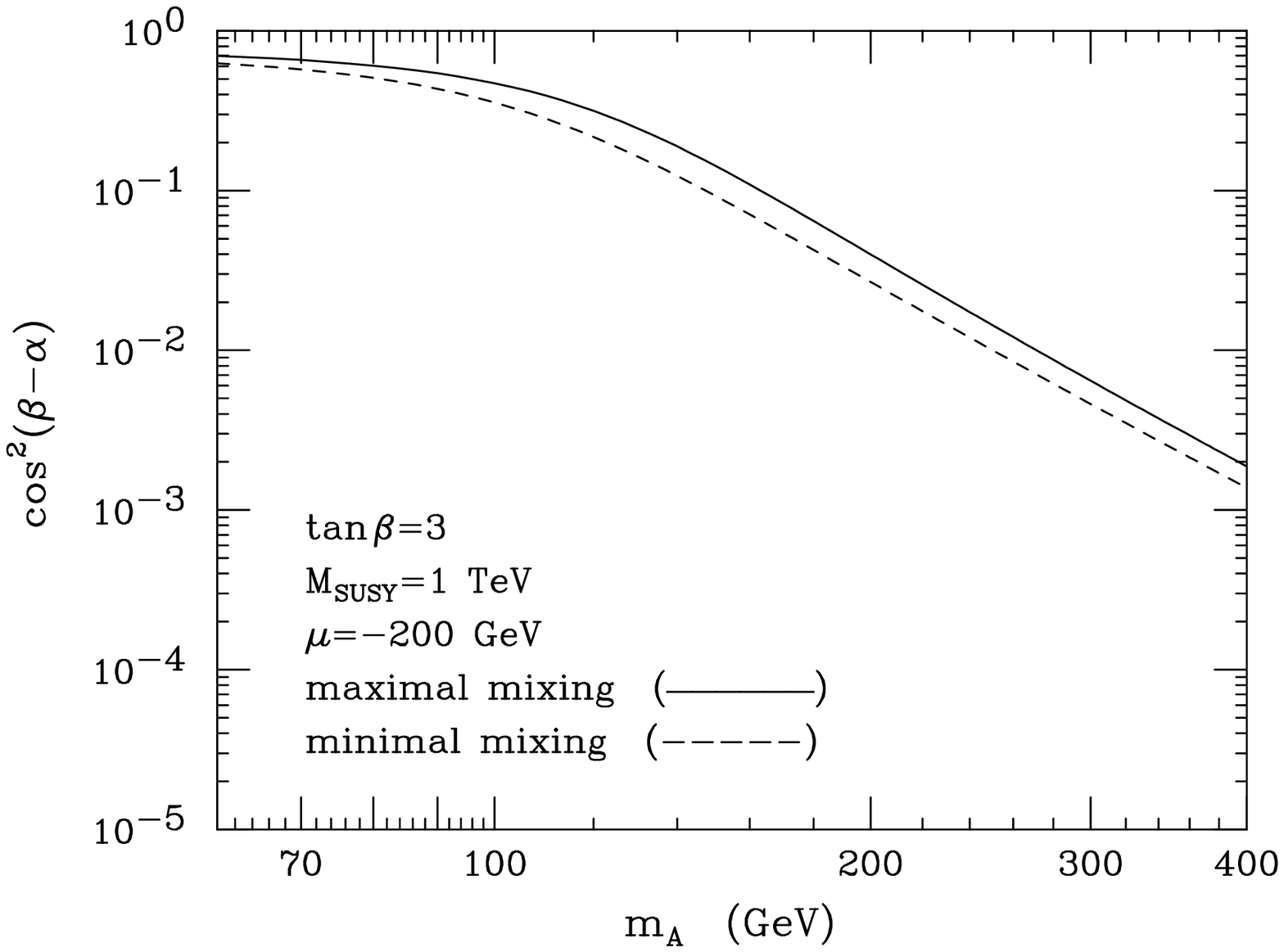}
\includegraphics[width=3.2in]{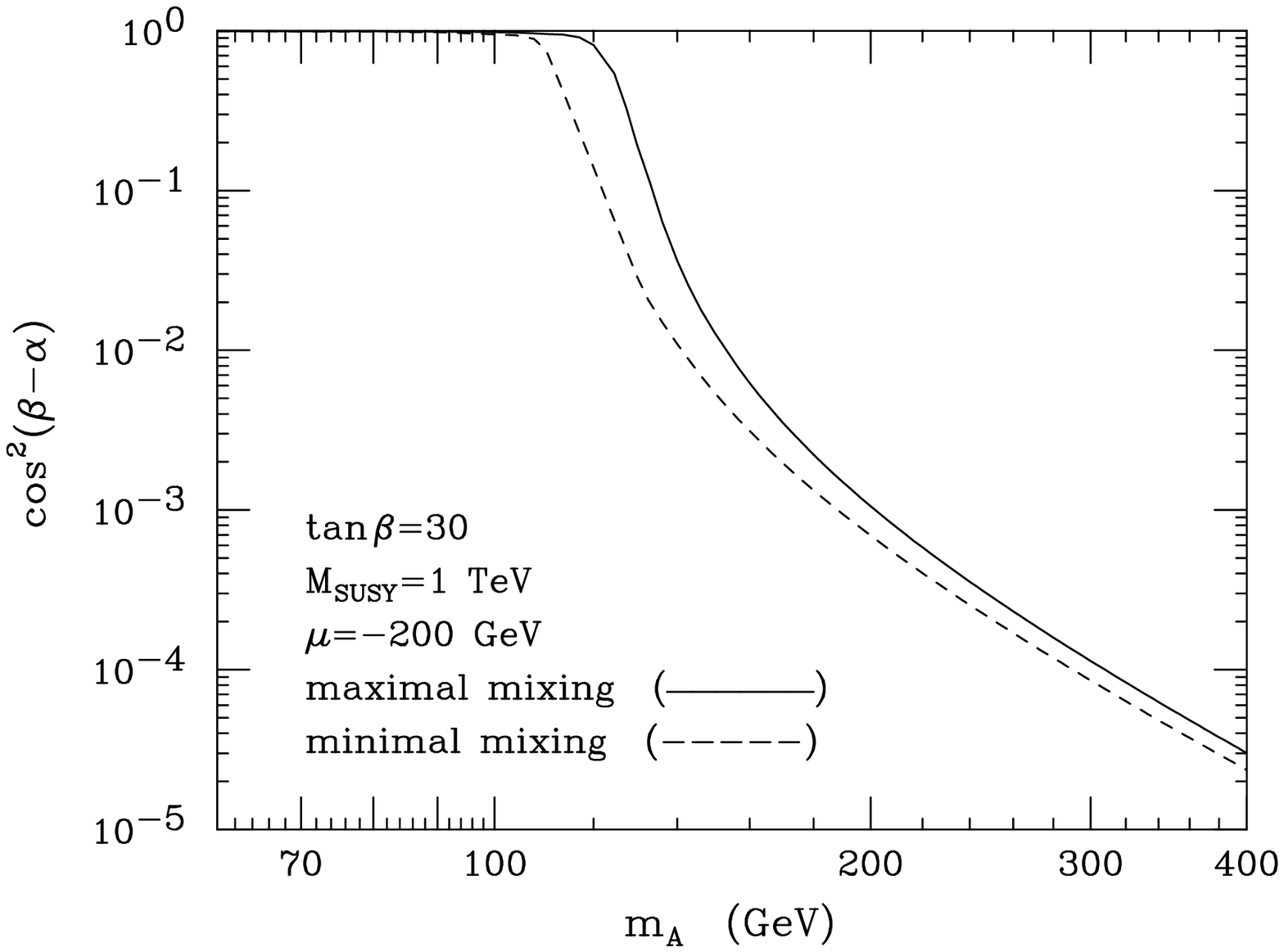}
\end{center}
\caption[0]{\label{cosgraph} \small The value of $\cos^2(\beta-\alpha)$
is shown as a function of
$\mha$ for two choices of $\tan\beta = 3$ and $\tan\beta = 30$.
When radiative-corrections are included, one can define an approximate
loop-corrected angle $\alpha$ as a function of $\mha$, $\tan\beta$ and
the MSSM parameters.  In the figures above, we
have incorporated radiative corrections, assuming that the top-squark
masses are given by $\msusy=1$~TeV.  In addition,
two extreme cases for the squark mixing parameters
are shown (see Section~\ref{seceb} for further discussion of the
radiative corrections and their dependence on the supersymmetric
parameters). The decoupling effect expected from
\eq{largema4}, in which $\cos^2(\beta-\alpha)\propto \mz^4/\mha^4$
for $\mha\gg m_Z$,
continues to hold even when radiative corrections are included.\idecoup\
}
\end{figure}

The region of MSSM Higgs sector parameter space in which the
decoupling limit applies is large, because $\cos(\beta-\alpha)$
approaches zero quite rapidly once $\mha$ is larger than about
200~GeV, as shown in \fig{cosgraph}.\idecoup\ As a result, over a
significant region of the MSSM parameter space, the search for the
lightest CP-even Higgs boson of the MSSM is equivalent to the
search for the \SM\ Higgs boson.\ihdiscovery\

\subsection{The radiatively corrected MSSM Higgs sector}
\label{seceb}

When one-loop radiative corrections are incorporated, the Higgs
masses and couplings depend on additional parameters of the
supersymmetric model that enter via the virtual loops.\iradiative\ 
One of the most
striking effects of the radiative corrections to the MSSM Higgs
sector is the modification of the upper bound of the light CP-even
Higgs mass, as first noted in \cite{hhprl}.\ihiggsbounds\ihmass\   
When $\tanb\gg 1$ and
$\mha\gg\mz$, the {\it tree-level} prediction of $\mhl=\mz$
corresponds to the theoretical upper bound for $\mhl$.\ihtheoryc\
Including radiative corrections, this theoretical upper bound is
increased, primarily because of an incomplete cancellation of the
top-quark and top-squark (stop) loops.  (These contributions would
cancel if supersymmetry  were exact.) The relevant parameters
that govern the stop sector are the average of the two stop
squared-masses: $M^2_{\rm SUSY}\equiv \half(\mstopa^2+\mstopb^2)$, and
the off-diagonal element of the stop squared-mass matrix: $m_t
X_t\equiv m_t(A_t-\mu\cot\beta)$, where $A_t$ is a soft supersymmetry-breaking 
trilinear scalar interaction term, and $\mu$ is the supersymmetric Higgs mass 
parameter. 
The qualitative behavior of the
radiative corrections can be most easily seen in the large top
squark mass limit, where, in addition, the splitting of the two
diagonal entries and the off-diagonal entry of the stop
squared-mass matrix are both small in comparison to $M^2_{\rm SUSY}$.  In
this case, the upper bound on the lightest CP-even Higgs mass is
approximately given by 
\begin{equation} \label{deltamh} 
\mhl^2  \lsim 
\mz^2+{3g^2\mt^4\over
8\pi^2\mw^2}\left[\ln\left({M^2_{\rm SUSY}\over\mt^2}\right)
+{X_t^2\over M^2_{\rm SUSY}} \left(1-{X_t^2\over
12M^2_{\rm SUSY}}\right)\right]\,. 
\end{equation} 
More complete treatments of the
radiative corrections include the effects of stop mixing,
renormalization group improvement, and the leading two-loop
contributions, and imply that these corrections somewhat overestimate
the true upper bound of $\mhl$ (see \cite{higgsrad} for the most
recent results).\iradiative\ihmass\irge\
Nevertheless, \eq{deltamh} correctly illustrates
some noteworthy features of the more precise result. 
First, the
increase of the light CP-even Higgs mass bound beyond $\mz$ can be
significant due to the $m_t^4$ enhancement of
the one-loop radiative correction. Second, the dependence of the
light Higgs mass on the stop mixing parameter $X_t$ implies that
(for a given value of $M_{\rm SUSY}$) the upper bound of the light Higgs
mass initially increases with $X_t$ and reaches its {\it maximal}
value at $X_t\simeq\sqrt{6}M_{\rm SUSY}$.  This point is referred to as the
{\it maximal mixing} case (whereas $X_t=0$ is called the {\it
minimal mixing} case).\ihtheoryc\

\begin{figure}[t!]
\begin{center}
\includegraphics*[width=0.75\textwidth]{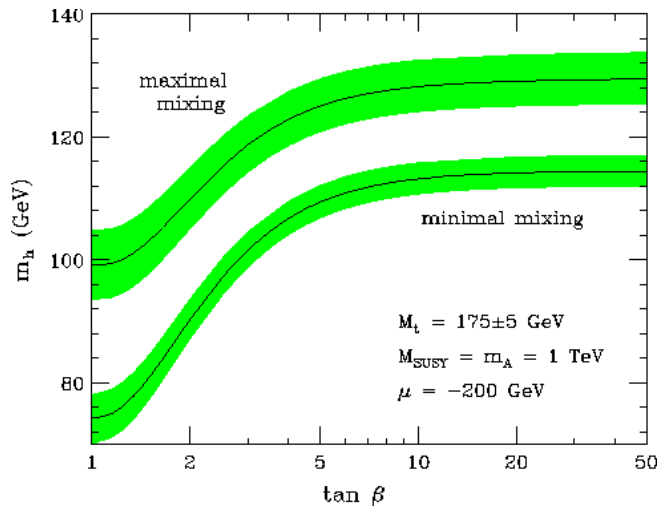}
\end{center}
\caption[0]{\label{mhtanb} \small
The radiatively corrected light CP-even Higgs mass is
plotted as a function of $\tanb$, for the maximal mixing [upper
band] and minimal mixing cases. The impact of the top quark mass
is exhibited by the shaded bands; the central value corresponds to
$m_t=175$~GeV, while the upper [lower] edge of the bands
correspond to increasing [decreasing] $m_t$ by 5~GeV.\ihmass}
\end{figure}
Taking $\mha$ large, \fig{mhtanb} illustrates that the maximal
value of the lightest CP-even Higgs mass bound is realized 
in the case of maximal mixing at large $\tan\beta$.\idecoup\ihtheoryc\ 
Allowing for the
uncertainty in the measured value of $\mt$ and the uncertainty
inherent in the theoretical analysis, one finds for $M_{\rm SUSY}\lsim
2$~TeV that $\mhl\lsim \mhmax$, where\ihiggsbounds\ihmass\ihtheoryc\
\begin{eqnarray} \label{mhmaxvalue}
\mhmax&\simeq&  122~{\rm GeV}, \quad
\mbox{minimal stop mixing,} \nonumber \\
\mhmax&\simeq&  135~{\rm GeV}, \quad \mbox{maximal stop mixing.}
\end{eqnarray}

The $\hl$ mass bound in the MSSM quoted above does not 
apply to non-minimal supersymmetric extensions of the Standard Model. 
If additional Higgs singlet and/or triplet fields  are
introduced, then new Higgs self-coupling parameters appear, which
are not significantly constrained by 
present data.\ihsinglets\ihtriplets\ihself\  For example,
in the simplest non-minimal supersymmetric extension of the
Standard Model (NMSSM), the addition of a complex Higgs singlet 
field $S$ adds a new
Higgs self-coupling parameter, $\lambda_S$ \cite{singlets}. The mass
of the lightest neutral Higgs boson can be raised arbitrarily by
increasing the value of $\lambda_S$, analogous to the behavior of
the Higgs mass in the Standard Model. Under the assumption that
all couplings stay perturbative up to the Planck scale,
one finds
in nearly all cases that $\mhl\lsim 200$~GeV, independent of
the details of the low-energy supersymmetric 
model~\cite{Espinosa:1998re}.\iperturbativity\impl\ihtheoryc\
Moreover, if
the perturbative unification scale is significantly lower than the
Planck scale, then the lower bound on the Higgs
mass can be substantially larger~\cite{Tobe:2002zj}.\impl\

\begin{figure}[t!]
\begin{center}
\includegraphics*[width=0.75\textwidth]{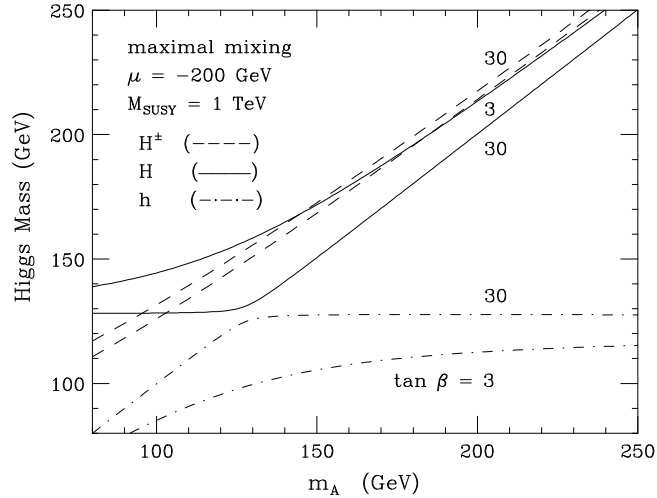}
\end{center}
\caption[0]{\label{massvsma} \small
Lightest CP-even Higgs mass ($\mhl$),
heaviest CP-even Higgs mass ($\mhh$) and charged Higgs mass ($\mhpm$) as
a function of $\mha$ for two choices of $\tan\beta=3$ and
$\tan\beta=30$. 
The slight increase in the charged Higgs mass as
$\tan\beta$ is increased from 3 to 30 is a consequence of the
radiative corrections.\ihmass\ihtheoryc\
}
\end{figure}

Radiative corrections also modify the tree-level expressions for
$\mhh$ and $\mhpm$ [\eqns{kviii}{susymhpm}].\iradiative\  However, since 
$\mhh\simeq\mhpm\simeq\mha$ when $\mha\gg\mz$, the impact of the
radiative corrections on the heavier Higgs masses
is less pronounced that those described above.
In \fig{massvsma}, we exhibit the masses of
the CP-even neutral and the charged Higgs masses as a function of
$\mha$.  
Note that $\mhh\geq\mhmax$ for all values of
$\mha$ and $\tan\beta$, where $\mhmax$ is to be evaluated depending on
the top-squark mixing, as indicated in \eq{mhmaxvalue}.\ihmass\ihtheoryc\

Radiative corrections also significantly modify the tree-level
values of the Higgs boson couplings to fermion pairs and to vector
boson pairs.\ihcoups\ As discussed above, the tree-level Higgs couplings
depend crucially on the value of $\cosbma$.  In the first
approximation, when radiative corrections of the Higgs
squared-mass matrix are computed, the diagonalizing angle $\alpha$
is modified.  This provides one important source of the
radiative corrections of the Higgs couplings.   In \fig{cosgraph},
we show the effect of radiative corrections on the value of
$\cosbma$ as a function of $\mha$  for different values of the
squark mixing parameters and $\tanb$.  One can then insert
the radiatively corrected value of $\alpha$ into
eqs.~(\ref{vvcoup})--(\ref{qqcouplings}) to
obtain radiatively improved couplings of Higgs bosons to vector
bosons and to fermions.\iradiative\

At large $\tanb$, there is another potentially important class of radiative
corrections in addition to those that enter through the modified 
value of $\alpha$.\iradiative\
These corrections arise in the relation between $m_b$ and $\tanb$ and 
depend on the details of the MSSM spectrum (which
enter via loop-effects).   
At tree level, the Higgs couplings to
$b\bar b$ are proportional to the Higgs--bottom-quark Yukawa 
coupling, $h_b$.\iyuk\ihcoups\ Deviations from the tree-level relation
between $h_b$ and $m_b$ [\eq{hbdef}]
due to radiative corrections are calculable and finite
\cite{hffsusyqcd,deltamb,deltamb1,deltamb2,hffsusyprop}.  One of the
fascinating properties of such corrections is that in certain
cases the corrections do {\it not} vanish in the limit of large
supersymmetric mass parameters.  These corrections grow with
$\tanb$ and therefore can be significant in the large $\tanb$
regime.\inondecoupling\iradiativesusy\  
In the supersymmetric limit, bottom quarks only couple to
$\Phi_d^0$. However, supersymmetry is broken and the bottom quark
will receive a small coupling to $\Phi_u^0$ from radiative
corrections,\iradiative\ihcoups\
\begin{equation}
-\mathcal{L}_{\rm Yukawa} \simeq h_b \Phi_d^0 b \bar{b} + (\Delta h_b)
\Phi_u^0 b\bar{b}\,.
\label{couplings}
\end{equation}
Because the Higgs doublet acquires a vacuum
expectation value, the bottom quark mass receives
an extra contribution equal to
$(\Delta h_b) v_u$.  Although $\Delta h_b$ is
one--loop suppressed with respect to $h_b$,
for sufficiently large values of $\tan\beta$ ($v_u \gg v_d$)
the contribution to the bottom quark mass of both terms in
\eq{couplings} may be comparable in size. This induces a
large modification in the tree level relation,
\begin{equation}
m_b = {h_b v_d\over\sqrt{2}} (1+\Delta_b)\,, \qquad
\label{yukbmass}
\end{equation}
where $\Delta_b \equiv (\Delta h_b)\tan\beta/h_b$.\ihcoups\
The function $\Delta_b$ contains two main
contributions: one from a bottom squark--gluino loop
(depending on the two bottom squark masses $M_{\tilde b_1}$
and $M_{\tilde b_2}$ and the gluino mass $M_{\tilde g}$) and another one
from a
top squark--higgsino loop (depending on the two top squark masses
$M_{\tilde t_1}$ and $M_{\tilde t_2}$ and the higgsino mass parameter
$\mu$).  The explicit form of $\Delta_b$ at one--loop in the limit of
$M_S \gg m_b$ is given by \cite{deltamb,deltamb1,deltamb2}:
\begin{equation}
\!\!\Delta_b \simeq {2\alpha_s \over 3\pi}
M_{\tilde g}\mu\tan\beta~I(M_{\tilde b_1},
M_{\tilde b_2},M_{\tilde g})
 + {Y_t \over 4\pi} A_t\mu\tan\beta~I(M_{\tilde t_1},M_{\tilde t_2},\mu),
\label{deltamb}
\end{equation}
where $\alpha_s=g_s^2/4\pi$,
$Y_t\equiv h_t^2/4\pi$, and contributions proportional to the
electroweak gauge couplings have been neglected.  In addition,
the function $I$ is defined by
\begin{equation}
I(a,b,c) = {a^2b^2\ln(a^2/b^2)+b^2c^2\ln(b^2/c^2)+c^2a^2\ln(c^2/a^2) \over
(a^2-b^2)(b^2-c^2)(a^2-c^2)},
\end{equation}
and is manifestly positive.
Note that the Higgs coupling proportional to $\Delta h_b$ is a
manifestation of the broken supersymmetry in the low energy theory;
hence, $\Delta_b$ does not decouple
in the limit of large values of the supersymmetry breaking masses. Indeed,
if all supersymmetry breaking mass parameters (and $\mu$)
are scaled by a common factor, the correction
$\Delta_b$ remains constant.\ihcoups\iradiative\inondecoupling\

Similarly to the case of the bottom quark, the relation between $m_\tau$ and
the Higgs--tau-lepton Yukawa 
coupling $h_\tau$ is modified:\iyuk\ihcoups\iradiative\inondecoupling\
\begin{equation}
m_\tau = {h_\tau v_d\over\sqrt{2}} (1+\Delta_\tau).
\end{equation}
The correction $\Delta_\tau$ contains a contribution from a
tau slepton--neutralino loop (depending on the two stau masses
$M_{\tilde \tau_1}$ and $M_{\tilde \tau_2}$ and the
mass parameter of the $\widetilde B$ (``bino'') component
of the neutralino, $M_{1}$) and a
tau sneutrino--chargino loop (depending on the tau sneutrino mass
$M_{\tilde \nu_\tau}$, the mass parameter of the $\widetilde W^\pm$
component of the chargino, $M_{2}$, and $\mu$).
It is given by \cite{deltamb1,deltamb2}:
\begin{equation}
\Delta_\tau = {\alpha_1 \over 4\pi} M_1\mu\tan\beta\,
I(M_{\tilde\tau_1},
M_{\tilde\tau_2},M_1) - {\alpha_2 \over 4\pi} M_2\mu\tan\beta \,
I(M_{\tilde\nu_\tau},M_2,\mu),
\end{equation}
where $\alpha_2\equiv g^2/4\pi$ and $\alpha_1\equiv g^{\prime 2}/4\pi$
are the electroweak gauge couplings.
Since corrections to $h_\tau$ are proportional to $\alpha_1$ and
$\alpha_2$, they are typically smaller than the corrections to $h_b$.

From \eqns{couplings}{yukbmass} 
we can obtain the couplings of the physical neutral Higgs
bosons to $b\bar b$.\ihcoups\
At large $\tanb$, the dominant corrections
to \eq{qqcouplings} are displayed below:\footnote{A better
approximation in which non-leading terms at large $\tanb$ are kept can
be found in \cite{Carena:2001bg}.}\ihcoups\iradiative\
\begin{eqnarray} \label{bbcouplings}
\hl b\bar b& :&~~~ -
{\sin\alpha\over\cos\beta}{1\over 1+\Delta_b}
\left[ 1 - \frac{\Delta_b\cot\alpha}{\tan\beta} \right]
\,,\nonumber\\[3pt]
\hh b\bar b& :&~~~
\phm{\cos\alpha\over\cos\beta}{1\over 1+\Delta_b}
\left[ 1 +\frac{\Delta_b\tan\alpha}{\tan\beta} \right]
\,,\nonumber\\[3pt]
\ha b \bar b& :&~~~\phm\gamma_5\,\frac{\tan\beta}{1+\Delta_b}\,,
\end{eqnarray}
where $\Delta_b\propto\tan\beta$ [see \eq{deltamb}].
Similarly, the neutral Higgs couplings to $\tau^+\tau^-$ are modified
by replacing $\Delta_b$ in \eq{bbcouplings} with $\Delta_\tau$
\cite{deltamb1,deltamb2}.  The corresponding
radiative corrections to the couplings
of the neutral Higgs bosons to $t\bar t$ are not $\tanb$-enhanced
(see \cite{Carena:2001bg} for a useful approximation for these corrections).
One can also derive radiatively corrected couplings
of the charged Higgs boson to fermion pairs \cite{chhiggstotop2,eberl}.
The tree-level couplings
of the charged Higgs boson to fermion pairs
are modified accordingly by replacing
$m_b \rightarrow m_b/(1 + \Delta_b)$ and
$m_{\tau} \rightarrow m_{\tau}/(1 + \Delta_{\tau})$,
respectively.\iradiative\inondecoupling\ihcoups\

One consequence of the above results is that the neutral Higgs
coupling to $b\bar b$ (which is expected to be the dominant decay
mode over nearly all of the MSSM Higgs parameter space), can be
significantly suppressed at large $\tan\beta$~\cite{CMW,Wells,bdhty}
if $\Delta_b\simeq \mathcal{O}(1)$.\ihcoups\iradiative\ For example,
\eq{bbcouplings} implies that $g_{\hl b\bar b}\simeq 0$ if
$\tan\alpha=\Delta_b/\tanb$. Inserting this result into the
corresponding expression for the $\hl\tau^+\tau^-$ coupling, it
follows that
\begin{equation}
g_{\hl\tau^+\tau^-} \simeq\frac{\sqrt{2}m_{\tau}\cosa}{v\sin\beta} \left(
\frac{ \Delta_{\tau} - \Delta_b}{ 1 + \Delta_{\tau}} \right)\,,
\qquad [{\rm if}~g_{\hl b\bar b}\simeq 0]\,.
\end{equation}
Similarly, $g_{\hh b\bar b}\simeq 0$ if $\tan\alpha=-\tanb/\Delta_b$.
Inserting this result into the corresponding expression for the
$\hh\tau^+\tau^-$ coupling, it follows that
\begin{equation}
g_{\hh\tau^+\tau^-} \simeq\frac{\sqrt{2}m_{\tau}\sina}{v\sin\beta}\left(
\frac{ \Delta_{\tau} - \Delta_b}{ 1 + \Delta_{\tau}} \right)\,,
\qquad [{\rm if}~g_{\hh b\bar b}\simeq 0]\,.
\end{equation}
In both cases, we see that although the Higgs coupling to $b\bar b$ can
be strongly suppressed for certain parameter choices, the corresponding
Higgs coupling to $\tau^+\tau^-$ may be unsuppressed.\ihcoups\  In such
cases, the $\tau^+\tau^-$ decay mode can be the dominant Higgs decay
channel for the CP-even Higgs boson with SM-like couplings to gauge
bosons.\ihbrs\

Near the decoupling limit,
$\cot\alpha\cot\beta=-1+\mathcal{O}(m_Z^2/\mha^2)$
[after an appropriate manipulation of \eq{largema4}].\idecoup\  
Inserting this result into \eq{bbcouplings}, one can check that the 
$\hl b\bar b$ coupling does indeed approach its Standard Model value.
However, because $\Delta_b\propto\tan\beta$, the deviation of
the $\hl b\bar b$ coupling from the corresponding SM result is of 
$\mathcal{O}(m_Z^2\tan\beta/\mha^2)$.  That is, at large $\tan\beta$,
the approach to decoupling may be ``delayed'' 
\cite{loganetal}, depending on the
values of other MSSM parameters that enter the radiative
corrections.\idecoupdelayed\

\subsection{MSSM Higgs boson decay modes}
\label{secec}

In this section, we consider the decay properties of
the three neutral Higgs bosons
($\hl$, $\hh$ and $\ha$) and of the charged Higgs pair ($\hpm$).\ihbrs\
Let us start with the lightest state, $\hl$.
When $\mha\gg m_Z$, the decoupling limit
applies, and the couplings of $\hl$ to SM particles are nearly
indistinguishable from those of $h_{\rm SM}$.
If some superpartners are light, there may be some additional decay
modes, and hence the $\hl$ branching ratios would be different
from the corresponding Standard Model values, even though
the partial widths to Standard Model particles are the same.
Furthermore, loops of light charged or colored superpartners could modify
the $\hl$ coupling to photons and/or gluons, in which case the one-loop
$gg$ and $\gamma\gamma$ decay rates would also be different.
On the other hand, if all superpartners are heavy, all the decay
properties of $\hl$ are essentially those of the SM Higgs boson, and the
discussion of Section~\ref{secca} applies.\ihbrssusymod\

For $\mha\gg\mz$,
the heavier Higgs states, $\hh$, $\ha$ and $\hpm$, are roughly 
mass-degenerate and have negligible couplings to vector boson pairs.\idecoup\
In particular, $\Gamma(\hh\to VV)\ll\Gamma(h_{\rm SM}\to VV)$, while the
couplings of $\ha$ and $\hpm$ to the gauge bosons are loop-suppressed.
The couplings of $\hh$, $\ha$ and $\hpm$
to down-type (up-type) fermions are significantly
enhanced (suppressed) relative to those of $h_{\rm SM}$ if $\tanb\gg 1$.
Consequently, the decay modes $\hh,\ha \to b\bar b$,
$\tau^+\tau^-$ dominate the neutral Higgs decay modes for 
moderate-to-large values of
$\tanb$ below the $t\bar t$ threshold,
while $H^+\to\tau^+\nu$  dominates the charged Higgs decay below the
$t\bar b$ threshold.\ihbrs\

For values of $\mha$ of order $\mz$, all Higgs boson states lie
below 200~GeV in mass and would be accessible at the LC.
In this parameter
regime, there is a significant area of the parameter space in which
none of the neutral Higgs boson decay properties approximates those of
$h_{\rm SM}$.\ihbrs\ For example, when
$\tan\beta$ is large, supersymmetry-breaking effects can significantly
modify the $b\bar b$ and/or the $\tau^+\tau^-$ decay rates with
respect to those of $h_{\rm SM}$.\ihbrssusymod\
Additionally, the heavier Higgs bosons can decay into lighter
Higgs bosons.  Examples of such decay modes are: $\hh\to \hl\hl$,
$\ha\ha$, and $Z\ha$, and $\hpm\to W^\pm\hl$, $W^\pm\ha$ (although in
the MSSM, the Higgs branching ratio into vector boson--Higgs boson
final states, if kinematically allowed, rarely exceeds a few percent).  
The decay of the heavier Higgs boson into two lighter Higgs bosons can
provide information about Higgs self-couplings.\ihself\  
For values of $\tanb\lsim 5$, the branching ratio of
$\hh\to \hl\hl$ is dominant for a Higgs mass range of $200~{\rm
GeV}\lsim\mhh\lsim 2m_t$. The dominant radiative corrections to
this decay arise from the corrections to the self-interaction
$\lambda_{\hh\hl\hl}$ in the MSSM and are large \cite{7}.\iradiative\

The phenomenology of charged Higgs bosons is less model-dependent,
and is governed by the values of $\tanb$ and $\mhpm$. Because
the $\hpm$ couplings are proportional to fermion masses, 
the decays to third-generation quarks and leptons are
dominant.  In particular, for $\mhpm<m_t+m_b$ (so that the channel
$H^+\to t\bar b$ is closed), $H^+\to\tau^+\nu_\tau$ is favored if
$\tan\beta\gsim 1$, while $H^+\to c\bar s$ is favored if
$\tan\beta$ is small. Indeed, ${\rm BR}(H^+\to\tau^+\nu_\tau)
\simeq 1$ if $\tan\beta\gsim 5$.\ihbrs\ These results apply
generally to Type-II two-Higgs doublet models.\itwohdm\ For $\mhpm\gsim
200$~GeV (the precise value depends on $\tanb$), 
the decay $H^+\to t\bar b\to W^+b \bar b$ is the
dominant decay mode.

In addition to the above decay modes, there exist new Higgs decay
channels that involve supersymmetric final states.\ihbrssusymod\ Higgs decays
into charginos, neutralinos and third-generation squarks and
sleptons can become important, once they are kinematically allowed
\cite{13a}. For Higgs masses below 130~GeV, the range of
supersymmetric parameter space in which supersymmetric decays are
dominant is rather narrow when the current bounds on
supersymmetric particle masses are taken into account.  One
interesting possibility is a significant branching ratio of
$\hl\to\widetilde\chi^0 \widetilde\chi^0$, which could arise  for
values of $\mhl$ near its upper theoretical limit.  
Such an invisible decay mode
could be detected at the LC by searching for the missing mass
recoiling against the $Z$ in $e^+e^-\to\hl Z$.

\subsection{MSSM Higgs boson production at the LC}
\label{seced}

For $\mha\gsim 150$~GeV, \fig{cosgraph} shows that the MSSM Higgs
sector quickly approaches the decoupling limit, where
the properties of $\hl$ approximately coincide with
those of $h_{\rm SM}$.\idecoup\  
In contrast, $|\cos(\beta-\alpha)|\ll 1$  
implies that the $\hh VV$ 
couplings are highly suppressed [see \eq{vvcoup}], and the $\ha VV$
couplings are loop-suppressed.\ihcoups\ihddiff\
Thus, the Higgsstrahlung and 
vector-boson-fusion cross-sections 
for $h_{\rm SM}$ production also apply to $\hl$ production, but
these are not useful for $\hh$ and $\ha$
production.\ihprod\ihddiff\ 
For the heavier neutral Higgs bosons,
the most robust production mechanism is $e^+e^-\to 
Z^*\to \hh\ha$, which is not suppressed since the $Z\hh\ha$ coupling
is proportional to $\sin(\beta-\alpha)$, as indicated in
\eq{hvcoup}.\ihprod\   
Radiatively corrected cross-sections  
for $e^+e^-\to Z\hl$, $Z\hh$, $\hh\ha$, and
$\hl\ha$ have been obtained in \cite{rosiek}.\iradiative\  The
charged Higgs boson is also produced in pairs via $s$-channel photon
and $Z$ exchange.\ihprod\
However, since $\mhh\simeq\mha\simeq\mhpm$ in the decoupling limit, 
$\hh\ha$
and $\hp\hm$ production are kinematically allowed only when 
$\mha\lsim\sqrt{s}/2$\idecoup\ihddiff.\footnote{The pair 
production of scalars is
P-wave suppressed near threshold, so in practice the corresponding
Higgs mass reach is likely to be somewhat lower than $\sqrt{s}/2$.}
In $\gamma\gamma$ collisions, one can extend the Higgs mass reach for
the neutral Higgs bosons.\ihprod\igamc\ihnolose\  As described
in Section~\ref{secj}, the $s$-channel resonant production of
$\hh$ and $\ha$ (due primarily to the top and bottom-quark loops in
the one-loop Higgs--$\gamma\gamma$ triangle)
can be detected for some choices of $\mha$ and $\tanb$ 
if the heavy Higgs masses are less than
about 80\% of the initial $\sqrt{s}$ of the primary $e^+e^-$ system.
The corresponding cross sections are a 
few~fb \cite{Gunion:1993ce,Muhlleitner:2001kw}.\igamc\ihnolose\

If $\mha\lsim 150$~GeV, deviations from the decoupling limit become
more pronounced, and $\hh$ can now be produced via Higgsstrahlung and
vector boson fusion at an observable rate.  In addition, 
the factor of $\cos(\beta-\alpha)$ in the $Z\hl\ha$ coupling
no longer significantly suppresses $\hl\ha$ production.
Finally, if $\mhpm\lsim 170$~GeV, the 
charged Higgs boson will also be produced in $t\to H^+ b$.\ihprod\
In the non-decoupling regime, all non-minimal
Higgs states can be directly produced and studied at the LC.\ihprod\ihnolose\

Higgs boson production in association with a
fermion-antifermion pair can also be considered.\ihprod\  Here, the new
feature is the possibility of enhanced Higgs--fermion Yukawa
couplings.\iyuk\
Consider the behavior of the Higgs couplings at large
$\tan\beta$, where some of the Higgs couplings to down type
fermion pairs (denoted generically by $b\bar b$) 
can be significantly enhanced.\ihcoups\footnote{We do not consider
the possibility of $\tan\beta\ll 1$, which would lead to enhanced Higgs
couplings to up-type fermions.  As previously noted,
models of low-energy supersymmetry 
tend to favor the parameter regime of $1\lsim\tanb\lsim m_t/m_b$. 
Moreover, some of the low $\tan\beta$ region is already
ruled out by the MSSM Higgs search (see Section~\ref{secfa}).}
Let us examine two
particular regions with large $\tan\beta$.  In the
decoupling limit (where $\mha\gg\mz$ and $|\cos(\beta-\alpha)|\ll
1$), it follows from
\eq{qqcouplings} that the $b\bar b\hh$
and $b\bar b\ha$ couplings have equal strength and are significantly
enhanced by a factor of
$\tanb$ relative to the $b\bar bh_{\rm SM}$ coupling, while the
$b\bar b\hl$ coupling is given by the corresponding Standard Model 
value.\ihcoups\idecoup\
If $\mha\lsim \mz$ and $\tanb\gg 1$, then $|\sin(\beta-\alpha)|\ll
1$, as implied by \fig{cosgraph}, and $\mhl\simeq\mha$.
In this case, the $b\bar b\hl$
and $b\bar b\ha$ couplings have equal strength and are significantly
enhanced (by a factor of
$\tanb$)\footnote{%
However in this case, the value of the $b\bar b\hh$ coupling can differ
from the corresponding $b\bar bh_{\rm SM}$ coupling when $\tanb\gg 1$,
since in case (ii), where $|\sin(\beta-\alpha)|\ll 1$,
the product $\tan\beta\sin(\beta-\alpha)$ need not
be particularly small.}    
relative to the $b\bar bh_{\rm SM}$ coupling.
Note that in both cases above,
only two of the three neutral Higgs bosons have enhanced
couplings to $b\bar b$.   If  $\phi$ is one of the two neutral Higgs 
bosons with enhanced $b\bar b\phi$ couplings,
then the cross-section for $e^+e^-\to f\bar f\phi$ 
($f=b$ or $\tau$) will be
significantly enhanced relative to the corresponding Standard Model
cross-section by a factor of $\tan^2\beta$.\ihddiff\  The phase-space
suppression is not as severe as in $e^+e^-\to t\bar t\phi$ (see
\fig{ttbarhiggs}), so this process could extend the mass reach of the
heavier neutral Higgs states at the LC given sufficient luminosity. 
The production of the charged Higgs boson via  $e^+e^-\to t\bar b H^-$
is also enhanced by $\tan^2\beta$, although this process has a more
significant phase-space suppression because of the final state top quark.
If any of these processes can be observed, it would allow for a direct
measurement of the corresponding Higgs--fermion Yukawa coupling.\ihmeas\iyuk\

\section{MSSM Higgs boson searches before the LC}
\label{secf}

\subsection{Direct search limits from LEP}
\label{secfa}

Although no direct experimental evidence for the Higgs boson yet exists,
there are both experimental as well as theoretical constraints on
the parameters of the MSSM Higgs sector. Experimental limits on
the charged and neutral Higgs masses have been obtained at LEP.
For the charged Higgs boson, $\mhpm>78.6$~GeV \cite{LEPHIGGS}.\ihdirect\
This is the most model-independent bound.  It is valid for more
general non-supersymmetric two-Higgs doublet models and assumes
only that the $H^+$ decays dominantly into $\tau^+\nu_\tau$ and/or
$c \bar s$. The LEP limits on the masses of $\hl$ and $\ha$ are
obtained by searching simultaneously for $e^+e^- \to Z \to Z\hl$
and $e^+e^- \to Z \to\hl\ha$.  
Radiative corrections can be significant,
as shown in Section~\ref{seceb},
so the final limits depend on the choice of MSSM parameters that
govern the radiative corrections.\iradiative\  The third generation
squark parameters are the most important of these.\iradiativesusy\ 
The LEP Higgs working group \cite{LEPHIGGSgroup} quotes limits for
the case of $M_{\rm SUSY}=1$~TeV in the maximal-mixing scenario, which
corresponds to the choice of third generation squark parameters
that yields the largest corrections to $\mhl$.  The present LEP
95\%~CL lower limits are $\mha>91.9$~GeV and $\mhl>91.0$~GeV.  The
theoretical upper bound on $\mhl$ as a function of $\tanb$,
exhibited in \fig{mhtanb}, can then be used to exclude a region of
$\tanb$ in which the predicted value of $\mhl$ lies below the
experimental bound.  Under the same MSSM Higgs parameter
assumptions stated above, the LEP Higgs search excludes the region
$0.5<\tanb<2.4$ at 95\%~CL.\ihdirect\

In discussing Higgs discovery 
prospects at the Tevatron and LHC, we shall quote
limits based on the assumption of $M_{\rm SUSY}=1$~TeV and maximal squark
mixing.  This tends to be a conservative assumption;  that is,
other choices give sensitivity to {\it more} of the [$\mha$, $\tan\beta]$ 
plane.  However, there are other parameter regimes  
where certain Higgs search strategies become more difficult.
While these issues are of vital importance to the Tevatron and LHC
Higgs searches, they are much less important at the LC.

\subsection{MSSM Higgs searches at the Tevatron}
\label{secfb}

The Tevatron SM Higgs search can be reinterpreted in terms of
the search for the CP-even Higgs boson of the MSSM. Since the theoretical
upper bound was found to be $\mhl\lsim 135$~GeV  (for $M_{\rm SUSY}<2$~TeV),
only the Higgs search of the low-mass region, 100~GeV $\lsim\mhl\lsim
135$~GeV,  applies.  In the MSSM at large $\tanb$, the enhancement of
the $\ha b\bar b$ coupling (and a similar enhancement of either the
$\hl b\bar b$ or $\hh b\bar b$ coupling)  provides a new search
channel: $q\bar q$, $gg\to b\bar b\phi$, where $\phi$ is a 
Higgs boson with enhanced couplings to $b\bar b$.  Combining both
sets of analyses, the Tevatron Higgs Working Group~\cite{tevreport} 
obtained the
expected 95\% CL exclusion and 5$\sigma$ Higgs discovery contours for 
the maximal mixing scenario as a function of total integrated
luminosity per detector (combining both CDF and D0 data sets).\ihnolose\

\begin{figure}[t!]
\begin{center}
\includegraphics*[width=0.75\textwidth]{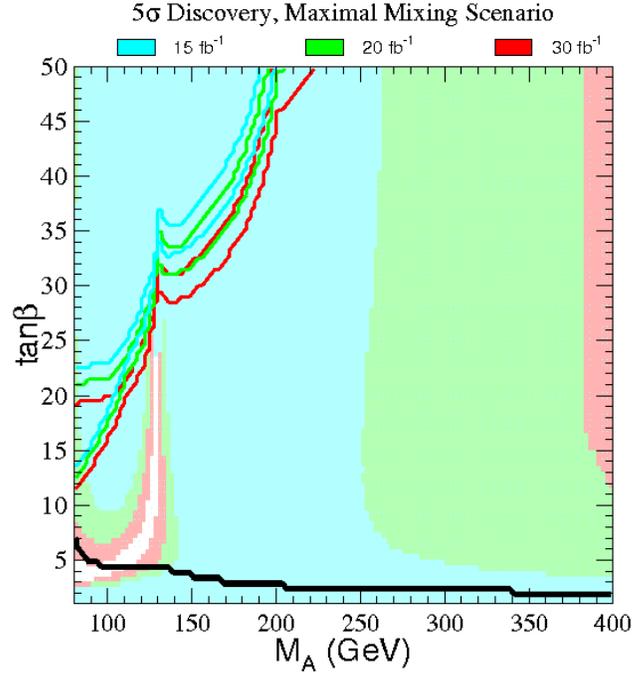}
\end{center}
\caption[0]{\label{fullmhmax95} \small
The $5\sigma$ discovery region in the $\mha$--$\tan \beta$
plane, for the maximal mixing scenario
and two different search channels:
$q\bar q\to V\phi$ ($\phi=\hl$, $\hh$), $\phi\to b\bar b$
(shaded regions) and 
$gg$, $q\bar q\to b\bar b\phi$ ($\phi=\hl$, $\hh$, $\ha$),
$\phi\to b\bar b$ (region in the upper left-hand corner bounded by the
solid lines).  Different integrated 
luminosities are explicitly shown by the color coding.
The two sets of lines (for a given color) 
correspond to the CDF and D\O\ simulations, 
respectively.  The region below the solid black line near the bottom
of the plot is excluded by the absence of observed $e^+e^-\to Z\phi$
events at LEP.  Taken from \protect\cite{tevreport}.\ihnolose\
}
\end{figure}

With 5~fb$^{-1}$ of integrated luminosity per experiment, it is
possible to 
test nearly the entire MSSM Higgs parameter space, and obtain a 95\% CL 
exclusion limit if no signal is observed.\icollparam\ihnolose\
To assure discovery of a CP-even Higgs boson at the 5$\sigma$ level,
the luminosity requirement becomes very important.
\Fig{fullmhmax95} shows that 
a total integrated luminosity of about 20~fb$^{-1}$ per experiment is 
necessary in order to assure a significant, although not exhaustive,
coverage of the MSSM parameter space.\icollparam\ 
If the anticipated 15~fb$^{-1}$
integrated luminosity is achieved, the discovery reach will
significantly extend beyond that of LEP.  
Nevertheless, the MSSM Higgs boson could still evade capture at the
Tevatron.\ihddiff\ One would then turn to the LHC to try to obtain a definitive
Higgs boson discovery.

\subsection{MSSM Higgs searches at the LHC}
\label{secfc}

The potential of the LHC to discover one or more of the MSSM Higgs
bosons has been exhaustively studied for the minimal and maximal
mixing scenarios described above. One of the primary goals of these
studies has been to demonstrate that at least one of the MSSM Higgs
bosons will be observable by ATLAS and CMS
for any possible choice of $\tanb$ and $\mha$
consistent with bounds coming from current LEP data. In order to establish
such a ``no-lose'' theorem, 
an important issue is whether or not the Higgs bosons have substantial
decays to supersymmetric particle pairs.\ihnolose\ 
It is reasonable to suppose
that these decays will be absent or relatively insignificant
for the light $\hl$.  Current mass
limits on supersymmetric particles are such that 
only $\hl\to\widetilde\chi^0_1\widetilde\chi^0_1$ is not
kinematically excluded, and this possibility arises only
in a very limited class of models.\ihbrssusymod\ For $\mha\gsim 200\gev$,
decays of the $\ha,\hh,\hpm$ to supersymmetric particles
(especially pairs of light charginos/neutralinos)
are possible, but the branching
ratios are generally not significantly large.  
In all such cases, the discovery
limits we discuss below would only be slightly weakened.
Furthermore, at high $\tanb$ the enhancement
of the $b\anti b$ and $\tau^+\tau^-$ couplings of the heavy $\ha$
and $\hh$ imply that supersymmetric decay
modes will not be important even when $\mha\sim\mhh\sim\mhpm\gg\mz$.
We will summarize the LHC discovery prospects for the MSSM
Higgs bosons assuming that supersymmetric decays are not significant.

\begin{figure}[t!]
\begin{center}
\includegraphics*[width=0.85\textwidth]{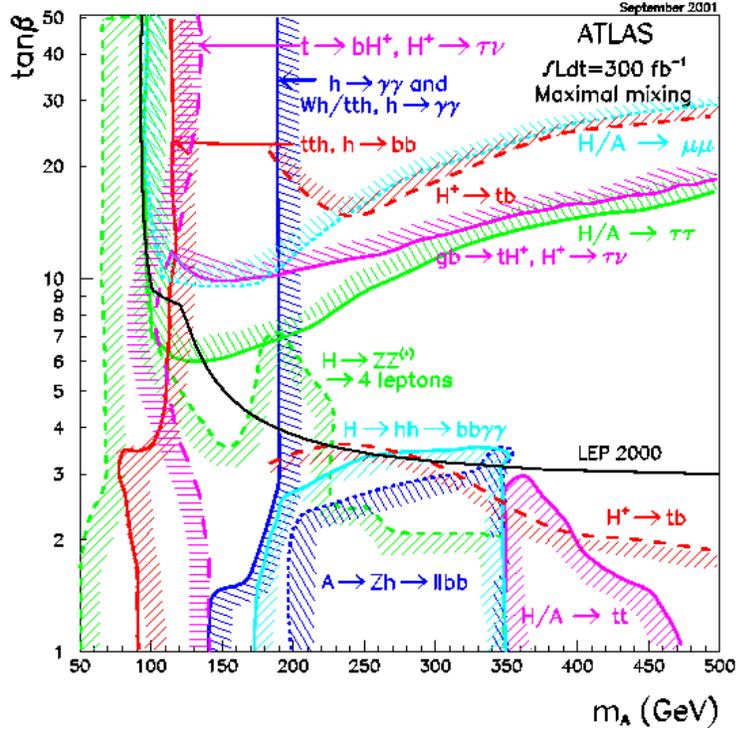}%
\end{center}
\caption[0]{\label{f:atlasmssm} \small
$5\sigma$ discovery contours for MSSM Higgs boson detection
in various channels are shown in the $[\mha,\tanb]$ parameter space, 
assuming maximal mixing and an integrated luminosity of $L=300\fbi$
for the ATLAS detector.  Taken from \protect\cite{atlasmaxmix}.\ihnolose}
\end{figure}

One of the primary Higgs discovery modes is detection of the relatively
SM-like $\hl$ using the same
modes as employed for a light $h_{\rm SM}$.\ihdiscovery\ 
Based on \fig{f:atlasmssm} (which assumes $L=300\fbi$)
\cite{atlasmaxmix}, we see that for $\mha\gsim 180\gev$,
the $\hl$ will be detected via  $gg,WW\to\hl$ and $W\hl,t\anti t\hl$ with
$\hl\to \gam\gam$, while 
the $t\anti t\hl$ with $\hl\to b\anti b$ mode is viable
down to $\mha\gsim 100$--$120\gev$, depending on $\tanb$. 
Similar results have been obtained by CMS~\cite{cmsmssmhref}.
There are also many possibilities for detecting the other MSSM Higgs
bosons.  First, there is a small domain
in which $\mha\lsim 130\gev$, but yet $\mha$ is still large enough
for consistency with LEP limits, in which $t\to b\hpm$ discovery will
be possible.  However, the most interesting alternative
detection modes are based on $gg\to\ha,\hh$ and $gb\to \hpm t$
production. We focus first on the former. For low-to-moderate $\tanb$ values,
the channels $\hh\to ZZ^{(*)}\to 4\ell$, $\hh\to \hl\hl\to
b\anti b\gam\gam$ and $\ha\to Z\hl\to\ell\ell b\anti b$ 
are viable when $\mha\lsim 2m_t$, whereas the 
$\ha,\hh\to t\anti t$ modes are viable for $\mha>2m_t$.
For large enough $\tanb$, the
$gg\to \ha,\hh\to \tau^+\tau^-,\mu^+\mu^-$ discovery modes become viable.  
For the $gb\to \hpm t$ process, the
$\hpm\to t b$ decays provide a $5\sigma$ signal
both for low-to-moderate $\tanb\lsim 2$--3 and for high $\tanb\gsim
15$--25, depending upon mass. In addition, the $\hpm\to \tau^\pm \nu$
decay mode yields a viable signal for 
$\tanb\gsim 7$--12.\ihdiscovery\ihnolose\ Of course,
if the plot were extended to higher $\mha$, the minimum $\tanb$
value required for $\hh,\ha$ or $\hpm$ detection would gradually
increase.\ihddiff\

We noted earlier that the present LEP limits imply that $\tanb>2.4$ in
the case of maximal mixing and $\msusy=1$~TeV (with even more
stringent limits possible 
in other regions of the supersymmetric parameter space).\ihdirect\ 
Thus, it is very likely that $\tanb$ and $\mha$
will be in one of two regions: (a)~the increasingly large (as
$\mha$ increases) wedge of moderate $\tanb>3$ in which only the $\hl$
will be detected; or (b)~the high $\tanb$ region for which
the $gg\to\hh,\ha\to \tau^+\tau^-,\mu^+\mu^-$ and 
$gb\to \hpm t\to \tau^\pm \nu t,tbt$ modes are 
viable as well.\ihddiff\ihnolose\
If the $\hh,\ha,\hpm$ are heavy and cannot be detected 
either at the LHC (because $\tanb$ is not large enough) or at the LC
(because they are too heavy to be pair-produced),  precision measurements
of the $\hl$ branching ratios and other properties will 
be particularly crucial.\ipew\ The precision measurements might provide the
only means for constraining or approximately determining the value
of $\mha$ aside from possible direct detection in $\gam\gam\to \hh,\ha$
production. Expected LC precisions are such that deviations of
$\hl$ branching ratios from the predicted SM values can be detected 
(in the maximal mixing scenario)
for $\mha\lsim 600\gev$~\cite{Carena:2001bg}.

At the LHC there is another important possibility for 
$\hl$ detection.\ihdiscovery\
Provided that the mass of the second-lightest neutralino exceeds that of
the lightest neutralino (the LSP) by at least $\mhl$, gluino
and squark production will lead to chain decays in which 
$\widetilde\chi_2^0\to\hl\widetilde\chi_1^0$
occurs with substantial probability.  In this way, an enormous number
of $\hl$'s can be produced, and the $\hl\to b\anti b$
decay mode will produce a dramatic signal~\cite{susytohiggs}.\ihnolose\ 

\section{Non-exotic extended Higgs sectors}
\label{secg}

In this section, we consider the possibility of extending
only the Higgs sector of the SM, leaving unchanged the gauge
and fermionic sectors of the SM. We will also consider
extensions of the two-doublet Higgs sector of the MSSM.\ihextended\itwohdm\

The simplest extensions of the minimal one-doublet Higgs sector
of the SM contain additional doublet and/or singlet Higgs fields.
\ihsinglets\
Such extended Higgs sectors will be called non-exotic (to distinguish
them from exotic Higgs sectors with higher representations, which will be
considered briefly in Section~\ref{seck}).
Singlet-only extensions have the advantage of not introducing
the possibility of charge violation, since there are no charged Higgs bosons.
In models with more than one Higgs doublet, tree-level Higgs-mediated 
flavor-changing neutral currents are present unless additional symmetries 
({\it e.g.}, discrete symmetries or supersymmetry) 
are introduced to restrict the 
form of the tree-level Higgs-fermion interactions \cite{gwp}.  
Extensions containing additional doublet fields allow for spontaneous
and explicit CP violation within the Higgs sector.\ihcpviol\ These could be
the source of observed CP-violating phenomena. 
Such models require positive squared-masses for the charged Higgs boson(s)
in order to avoid
spontaneous breaking of electric charge conservation.

Extensions of the 
SM Higgs sector containing doublets and singlets can certainly be considered
on a purely {\it ad hoc} basis.\ihsinglets\ But there are also many dynamical 
models in which the effective low-energy sector
below some scale $\Lambda$ of order 1 to 10 TeV, or higher, 
consists of the SM fermions and gauge bosons plus an extended Higgs
sector.  Models with an extra doublet of Higgs fields 
include those with new strong forces, in which the effective Higgs
doublet fields are composites containing new heavier fermions.\ieft\iewsb\
The heavy fermions
should be vector-like to minimize extra contributions to precision
electroweak observables. In many of these models,
the top quark mixes with the right-handed component of a new 
vector-like fermion. The top quark could also mix with the right-handed
component of a Kaluza-Klein (KK) excitation of a fermion field, so
that Higgs bosons would be composites of the top quark and fermionic
KK excitations. (For a review and references to the literature,
see \cite{Dobrescu:1999cs}.)
Although none of these
(non-perturbative) models have been fully developed, 
they do provide significant motivation for
studying the Standard Model with a Higgs sector containing
extra doublets and/or singlets if only as 
the effective low-energy theory below a scale $\Lambda$ in the 
TeV range.\itwohdm\ihsinglets\

When considering Higgs sectors in the context
of a dynamical model with strong couplings 
at scale $\Lambda$, restrictions on Higgs self-couplings and Yukawa
couplings, which would arise by requiring perturbativity for such couplings
up to some large GUT scale, do not apply.\iewsb\iyuk\iperturbativity\ 
At most, one should only demand 
perturbativity up to the scale $\Lambda$ at which the new (non-perturbative)
dynamics enter and the effective theory breaks down.

The minimal Higgs sector of the MSSM
is a Type-II two-doublet model, where one neutral Higgs doublet ($\Phi^0_d$)
couples at tree level
only to down quarks and leptons while the other ($\Phi^0_u$) couples
only to up quarks.\ihcoups\itwohdm\  Non-minimal extended Higgs sectors 
are also possible in low-energy supersymmetric models.
Indeed, string theory realizations of low-energy supersymmetry often
contain many extra singlet, doublet and even higher representations, some of
which can yield light Higgs bosons  
(see, {\it e.g.}, \cite{Cvetic:2000nc}).\ihextended\
However, non-singlet Higgs representations spoil gauge coupling unification,
unless additional intermediate-scale matter fields 
are added to restore it.\icoupu\
A particularly well-motivated extension is the inclusion of 
a single extra complex singlet Higgs field, often denoted $S$.\ihsinglets\
Including $S$, the superpotential
for the theory can contain the term $\lambda_S H_u H_d S$, which can then
provide a natural source of a weak scale value for the $\mu$
parameter appearing in the bilinear superpotential form $\mu H_u H_d$
required in the MSSM. A weak-scale value for
$s\equiv \VEV{S^0}$, where $S^0$ is the scalar component of the superfield $S$,
is natural and yields an effective $\mu=\lambda_S s$.
This extension of the MSSM is referred to as the next-to-minimal supersymmetric
model, or NMSSM, and has received considerable attention. For
an early review and references, see \cite{hhg}.

\subsection{The decoupling limit}
\label{secga}

In many extended Higgs sector models, the most natural parameter
possibilities correspond to a decoupling limit in which there is only
one light Higgs boson, with Yukawa and vector boson couplings close to
those of the SM Higgs boson.\idecoup\iyuk\   
In contrast, all the other Higgs bosons
are substantially heavier (than the $Z$) with negligibly small relative mass
differences, and with suppressed vector boson couplings (which vanish in
the exact limit of decoupling).  By assumption, the decoupling limit
assumes that all Higgs 
self-couplings are kept fixed and perturbative 
in size.~\footnote{In the decoupling limit, the heavier Higgs bosons
may have enhanced couplings to fermions ({\it e.g.}, at large
$\tan\beta$ in the two-Higgs doublet model).
We assume that these couplings also remain
perturbative.}\iperturbativity\  
In the MSSM, such a decoupling limit arises for
$\mha\gg\mz$, and quickly becomes a very good approximation for
$\mha\gsim 150$~GeV.

The decoupling limit can be evaded in special cases, in which the
scalar potential exhibits a special form ({\it e.g.}, a discrete
symmetry can forbid certain terms).  In such models, there could exist
regions of parameter space in which all but one Higgs boson are
significantly heavier than the $Z$, but the light scalar state does
{\it not} possess SM-like properties~\cite{Chankowski:2000an}. A
complete exposition regarding the decoupling limit in the 
two-Higgs doublet model, and
special cases that evade the limit can be 
found in \cite{gunhabdecoup}.\ihextended\ihnondecoup\

\subsection{Constraints from precision electroweak data and LC implications}
\label{secgb}

In the Standard Model, precision electroweak
constraints require $m_{h_{\rm SM}}\lsim 193\gev$ at 95\% CL.\ipew\
This is  precisely
the mass region preferred in the MSSM and its extensions.\ipew\
However, in models of general extended Higgs sectors with
only weak doublet and singlet representations,
there are more complicated possibilities.\ihextended\  First, it could
be that there are several, or even many, 
Higgs bosons that couple to vector bosons and it is only 
their average mass weighted by the square of their $VV$ coupling
strength (relative to the SM strength) that must obey 
the SM Higgs mass limit.\ihextended\ihsinglets\itwohdm\
Second, there can be weak isospin violations either within the
Higgs sector itself or involving extra dynamics (for example related
to the composite Higgs approach)  can compensate for 
the excessive deviations predicted if there is a SM-like
Higgs boson with mass substantially above $200\gev$.  

\begin{figure}[t!]
\begin{center}
\includegraphics[width=12cm]{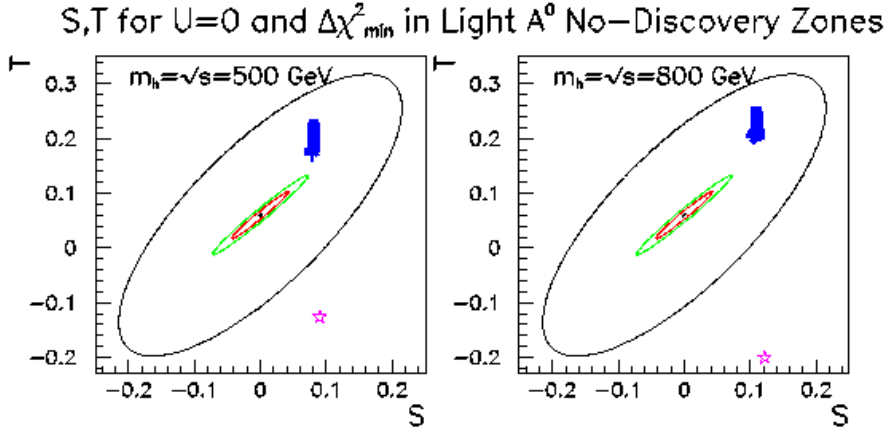}
\caption{\label{dsdt}
The outer ellipses show the 90\% CL region from current
precision electroweak data in the $S,T$ plane
for $U=0$ relative to a central point defined
by the SM prediction with $\mhsm=115$ GeV. 
The blobs of points show the $S,T$ predictions for 2HDM  models 
with a light $\ha$ and with
$\tanb$ such that the $\ha$ cannot be detected
in $b\anti b\ha$ or $t\anti t\ha$ production
at either the LC or the LHC; the mass of
the SM-like $\hl$ is set equal to $\rts=500\gev$ (left)
or $800\gev$ (right) and $\mhpm$ and $\mhh$ have been
chosen to minimize the $\chi^2$ of the full precision electroweak fit.
The innermost (middle) ellipse shows the  90\% (99.9\%) CL
region for $\mhsm=115$ GeV after
the Giga-$Z$ operation {\it and} a $\Delta m_W\protect\lsim 6$ MeV threshold
scan measurement. The stars to the bottom right show the $S,T$ predictions
in the case of the SM with $\mhsm=500\gev$ (left) or $800\gev$ (right).
This figure is from \protect\cite{Gunion:2000ab}.} 
\end{center}
\end{figure}

A particularly simple
example of this latter possibility arises in the context of the 
two-Higgs doublet model (2HDM) \cite{Chankowski:2000an}
and is illustrated in \fig{dsdt}.
\itwohdm\ihextended\
Consider a 2HDM in which one of the CP-even neutral
Higgs bosons has SM-like couplings but has mass
just above a particular presumed value of $\sqrt s$ ($500$ or $800\gev$)
for the linear collider. In addition, focus on cases in which
there is a lighter $\ha$ or $\hl$ 
with no $VV$ coupling (for either, we use the notation $\what h$) 
and in which all other Higgs bosons have
mass larger than $\sqrt s$. Next, isolate mass and $\tanb$
choices for which detection of the $\what h$ 
will also be impossible at the LC. Finally, scan over masses of the
heavy Higgs bosons so as to achieve the smallest
precision electroweak $\Delta\chi^2$ relative to that found
in the Standard Model 
for $m_{h_{\rm SM}}=115\gev$~\cite{Gunion:2000ab}.\ipew\  
For $\what h=\ha$, the blobs of overlapping points in \fig{dsdt} 
indicate the $S,T$ values for the optimal choices
and lie well within the current 90\% CL ellipse.
The heavy Higgs boson with SM couplings gives a large positive contribution
to $S$ and large negative contribution to $T$, and in the absence
of the other Higgs bosons would give the $S,T$ location indicated
by the star.  However, there is an additional positive contribution
to $T$ arising from a slight mass non-degeneracy among the
heavier Higgs bosons.  For instance, for the case of a light $\what h=\ha$,
the $\hl$ is heavy and SM-like and  $\Delta\rho\equiv\alpha \Delta T$, where
\beqa
\Delta T &=&
\frac{1}{16 \pi m_W^2 \cos^2\theta_W}\left\{\frac{\cos^2\theta_W}
{\sin^2\theta_W}
   \left(\frac{m_{\hpm}^2-m_{\hh}^2}{2}\right)\right. \nonumber \\[6pt]
&&\qquad\qquad\qquad \left. -3m_W^2\left[\ln\frac{m_{\hl}^2}{m_W^2}
   +\frac{1}{6}+\frac{1}{\sin^2\theta_W}\ln\frac{m_W^2}{m_Z^2}\right]\right\}
\label{drhonew}
\eeqa
can be adjusted by an appropriate choice of $\mhpm^2-\mhh^2$
to place the prediction in the $S$--$T$ plane at the location of the 
blob in \fig{dsdt}.
Indeed, even if the ``light'' decoupled Higgs boson is not so light,
but rather has mass equal to $\rts$ (and is therefore unobservable),
one can still obtain adequate agreement with current
precision electroweak data.\ihddiff\ipew\ Fortunately, one can only
push this scenario so far. To avoid moving beyond the current 90\%
ellipse (and also
to maintain perturbativity for the Higgs self-couplings), 
the Higgs with SM-like $VV$ coupling must have mass $\lsim 1\tev$.

In composite Higgs models with extra fermions, there are similar
non-degeneracies of the fermions that can yield a similar positive contribution
to $\Delta T$.\iewsb\ihextended\ As reviewed in \cite{peskinwells},
consistency with current precision electroweak data inevitably constrains
parameters so that some type of new physics (including a possible heavy
scalar sector) would again have to lie below a TeV or so.  
Future Giga-$Z$ and $W$-threshold measurements
could provide much stronger constraints
on these types of models, as discussed in Section~\ref{seci}.\igigaz\
Moreover, such measurements
would become a priority if a heavy SM Higgs boson is found at the LHC,
but no other new physics (needed for a consistent explanation of the
precision electroweak data) is discovered.

\subsection{Constraints on Higgs bosons with $VV$ coupling}
\label{secgc}

In the MSSM, we know that the Higgs boson(s) that carry the $VV$ coupling must
be light: if $\mha$ is large (the decoupling limit\idecoup) then it is
the mass-bounded $\hl$ that has all the $VV$ coupling strength; if
$\mha\lsim 2\mz$, then the $\hh$ can share the $VV$ coupling with
the $\hl$, but then $\mhh$ cannot be larger than  about $2\mz$.\ihcoups\ipew\
In the NMSSM, assuming Higgs-sector
CP conservation, there are 3 neutral CP-even Higgs 
bosons, $h_{1,2,3}$ ($m_1<m_2<m_3$), 
which can share the $VV$ coupling strength. One can show 
(see \cite{Ellwanger:1999ji} for a recent update)
that the masses of the $h_i$ with substantial $VV$
coupling are strongly bounded from above.\ihiggsbounds\
This result generalizes to the most general supersymmetric Higgs 
sector as follows.  Labeling the neutral Higgs bosons by $i$
with masses $m_{h_i}$ and denoting the $ZZ$ squared-coupling relative to the SM
by $K_i$, it can be shown that~\cite{Espinosa:1999xj}
\begin{equation} 
\sum_i K_i \geq 1\,,\quad\qquad \sum_i K_i m_{h_i}^2\leq m_B^2\,.
\label{hsumrules}
\end{equation}
where the value of $m_B$ depends somewhat on the low-energy
supersymmetric model.\ihcoups\
That is, the aggregate strength of the $VV$ squared-coupling
of all the neutral Higgs bosons is at least that of the SM, and the
squared-masses of the neutral $h_i$ weighted by the squared-couplings
must lie below a certain bound.  A value of
$m_B\sim 200\gev$ in \eq{hsumrules} 
is obtained \cite{Espinosa:1998re} by 
assuming that the low-energy supersymmetric theory remains
perturbative up to the GUT scale of order $10^{19}\gev$.\iperturbativity\
This value of $m_B$ applies for the most general 
possible Higgs representations (including triplets) in
the supersymmetric Higgs sector and for arbitrary numbers
of representations.\ihtriplets\  If only doublet and singlet
representations are
allowed, the bound would be lower.\ihsinglets\  
The value of $m_B\sim 200\gev$ 
also applies to general Higgs-sector-only extensions
of the SM by requiring consistency with precision electroweak 
constraints {\it and} assuming the absence of a large contribution
to $T$ from the Higgs sector itself or from new physics,
such as discussed in \sect{secgb}.\ipew\

\subsection{Detection of non-exotic extended Higgs sector scalars
at the Tevatron and LHC}
\label{secgd}

In the case of extended Higgs sectors,  all of
the same processes as discussed for the SM and MSSM will again
be relevant.\ihdiscovery\ihextended\ 
However, one can no longer guarantee Higgs discovery
at the Tevatron and/or LHC.\ihddiff\
In particular, if there are many Higgs bosons sharing
the $WW,ZZ$ coupling, Higgs boson discovery 
based on processes that rely on the $VV$ coupling could be much more difficult 
than in models with just a few light Higgs bosons with substantial $VV$
coupling.\ihextended\ 
This is true even if the sum rule of \eq{hsumrules} applies.
For example, in the NMSSM, the addition
of one singlet to the minimal two-doublet Higgs structure
allows for parameter choices such that no Higgs boson can be
discovered at the LHC
\cite{Gunion:1996fb,Ellwanger:2001iw} using any of the processes
considered for SM Higgs and MSSM Higgs detection.\ihsinglets\ihddiff\ihnolose\
The problematical
regions of parameter space originally discussed in \cite{Gunion:1996fb}
were those in which the $\gam\gam$ decay channel 
signals for the CP-even Higgs bosons are 
decreased (because of decreased $W$-loop
contribution to the coupling) and a moderate value
of $\tanb$ was chosen so that 
$t\anti t+$Higgs processes are weak and $b\anti b+$Higgs
processes are insufficiently enhanced.\ihddiff\
However, as shown
in \cite{Ellwanger:2001iw}, the recent addition of the 
$t\anti t \h\to t\anti t b\anti b$ and, especially, the 
$W^*W^*\to \h\to\tau^+\tau^-$ discovery modes to the list of viable
channels by the CMS and ATLAS collaborations means that 
discovery of at least one NMSSM Higgs bosons would be possible in
all the ``bad'' parameter regions found in \cite{Gunion:1996fb}.
However, \cite{Ellwanger:2001iw} also finds that there
are parameter choices (explicitly excluded in 
the study of \cite{Gunion:1996fb}) such that the CP-even Higgs bosons
decay primarily to a pair of CP-odd Higgs bosons. Discovery
techniques for this type of final state have not been developed
for the LHC.\ihddiff\ihnolose\

However, in other cases, the Tevatron and LHC could observe
signals not expected in an approximate decoupling limit.  For example,
in the 2HDM model discussed earlier the light $\what h$
with no $VV$ couplings  decays via $\what h \to b\anti b,\tau^+\tau^-$
and discovery in  $t\anti t \what h$, $b\anti b \what h$
and even $gg\to \what h$ \cite{gungrnew} is  possible, 
though certainly not guaranteed. Further, in these models
there is a heavy neutral Higgs boson having the bulk of the $VV$ coupling
and (for consistency with current precision
electroweak constraints or with perturbativity) a mass 
less than about 1~TeV.  This latter
Higgs boson would be detected at the LHC 
using $gg,WW$ fusion production and $ZZ\to 4\ell,WW\to 2j\ell\nu,\ldots$
decay modes, just like a heavy SM Higgs boson.\ihnolose\

\subsection{LC production mechanisms for non-exotic extended 
Higgs sector scalars} 
\label{secge}

Any physical Higgs eigenstate  with substantial $WW$ and $ZZ$ coupling
will be produced in Higgsstrahlung and $WW$ fusion at the LC.\ihprod\
Although there could be considerable cross section dilution and/or 
resonance peak overlap, the LC will nonetheless
always detect a signal.  This has been discussed for the MSSM
in Section~\ref{seced}.  In the NMSSM,
if one of the heavier CP-even $h_i$ has most
of the $VV$ coupling, the strong bound on its mass~\cite{Ellwanger:1999ji}
noted earlier implies that it will be detected at any LC with 
$\sqrt s>350\gev$ within a small fraction of a 
year when running at planned 
luminosities~\cite{Kamoshita:1994iv}.\ihsinglets\ihnolose\
The worst possible case is that
in which there are many Higgs bosons 
with $VV$ coupling with masses spread
out over a large interval with separation smaller than the mass
resolution.\ihddiff\
In this case, the Higgs signal becomes a kind of continuum
distribution. Still, in \cite{Espinosa:1999xj} it is shown
that the sum rule of \eq{hsumrules} guarantees 
that the Higgs continuum signal will still be detectable for sufficient
integrated luminosity, $L\gsim 200\fbi$, as a broad excess in the recoil
mass spectrum of the $e^+e^-\to ZX$ process.\ihnolose\
(In this case, $WW$ fusion events
do {\it not} allow for the reconstruction of Higgs events independently
of the final state Higgs decay channel.)
As already noted, the value of $m_B\sim 200\gev$ 
appearing in \eq{hsumrules} 
can be derived from the constraints of 
perturbative evolution up to some high energy (GUT or Planck) scale
for the most general Higgs sector
in supersymmetric theories.  The same bound on $m_B$ is also required 
by precision electroweak data for general SM Higgs sector
extensions, at least in theories that do not have a large positive contribution
to $T$ from a non-decoupling structure in the Higgs sector
or from new physics not associated with the Higgs sector.

Other production modes of relevance include 
Higgs pair production~\cite{hpair}, 
single Higgs production in association with
$t\anti t$~\cite{Djouadi:1992tk,Bar-Shalom:1995jb,Gunion:1996vv,%
Dittmaier:1998dz,dawsontth,dawsonbbh,Dittmaier:2000tc,Grzadkowski:2000wj} 
and with 
$b\anti b$~\cite{hbb,dawsonbbh,Dittmaier:2000tc,Grzadkowski:2000wj}.\ihprod\
In multi-doublet models, 
$t\anti b H^-$/$b\anti t H^+$~\cite{Djouadi:1992tk,Kanemura:2000cw,htb}
and $\tau\nu\hpm$~\cite{Kanemura:2000cw,htaunu,cpyuan}
production are also present (and in some cases $c\anti bH^-$/$b\anti c H^+$
production is competitive~\cite{cpyuan}).  The
associated production of a charged Higgs boson and gauge boson
has also been studied~\cite{logansu}.
However, none of these modes are guaranteed 
to be either kinematically accessible at the LC or to 
have a sufficiently high event rate to be observed
(if kinematically allowed).\ihddiff\
Regardless of the production process,
relevant decay channels could include cases where 
heavier Higgs bosons decay to lighter ones.
If observed, such decays would provide vital information on
Higgs self-couplings~\cite{hhcouplings}.

\begin{figure}[t!]
\begin{center}
\includegraphics*[width=0.7\textwidth]{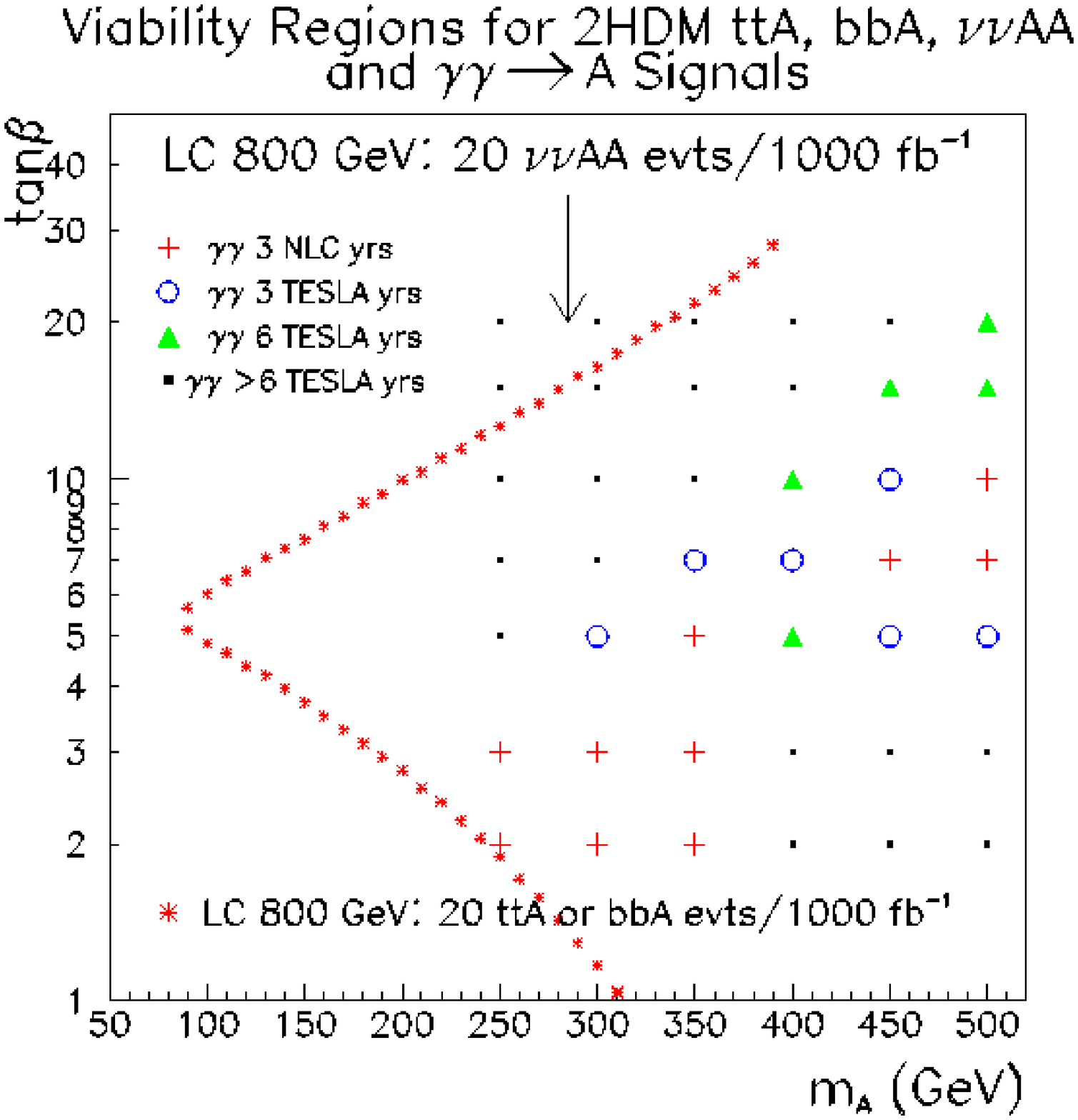}
\end{center}
\caption{\label{wedge} The stars form an outline of a wedge of $[\mha,\tanb]$
parameter space inside which the LC operating at $\rts=800\gev$ yields
fewer than 20 events per $1000\fbi$ 
in both the $t\anti t \ha$ and $b\anti b \ha$
production modes. Also shown by the arrow is the $\mha$ value above
which the process $\epem\to \nu\anti\nu \ha\ha$ yields fewer than 20
events per $1000\fbi$. The $+$ symbols on the grid of $[\mha,\tanb]$ values
show the points for which a $4\sigma$ signal for $\gam\gam\to\ha$
would be achieved using LC operation at $\rts=630\gev$ after 
three years  of operation assuming running conditions 
and strategies as specified in \protect\cite{gunasner}
using the NLC-based design of the $\gam\gam$ collider.
(The nominal $\epem$ luminosity for the underlying
NLC design employed is $220~{\rm fb}^{-1}$ per 
year at $\sqrt s=500$~GeV.)
The circles (triangles) show the additional points that would yield
a $4\sigma$ signal for twice (four times)
the integrated luminosity of the current
NLC $\gam\gam$ interaction region design.
The small squares show the additional points
sampled in the study of \protect\cite{gunasner}.  
This figure was taken from \protect\cite{gunfarris}.
}
\end{figure}

It is also important to consider the production processes that
are most relevant for those Higgs bosons (denoted $\what h$)
which do not have substantial $VV$ couplings.\ihprod\
Such processes have particular relevance in the non-decoupling scenario
for the general 2HDM model discussed earlier.\ihextended\  In this scenario,
$\what h$ is the only Higgs boson light enough to be produced at an
LC with $\rts\lsim 1\tev$, but it
cannot be produced and detected in $WW$ fusion or Higgsstrahlung.\ihddiff\
Since the other Higgs bosons are heavy, the $\what h$ also cannot
be produced in association with another Higgs boson.
As shown in \cite{Grzadkowski:2000wj,Chankowski:2000an}, the
$b\anti b\what h$ and $t\anti t\what h$ processes will also
not be detectable at the LC if $\tanb$ is moderate
in value. The most interesting tree-level processes
are then those based on the quartic couplings 
$WW\what h\what h$ and $ZZ\what h\what h$ required 
by gauge invariance \cite{Haber:1993jr}.
These couplings allow for $WW\to \what h\what h$ fusion and 
$Z^*\to Z\what h\what h$
production, respectively.  The exact cross sections
for these processes are only mildly sensitive to the 
masses of the other heavier Higgs bosons via 2HDM Higgs 
self-couplings. Of course, phase space restrictions
imply an upper limit on the $\what h$ masses that can 
be probed in this way.\ihnolose\ihddiff\

\begin{figure}[t!]
\begin{center}
\includegraphics*[width=0.8\textwidth]{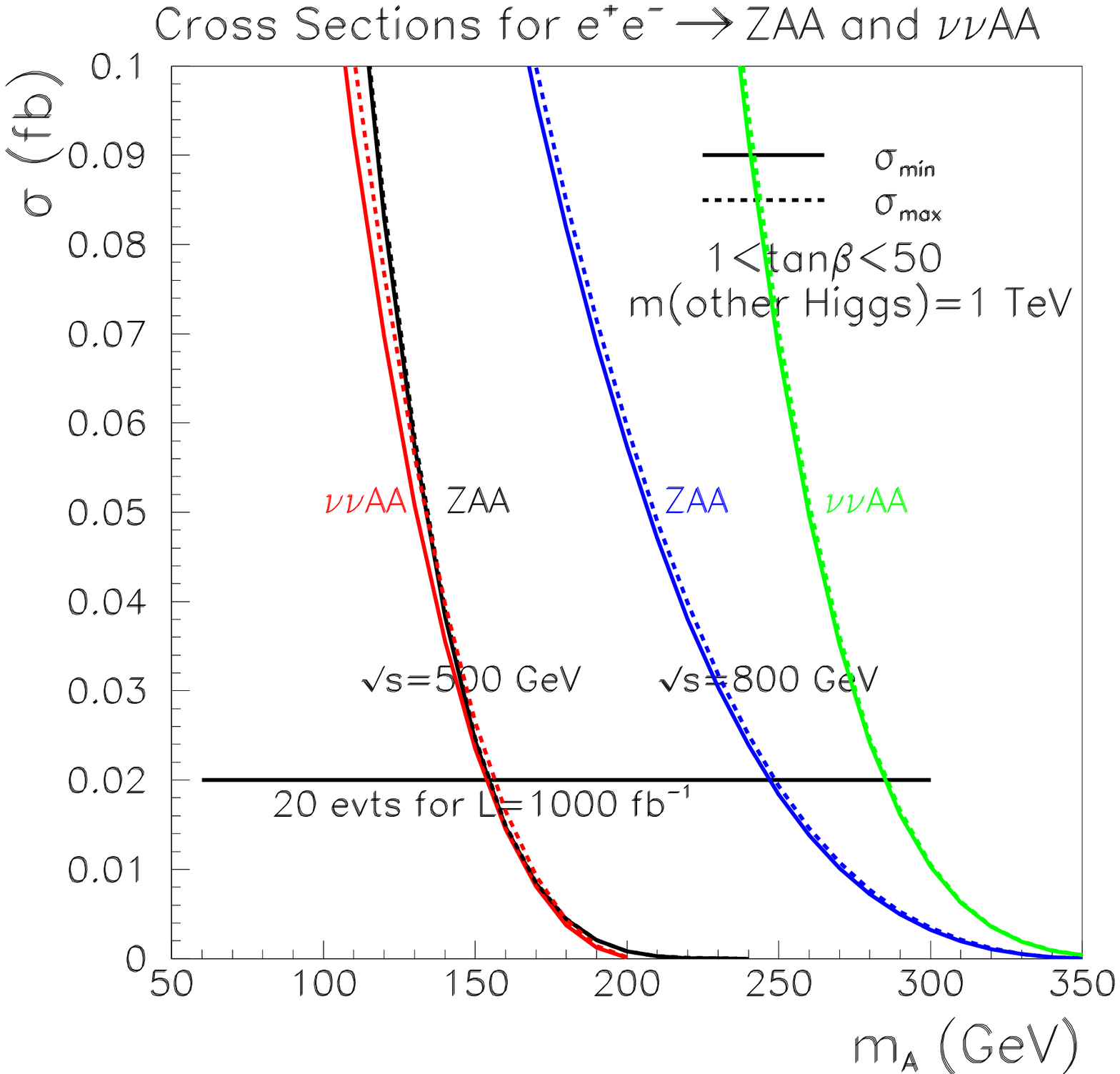}
\end{center}
\caption[0]{\label{aaz_nnaa_fig} 
The cross sections for $\epem\to Z\ha\ha$ and $\epem\to \nu\anti\nu
\ha\ha$ as a function of $\mha$ are shown, assuming a 2HDM model with a heavy
SM-like $\hl$. We have taken $\mhl=\mhh=\mhpm=1\tev$.
Maximum and minimum values found after scanning $1\leq\tanb\leq 50$
are shown for $\rts=500\gev$ and $800\gev$.  
The variation with $\tanb$ arises from small contributions
associated with exchanges of the heavy Higgs bosons. The 20 event level
for $L=1000\fbi$ is indicated.  Taken from \protect\cite{gunfarris}.}
\end{figure}

The case of $\what h=\ha$ is illustrated
in \figs{wedge}{aaz_nnaa_fig}~\cite{gunfarris}.\ihddiff\
\Fig{wedge} shows the wedge region in which
fewer than 20 events per $1000\fbi$ are obtained in $b\anti b \ha$
and $t\anti t \ha$ production at $\rts=800\gev$. 
\Fig{aaz_nnaa_fig} shows the $\mha$ values for which fewer than
20 events are obtained in $Z^*\to Z\ha\ha$ and $W^*W^*\to\ha\ha$
production at $\rts=500\gev$ and $800\gev$. Thus, single $\ha$
discovery is not possible via any of the above processes in the region of
the wedge with $\mha\gsim 300\gev$.  With
the $\gam\gam$ collider option at the LC, we note that
$\gam\gam\to \ha$ can provide a signal for the decoupled $\ha$
over a significant portion of the wedge region.\ihnolose\ The
results from a realistic study of \cite{gunasner} are
illustrated in \fig{wedge}, which focuses on
$\mha\geq 250\gev$.   In particular, the
operation of the $\gamma\gamma$ collider for two years using a polarization
configuration for the electron beams and laser photons 
yielding a broad $E_{\gam\gam}$ spectrum
and one year using a configuration yielding a peaked spectrum is 
assumed.\igamc\
The pluses indicate $4\sigma$ discovery points after three
years of operation (assuming that the machine operates for $10^7$~sec
during one calendar year) 
in the appropriate configurations at the NLC.
The results of \fig{wedge} employ the $E_{\gam\gam}$ luminosity 
spectra as computed in \cite{gunasner} based on the particular laser
and interaction region design that is discussed in more detail
in Section~\ref{secj}. A factor of two higher $\gam\gam$ luminosity,
as might be achievable at TESLA or through further design improvement
at the NLC, 
would allow $4\sigma$ discovery for the additional points indicated
by the circles.  The corresponding
results for a decoupled CP-even $\what h=\hl$ are similar.

\section{Exotic Higgs sectors and other possibilities}
\label{seck}

As we have seen, there are many scenarios and models in which the
Higgs sector is more complicated than the one-Higgs-doublet of the
Standard Model.\ihextended\ 
Supersymmetry requires at least two Higgs doublets. Even
in the absence of supersymmetry, a two-doublet 
Higgs sector allows for
CP-violating phenomena.\itwohdm\ 
Singlets can also be added without altering
the tree-level prediction of $\rho=1$.\ihsinglets\  However, the possibility of
Higgs representations with still higher weak isospin (left-handed, denoted
by $L$) and/or hypercharge should not be ignored. 
Indeed, for judicious choices
of the numbers and types of such representations, gauge coupling
unification (although at scales below $10^{15}\gev$)
is possible without introducing 
supersymmetry~\cite{Gunion:1996pq}.\icoupu\ihextended\
The main drawback of introducing scalars of this type is that, for
triplets and most higher representations, if the vacuum expectation
value of the neutral Higgs field member of the representation is
non-zero ($v_L\neq 0$) then $\rho$ becomes infinitely renormalized and
can no longer be computed \cite{gvw}.\ipew\  In this case, $\rho$ becomes an
independent input parameter to the theory.  Triplets  
have received the most attention, as they arise naturally in
left-right symmetric extensions of the Standard Model gauge group
\cite{leftright}.\ihtriplets\ 
(These and other models that utilize Higgs triplets are
reviewed in \cite{hhg}.)  In this section we will also briefly
consider the Higgs-like pseudo-Nambu-Goldstone bosons  that arise in
generic technicolor theories.\ipngb\

\subsection{A triplet Higgs sector} 

Including a single complex  SU(2)-triplet Higgs representation, in addition
to some number of doublets and singlets, 
results in six additional physical Higgs eigenstates:  
$H^{--,++}$, $H^{-,+}$, $H^0$ and $H^{0\,\prime}$.\ihtriplets\
All but the doubly-charged states
can mix with the doublet/singlet Higgs states 
under some circumstances.\ihdmm\  Even if $v_L\neq 0$
for the neutral field, $\rho=1$ can be preserved
at tree level if, in addition, a real triplet field
is also included \cite{Georgi:1985nv,Chanowitz:1985ug}.\ipew\
However, $\rho$ will still be infinitely renormalized
at one-loop unless $v_L=0$ is chosen.
Left-right (L-R) symmetric models capable of yielding
the see-saw mechanism for neutrino mass generation {\it require} two 
triplet Higgs representations (an L-triplet and an R-triplet).\ilrmodels\
The large see-saw mass entry, $M$, 
arises from a ``Majorana coupling'' which L-R symmetry requires
to be present for both the L-triplet and R-triplet
representations.~\footnote{The so-called Majorana couplings
correspond to the terms in the Yukawa interactions in which 
a fermion bilinear of lepton number $+2$ or $-2$ 
({\it e.g.}, the latter contains $\nu\nu$ and $e^-e^-$) couples 
to the triplet fields.\iyuk\   When the neutral component of the R-triplet
acquires a vacuum expectation value, a Majorana mass term for the
right-handed neutrino is generated, and lepton number is spontaneously
broken.  In this model ${\rm B}-{\rm L}$ is a spontaneously 
broken gauged U(1) symmetry.}\imajorana\ 
Again, $\rho$ will not be altered
if $v_L=0$, but $v_R$ must be non-zero and large for 
large $M$.
We will briefly discuss the phenomenology of an L-triplet;
the corresponding phenomenology of the R-triplet 
is quite different. (See \cite{hhg} for a review.)\ihtriplets\
 
The resulting Higgs phenomenology can be very complex.
We focus on the most unequivocal signal for a triplet 
representation, namely observation of a doubly-charged 
Higgs boson.\ihnolose\ihdmm\
Pair production, $Z^*\to H^{++}H^{--}$ would be visible at
the LHC for $m_{H^{--}}\lsim 1\tev$ \cite{Gunion:1996pq}, but has limited
mass reach, $m_{H^{--}}<\sqrt s/2$, at the LC.\ihddiff\
Fortunately, single production of a doubly-charged Higgs boson
at the LC is also 
generally possible.\ihprod\  In particular, the generically-allowed
Majorana coupling 
leads to an $e^-e^-\to H^{--}$ coupling and the possibility of $s$-channel
resonance production of the $H^{--}$ in $e^-e^-$ collisions.\imajorana\
Observation of this process 
would provide a dramatic confirmation of the presence of
the Majorana coupling and, in many cases, the ability to actually
measure its magnitude.\ihmeas\ For a discussion and review, see
\cite{Gunion:1996mq}. 
If the $H^{--}$ is heavy {\it and} has significant $W^-W^-$
coupling (requiring $v_L\neq 0$), then it can become broad and the $s$-channel
resonant production cross section is 
suppressed (see, \eg, \cite{Gluza:1997kg}) and might
not be observable. Another production mechanism
sensitive to the $e^-e^-\to H^{--}$ coupling 
that might be useful in such an instance is $e^-e^-\to H^{--} Z$, and
$e^-e^-\to H^{-}W^-$ will be sensitive to the $e^-\nu_e\to H^-$
coupling that would be present for the $H^-$ member of the triplet
representation \cite{Alanakian:1998ii}.\ihtriplets\
Using just the Majorana
coupling, doubly-charged Higgs bosons can also be produced
via $e^-\gam \to e^+ H^{--}$ and $e^+e^-\to 
e^+e^+ H^{--}$~\cite{Barenboim:1997pt} and the singly-charged
members of the same representation can be produced in 
$e^-e^-\to H^- W^-$ \cite{Alanakian:1998ii}.\ihprod\

Despite the loss of predictivity for $\rho$, it could be that
non-zero $v_L$ is Nature's choice.
In this case, the $e^-e^-$ collider option again has some unique advantages.
The neutral, singly-charged and doubly-charged Higgs bosons
of the triplet representation
can {\it all} be produced (via $ZZ$ fusion, $W^- Z$ fusion and $W^-W^-$
fusion, respectively).\ihdmm\  For example, \cite{Barger:1994wa}
studies $W^-W^-\to H^{--}$ fusion.\ihprod\

\subsection{Pseudo Nambu Goldstone bosons}

In the context of technicolor and related theories,
the lowest-mass states are typically a collection of
pseudo-Nambu-Goldstone bosons, of which the lightest
is very possibly a state $P^0$ which can have mass 
below $200\gev$ and couplings and other properties
not unlike those of a light SM-like Higgs boson.\ipngb\  Typically, its
$WW,ZZ$ coupling is very small (arising via loops or anomalies),
while its $b\anti b$ coupling can be larger.\ipngb\
The phenomenology of such a $P^0$ was studied in 
\cite{Casalbuoni:1999fs,Lane:2002wb}. The best modes
for detection of the $P^0$ at the LC are $e^+e^-\to\gam P^0\to \gam b\anti b$
and $\gam\gam\to P^0\to b\anti b$.\ipngb\ Since the $P^0$ is likely to
be discovered at the LHC in the $\gam\gam$ final state, we will know
ahead of time of its existence, and precision measurements of
its properties would be a primary goal of the LC.
High integrated luminosity would be required.

\section{LC Measurements of Higgs Boson Properties}
\label{sech}

In addition to the observation of one or more Higgs boson(s),
an essential part of the LC physics program consists of measuring
the mass, width, and couplings of the Higgs boson(s) with 
precision, and the determination of the scalar potential 
that gives rise to electroweak symmetry breaking.\ihmeas\

Determinations of the Higgs couplings are needed to demonstrate
that a Higgs boson generates mass for vector bosons, charged leptons, and
up- and down-type quarks.\ihcoups\
Whether there is only one Higgs doublet
can be checked if the ratios of measured branching ratios of directly coupled 
particles are indeed proportional to the ratios of the squared-masses. 
Small variations can distinguish between a SM Higgs and $\hl$ of the MSSM 
with couplings close to the SM Higgs boson.\idecoup\
Couplings are determined
through measurements of Higgs branching ratios and cross sections.
Higgs bosons are also expected to couple to themselves, and this self-coupling
$\lambda$ can only be explored through the direct production of two or
more Higgs bosons.\ihself\
It is in this category of {\it direct} and {\it model independent}
determination of many absolute couplings (and not just their ratios)
that the strength
of the LC Higgs physics program really stands out.

Details of some of the studies can be found in
\cite{Battaglia:2000jb} and a comprehensive description of
European studies using results of the simulated TESLA
detector can be found in \cite{tesla_report}.
North American studies consider simulations of detectors
with capabilities as described elsewhere~\cite{resourcebook}.
The program of measurements of Higgs boson properties strongly
impacts detector design. Measurement of branching ratios into fermions
requires sophisticated vertex detectors to separate $b$ from $c$
(and gluon) jets.\ihbrs\  Precise recoil mass measurements opposite 
leptons need excellent momentum resolution (particularly for $\mu^+\mu^-$)
from charged particle tracking. 
The performance of the combined tracking and calorimetry systems
needs to result in precise jet-jet invariant masses, missing mass
measurements, and the ability to separate hadronic $W$ from hadronic
$Z$ decays.

The specific measurements used to determine the Higgs couplings to 
vector bosons, fermions and scalars are significantly different
depending on the mass of the Higgs boson.\ihcoups\ 
A generic neutral CP-even 
Higgs boson will be denoted by $\h$  in this section.  We 
treat three cases separately: a light Higgs boson ($\mh < 2m_W$), an
intermediate mass Higgs boson ($2 m_W \leq \mh < 2m_t$), and a heavy
Higgs boson ($\mh \geq 2m_t$).

\subsection{Mass}
\label{secha}

In the Standard Model, the Higgs mass determines all its other
properties.\ihmeas\ Thus, the precision of the mass 
measurement affects the comparison of
theory and experiment, for example, in a global fit of cross sections,
branching ratios, and precision electroweak data.  Similarly, in the
MSSM or other models with extended Higgs sectors, the masses of all the
Higgs bosons are an important input to determining the underlying
model parameters.\ihmass\

For this fundamental mass measurement,
the LC can exploit recoils against a $Z$ (independently of Higgs decay),
and event reconstruction plus kinematic constraints can improve resolution
and clean up mass tails.\ihmass\
For a light or intermediate mass Higgs boson, 
optimal running conditions would have smaller
center-of-mass energy ({\it e.g.}, $\sqrt{s} = 350$~GeV for better 
momentum resolution), and possess 
as small a level of beamstrahlung as possible. 
Under such conditions, one can precisely measure the
recoil mass in $e^+e^- \rightarrow Z \h$ events opposite to the
reconstructed leptonic decay $Z \rightarrow e^+e^-$ or 
$\mu^+\mu^-$, with the additional advantage of decay mode independence.
Accuracy can be improved by reconstructing specific decay modes,
leading for example to a four-jet topology where effective (5-C)
kinematic constrained fits can be employed.

\begin{figure}[t!]
\begin{center}
\includegraphics*[width=\textwidth]{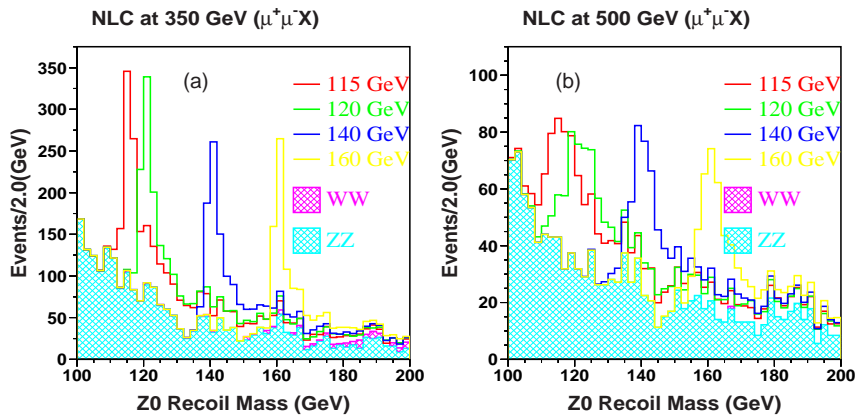}
\end{center}
\caption[0]{\label{fig:haijun_mass} \small
Recoil mass from a pair of leptons simulated in the LCD Large
detector for different Higgs masses
at (a) $\sqrt{s} = 350$~GeV and (b) 500 GeV.  Taken 
from \protect\cite{haijun_mass}.\ihmass\ihmeas}
\end{figure}

\Fig{fig:haijun_mass} shows the distribution of the recoil mass,
\begin{equation}
M_{\rm recoil} = \sqrt{s - 2 \sqrt{s} \cdot E_{\ell^+ \ell^-} + 
M^2_{\ell^+ \ell^-}}\,,
\end{equation}
in a 
simulation of the ``Large''
Linear Collider Detector (LCD)~\cite{LCD} for 
Higgs masses between 115 and 160 GeV~\cite{haijun_mass}. 
Using fits of Monte Carlo
shape templates and an integrated luminosity of 500 fb$^{-1}$, precisions 
of $\Delta \mhsm \simeq 60$~MeV at $\sqrt{s} = 350$~GeV and
$\Delta \mhsm \simeq 120$~MeV at $\sqrt{s} = 500$~GeV have been estimated 
for either the $e^+e^-$ or $\mu^+ \mu^-$ mode.\ihmass\ihmeas\

Realistic simulations have also been made of the 
LCD Large detector for the process $Z\h \rightarrow q\bar{q}\h$ resulting 
in four jets.  
\Fig{fig:juste_direct}(a) shows the jet-jet invariant mass
distribution for pairs of jets for Higgs with $\mhsm = 115$~GeV recoiling 
against a $Z$ that
has been reconstructed from its hadronic decay 
mode~\cite{Ronan}.\ihmass\ihmeas\   A clean Higgs
signal with a mass resolution of approximately 2~GeV is observed. The central
Higgs mass is shifted down due to the loss of low energy charged and
neutral particles in the simulated event reconstruction.  A low mass tail
of the Higgs signal arises from missing neutrinos in semi-leptonic $b$ and
$c$ quark decays.
Using neural net tags and full kinematic fitting~\cite{Juste:1999xv}, the
mass peak shown in \fig{fig:juste_direct}(b) 
is obtained for $\mhsm = 120$~GeV, 
$\sqrt{s} = 500$~GeV, and 500~fb$^{-1}$ resulting in 
$\Delta \mhsm \simeq 50$~MeV.
With the possibility of a second lower-energy interaction point,
scans across the $Z\h$ threshold may be attractive.  With a total integrated
luminosity of 100~fb$^{-1}$, $\Delta \mhsm \simeq 100$~MeV at $\mhsm = 150$~GeV
is achievable~\cite{Barger:1997pv}, competitive with the methods above.\ihmeas\

\begin{figure}[t!]
\begin{center}
\includegraphics*[width=\textwidth]{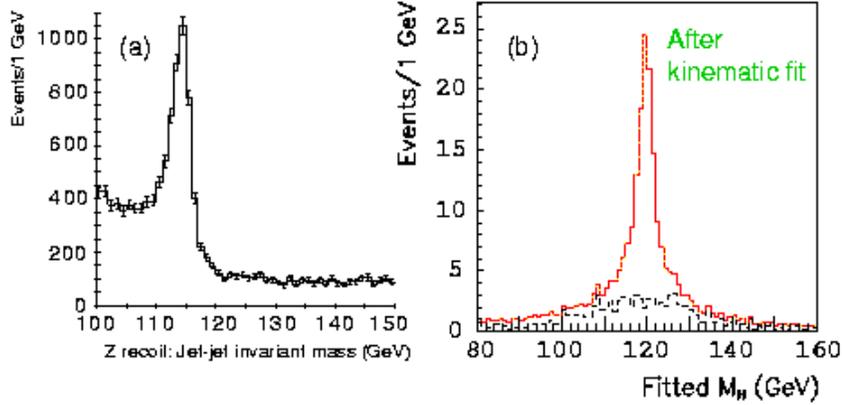}
\end{center}
\caption[0]{\label{fig:juste_direct} \small
(a) Jet-jet invariant mass of the jets recoiling
from a $Z$ reconstructed hadronically simulated in the
LCD Large detector, $\mhsm = 115$~GeV~\protect\cite{Ronan}.
(b) Direct reconstruction of the four-jet $q\bar{q}\hsm$ state
simulated in the LCD Large detector after fitting with full kinematic
constraints, $\mhsm = 120$~GeV~\protect\cite{Juste:1999xv}.\ihmass\ihmeas}
\end{figure}

More work is necessary to confirm analogous precisions for heavier Higgs
bosons and MSSM Higgs bosons with different decay modes and possible 
near-mass degeneracies.  The number of $Z\h$ events with 
$Z \rightarrow \ell^+ \ell^-$ for an intermediate mass 
($\mh > 2m_W$) or heavy Higgs ($\mh > 2m_t$) with SM 
coupling plummets 
quickly~\cite{fnal_report}.
In this case, and for the decays $\h \rightarrow ZZ$, hadronic decays
of the $Z$ would have to be considered to gain sufficient statistics.
For the heavier MSSM Higgs boson states,
European studies~\cite{kuskinen}
have shown typical mass precisions of
$\Delta m_{H^{\pm}}$ and $\Delta m_{\ha,\hh}$ of around 1~GeV for
500~fb$^{-1}$, but at $\sqrt{s} = 800$~GeV.

\subsection{Coupling Determinations -- Light Higgs Boson}
\label{sechb}

\subsubsection{Cross Sections}
\label{sechba}

For Higgs masses below $2 m_W$, the couplings $g_{\h ZZ}$ 
and $g_{\h WW}$ are best measured through measurements of the Higgsstrahlung
and $WW$ fusion cross sections, respectively.\ihcoups\ihmeas\ 
These cross sections are
also critical in the extraction of branching ratios since the experimental
measurement will be a product of cross section and branching ratio.

Measurement of the Higgsstrahlung cross section is best 
addressed via the recoil mass method 
outlined above~\cite{haijun_mass}.\ihmeas\  
To reduce the contribution from
the $WW$ fusion process, it is best to run at a lower energy, {\it i.e.}, 
$\sqrt{s} = 350$~GeV, and to examine recoil against $\mu^+\mu^-$ to avoid
large Bhabha backgrounds.\icollparam\
A simulation based on the LCD Large detector with 500~fb$^{-1}$ 
finds $\Delta\sigma / \sigma \simeq 3$\% [4.7\%]
at $\sqrt{s} = 350$ [500]~GeV, as shown in \fig{fig:haijun_cross}(a).
These results agree roughly with estimates from
European studies~\cite{Garcia-Abia:1999kv}.

\begin{figure}[t!]
\begin{center}
\includegraphics*[width=\textwidth]{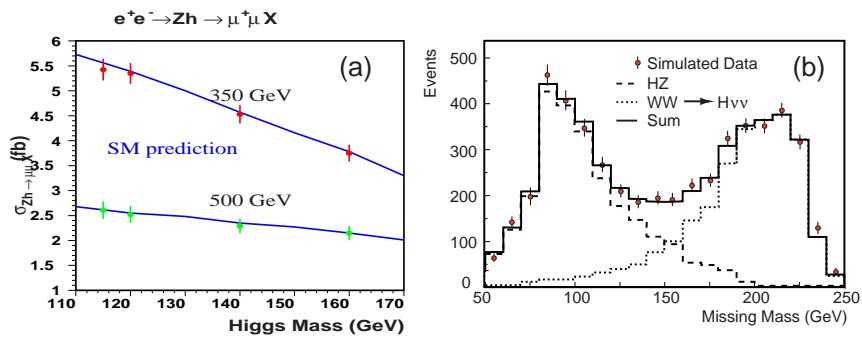}
\end{center}
\caption[0]{\label{fig:haijun_cross}
(a) Cross section measurement for 500~fb$^{-1}$~\protect\cite{haijun_mass} 
and (b)~separation of Higgsstrahlung and $WW$ fusion ($\sqrt{s} = 350$~GeV)
through a fit after background subtraction~\protect\cite{rvk_separate}, 
both simulated in the LCD Large detector.\ihmeas}
\end{figure}

With efficient and pure $b$-jet tagging, events due to the process
$e^+e^- \rightarrow W^+W^- \nu \bar{\nu} \rightarrow \nu \bar{\nu}\h
\rightarrow \nu \bar{\nu}b\bar{b}$ can be separated from those
due to Higgsstrahlung, 
$Z\h  \rightarrow  \nu \bar{\nu}\h \rightarrow \nu\bar{\nu}b\bar{b}$ by 
examining the missing mass distribution and
fitting to the expected shapes of a peak at $m_Z$ from Higgsstrahlung
and the higher missing masses from $WW$ fusion.  This technique
has been confirmed with simulations of the LCD Large detector as shown
in \fig{fig:haijun_cross}(b)~\cite{rvk_separate}.  
With 500~fb$^{-1}$ and a precision of
$\delta\BR(\hsm \rightarrow b \bar{b})/\BR \simeq 3$\% 
(see below), the fusion-process
cross section with this analysis can be found with a precision
$\Delta\sigma / \sigma$ = 3.5\% for $\mhsm = 120$~GeV.\ihmeas\ihcoups\

\subsubsection{Branching Ratios}
\label{sechbb}

A key advantage of the linear collider in Higgs studies is the attractive
situation of identifying Higgsstrahlung $Z\h$ events through the tag
of the $Z$ decays. This selection is largely independent of the decay
mode of the $\h$ and simplifies the measurement of Higgs boson branching
ratios.\ihmeas\ihbrs\

Small beam sizes, the possibility of a first measurement as close as
1~cm from the beam axis, and sophisticated pixel vertex detectors allow
for efficient and clean separation of quark flavors.  Separate tagging of
$b$ and $c$ jets is possible.\icollparam\  By assumption, jets that are not
identified as heavy flavor are assumed to be gluon jets, since the
Higgs partial width into $gg$ dominates the width into light quark pairs.

In a study~\cite{brau_vtx} of vertexing in a CCD vertex detector in a standard 
configuration of a LCD detector (C1 in \cite{brau_def}), 
topological vertexing~\cite{Jackson:1997sy} with neural net
selection [see \fig{fig:nnet_brs}(a)] was used for flavor (or anti-flavor,
{\it i.e.}, veto $WW^*$) tagging.  Assuming 500~fb$^{-1}$ and 
80\% polarization, results shown in Table~\ref{tab:higgs_br} 
were obtained (and include the
most recent updated branching ratio precisions  
of \cite{brau_update}).\icollparam\ 
These results have been checked to scale roughly with other
studies~\cite{hildreth,Borisov:1999mu,battaglia} as 
$( \sigma \int \L dt)^{-1/2}$, with the results of \cite{battaglia} 
(that includes $\hsm \nu \bar{\nu}$) being noticeably better 
for $c\bar{c}$ and $gg$ with errors as shown in \fig{fig:nnet_brs}.
As described in more detail in \sect{sechbe}, these  branching ratio
measurements can then be used to either distinguish a SM Higgs boson
from an MSSM Higgs boson, or to probe higher mass states and
extract MSSM parameters such as $\mha$ even if $\ha$ is not 
accessible.\ihmeas\ihbrs\
 
\begin{table}[t!]
\centering
\caption{Predicted branching ratio precisions in the LCD Large detector
and typical vertex detector
configuration for 500~fb$^{-1}$ and 
$\sqrt{s} = 500$~GeV~\cite{brau_update}.\ihmeas\ihbrs}
\label{tab:higgs_br}
\begin{tabular}{l||c|c||c|c}  \hline

    &  \multicolumn{2}{c||}{$\mhsm = 120$~GeV} & 
       \multicolumn{2}{c}{$\mhsm = 140$~GeV} \\ \hline
       
    & $\BR$ & $\delta\BR / \BR$ & $\BR$ & $\delta \BR /\BR$ \\ \hline

$\hsm \rightarrow b\bar{b}$ &  $(69 \pm 2.0)$\% & 2.9\% & 
                            $(34 \pm 1.3)$\% & 3.8\% \\
                           
$\hsm\rightarrow WW^*$     &  $(14 \pm 1.3)$\% & 10\% & 
                            $(51 \pm 1.5)$\% & 3.0\% \\

$\hsm\rightarrow c \bar{c}$ & $(2.8 \pm 1.1)$\% & 39\% & 
                            $(1.4 \pm 0.64)$\% & 44\% \\
                            
$\hsm\rightarrow gg$        & $(5.2 \pm 0.93)$\% & 18\% & 
                            $(3.5 \pm 0.79)$\% & 23\% \\  
                            
$\hsm\rightarrow \tau^+ \tau^-$   & $(7.1 \pm 0.56)$\% & 8.0\% & 
                            $(3.6 \pm 0.38)$\% & 10\% \\  \hline
                                                    
\end{tabular}
\end{table}

\begin{figure}[t!]
\begin{center}
\includegraphics*[width=\textwidth]{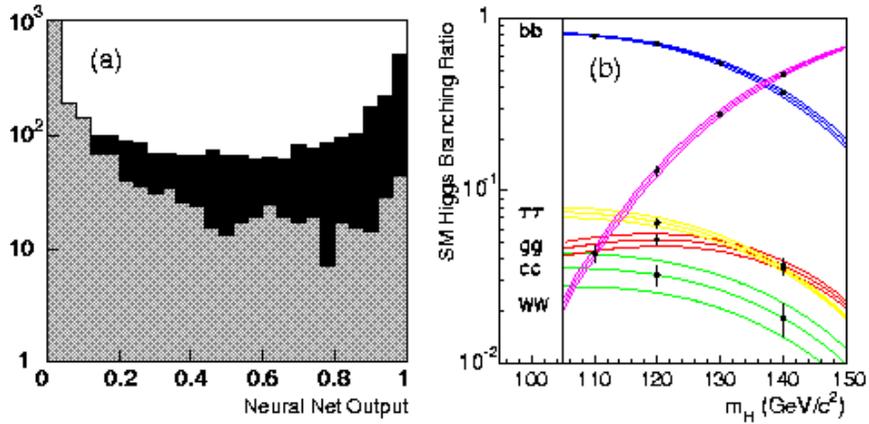}
\end{center}
\caption[0]{\label{fig:nnet_brs}
(a) For the simulated LCD Large detector with CCD vertex 
detector, neural net $\hsm \rightarrow c\bar{c}$ 
output for $\hsm\rightarrow c\bar{c}$ events
(dark) compared to output for $\hsm \rightarrow b\bar{b}$ events (gray).
Taken from \protect\cite{brau_def}.
(b) Variation of branching ratios with SM Higgs mass (bands are 1$\sigma$
uncertainties on the predictions) and measurement 
precisions in the TESLA detector (points with error bars).
Taken from \protect\cite{battaglia}.\ihmeas\ihbrs}
\end{figure}

With enough data and good charged particle momentum resolution, even
the rare decay mode into $\mu^+ \mu^-$ can be observed.  As shown in
\fig{fig:hmumu}, for a Higgs boson mass of 120~GeV where the 
Standard Model branching ratio is predicted to be only $3 \times 10^{-4}$,
the signal can be extracted from the underlying background with more than
5$\sigma$ significance with 1000~fb$^{-1}$~\cite{Battaglia:2001vf}.\icollparam\
For this channel, there are clear advantages to running at multi-TeV
energies where precisions could be reached to extend the test
of the Higgs mechanism of mass generation in the lepton sector by
checking if $g_{\hsm\mu\mu} / g_{\hsm \tau \tau} = m_{\mu} / m_{\tau}$.

\begin{figure}[t!]
\begin{center}
\includegraphics*[width=0.8\textwidth]{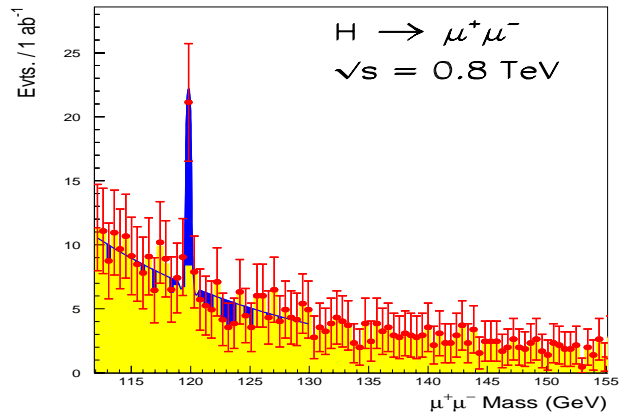}
\end{center}
\caption[0]{\label{fig:hmumu}
Distribution of the $\mu^+\mu^-$ invariant mass
for $\mhsm = 120$~GeV. The points with error bars
represent 1000~fb$^{-1}$ of data.
Taken from \protect\cite{Battaglia:2001vf}.\ihmeas}
\end{figure}

For lighter Higgs bosons, the coupling
to top quarks is still accessible via the radiative process $t\bar{t}\h$
described below or indirectly through 
$\BR(\h \rightarrow gg)$.\ihprod\ihcoups\
An accessible decay mode for light Higgs bosons is 
$\h \rightarrow \gamma \gamma$ requiring excellent electromagnetic
calorimetry in the detector.  For a SM
Higgs boson in a typical LCD detector, this is a difficult measurement 
requiring a great deal of luminosity, which is best for masses around 
120~GeV~\cite{Gunion:1996qg}, as shown in \fig{fig:hgamgam}.\ihmeas\ihbrs\ 
A higher luminosity study~\cite{Boos:2000bz} with 1000~fb$^{-1}$ and 
$\mhsm = 120$ GeV for the TESLA detector finds 
$\delta\BR/\BR = 14$\%.\icollparam\
A gamma-gamma collider, discussed in \sect{secj}, 
would be a more powerful tool for
determining the Higgs coupling to photons.\igamc\

\begin{figure}[t!]
\begin{center}
\includegraphics*[width=0.85\textwidth]{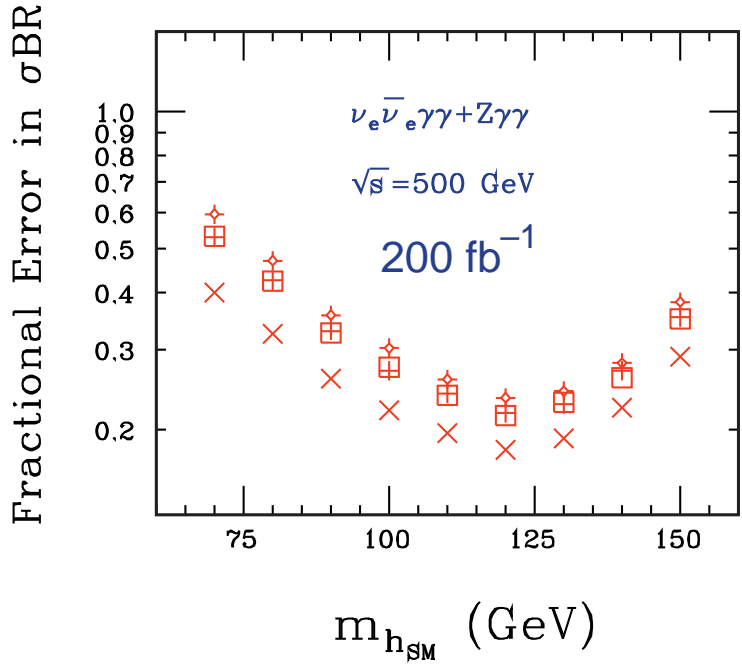}
\end{center}
\caption[0]{\label{fig:hgamgam}
Fractional error on the branching ratio 
$\BR(\hsm \rightarrow \gamma \gamma)$. The open squares are for a typical
LCD detector electromagnetic energy resolution of 
$\Delta E/E = 10\% / \sqrt{E} \oplus 1.0\%$.
Taken from \protect\cite{Gunion:1996qg}.\ihmeas}
\end{figure}

Invisible Higgs boson decay channels, if significant, 
pose a difficult challenge for the
LHC Higgs search.~\cite{Eboli:2000ze}.  
Possible invisible final states include neutralinos~\cite{higgschi},
majorons~\cite{higgsmaj}, heavy neutrinos~\cite{higgsnu}, or
the disappearance of the Higgs boson due to Higgs-radion 
mixing~\cite{higgsradion}.\ihbrssusymod\
Likewise, the LHC Higgs search will be difficult if the 
$b\bar b$, $WW^{*}$, $ZZ^*$ and $\gamma\gamma$
decay channels are significantly
suppressed (in which case, {\it e.g.}, 
Higgs decay into light hadronic jets could
be dominant~\cite{Berger:2002vs}).  The LC can close these loopholes
and measure the branching ratio easily even for branching ratios as small
as 5\% as shown in \fig{fig:brinvis} for a relatively narrow Higgs state using
the recoil mass method and demanding no detector activity opposite the
$Z$~\cite{Vankooten:2001a}, or by comparing the number of events tagged with 
$Z \rightarrow \ell^+ \ell^-$ with the total number of observed Higgs decays
into known states.\ihmeas\ihbrs\

\begin{figure}[htb!]
\begin{center}
\includegraphics*[width=0.95\textwidth]{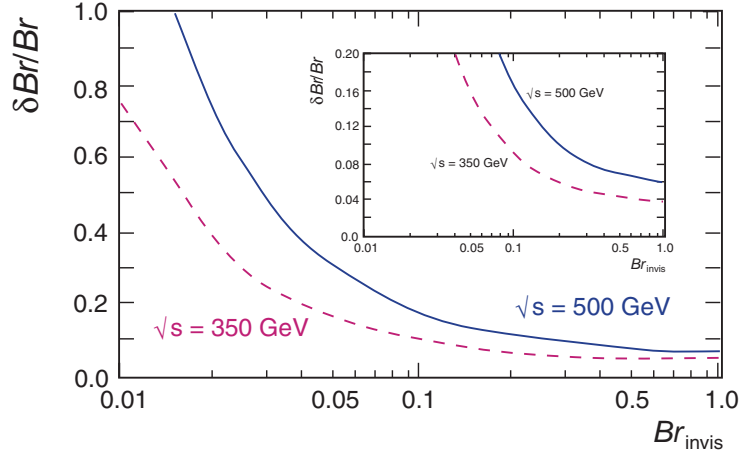}
\end{center}
\caption[0]{\label{fig:brinvis}
Fractional error on a Higgs invisible
branching ratio
for $\mh = 120$~GeV and 500~fb$^{-1}$\icollparam\ of data.
The asymptotic value as the branching ratio (Br) 
approaches unity is due solely
to the statistical uncertainty in $\sigma(hZ)$.
A blow-up of the region of $\delta{\rm Br}/{\rm Br}\leq 0.2$ is also shown.
Taken from \protect\cite{Vankooten:2001a}}
\end{figure}

\subsubsection{Radiative Production, $t\bar{t}\h$}
\label{sechbc}

For a light Higgs boson, production through radiation off a top quark is 
feasible~\cite{Djouadi:1992tk,Bar-Shalom:1995jb,Gunion:1996vv,%
dawsontth,Dittmaier:1998dz,dawsonbbh,Dittmaier:2000tc,Grzadkowski:2000wj}, 
resulting in a final state of $t\bar{t}\h$,
thus allowing for a determination of the Yukawa top quark coupling 
$g_{\h tt}$.\ihprod\iyuk\ 
For a SM-like Higgs boson with $\mh = 120$~GeV, the $t\bar{t}\h$ 
cross section is roughly
10 times larger at $\sqrt{s} = 700$--800 GeV than at 500~GeV.
At $\sqrt{s} = 800$~GeV, a statistical error of 
$\delta g_{\h tt} / g_{\h tt} \sim 5\%$ was 
estimated~\cite{Gunion:1996vv} for $L=500\fbi$ 
on the basis of an optimal-observable 
analysis~\cite{optimalobservables}.\icollparam\
At $\sqrt{s} = 500$~GeV, a statistical error of 
$\delta g_{\h tt} / g_{\h tt} \simeq 21$\% is 
estimated~\cite{dawsontth,dawsonbbh}
using a total integrated luminosity of 1000~fb$^{-1}$.
A more sophisticated analysis using
neural net selections, full simulation, and the same integrated luminosity
at $\sqrt{s} = 800$~GeV finds a total error of 6\% on the 
coupling~\cite{Juste:1999af}.\ihmeas\

\subsubsection{Self-Coupling}
\label{sechbd}

To fully delineate the Higgs sector, it is essential to measure
the shape of the Higgs potential.\ihself\  The cross section for double Higgs
production ({\it e.g.}, $Zhh$) is related to the triple Higgs coupling 
$g_{hhh}$, which in turn is related to the spontaneous
symmetry breaking shape of the Higgs potential~\cite{smhhcouplings}.
In the MSSM, a variety of double 
Higgs production processes~\cite{hhcouplings} 
would be required to determine $g_{hhh}$, $g_{hAA}$, {\it etc.}

\begin{figure}[t!]
\begin{center}
\includegraphics*[width=\textwidth]{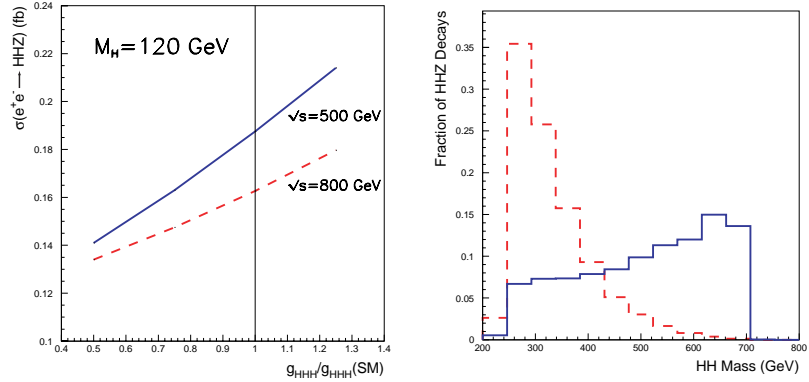}
\end{center}
\caption[0]{\label{fig:hself}
(a) Dependence of the $hhZ$ cross section
on the triple Higgs coupling, normalized to its SM
value for the indicated mass and $\sqrt{s}$ 
values.
(b) Distribution of the $hh$ invariant mass in $hhZ$
events from diagrams containing the triple Higgs
vertex (solid line) compared to that 
from other diagrams (dashed line).
Both plots have been taken from \protect\cite{Battaglia:2001nn}.\ihself}
\end{figure}

These cross sections are low, and 
high integrated luminosity is needed, bolstered by polarization
and neural net selections. Experimental 
studies~\cite{Miller:1999ji,Castanier:2001sf} indicate that
for a SM-like Higgs boson with $\mh = 120$~GeV at $\sqrt{s} = 500$~GeV and
1000~fb$^{-1}$, a precision of $\delta g_{\h\h\h} /g_{\h\h\h} = 23$\% is
possible.\icollparam\ihmeas\ihself\
A measurement of the double Higgs production cross section to 
extract $g_{\h\h\h}$ has to deal with dilution due to
the existence of diagrams that lead to the same final states but that
do not include a triple Higgs vertex [see~\fig{fig:hself}(a)], so additional 
kinematical
variables can be
considered to enhance sensitivity to the signal.
A study considering the $h$ decay angle in the $hh$ rest frame and
the $hh$ invariant mass [\fig{fig:hself}(b)] has shown improved separation and
an increased precision of 
$\delta g_{\h\h\h} /g_{\h\h\h} = 20$\%~\cite{Battaglia:2001nn} for
the same mass, energy, and integrated luminosity as above.
The regions of accessibility in the MSSM parameter space for the
MSSM Higgs self-couplings have also been 
performed~\cite{hhcouplings,mssmhhcouplings}.\ihself\

With the cross section for SM triple Higgs production so low
[$\sigma(Z\h\h\h) < 0.001$~fb], measurement of the quartic coupling
$g_{\h\h\h\h}$ is hopeless with currently envisioned luminosities.

\subsubsection{Implications for the MSSM Higgs Sector}
\label{sechbe}

The discussion of light Higgs coupling determinations has been based
on the assumption that the actual Higgs couplings to fermions, vector
bosons and scalars are close to the corresponding Standard Model
expectations.  In \sect{secga}, it was argued that such an expectation
is rather generic, and applies to the decoupling limit of models of
Higgs physics beyond the Standard Model.  In particular, the
decoupling limit of the MSSM Higgs sector sets in rather rapidly once
$\mha\gsim 150$~GeV [see \sect{secea}].\idecoup\  Since $\mhl\lsim 135$~GeV in
the MSSM [\eq{mhmaxvalue}], the precision study of $\hl$ using the
techniques discussed above can distinguish between $\hl$ and $\hsm$
with a significance that depends on how close the model is to the
decoupling limit.\idecoup\  Said another way, the detection of deviations in
the Higgs couplings from their Standard Model predictions
would yield evidence for the existence of the
non-minimal Higgs sector, and in the context of the MSSM would provide
constraints on the value of $\mha$ (with some dependence on
$\tan\beta$ and other MSSM parameters that enter in the Higgs
radiative corrections).\ihcoups\

In \cite{Carena:2001bg}, the potential
impact of precision Higgs measurements at the LC on
distinguishing $\hl$ from $\hsm$ was examined.  
The fractional deviation of the $\hl$ branching ratios into a
given final state from the corresponding result for $\hsm$ (assuming
the same Higgs mass in both cases) is defined as:
$\delta \BR \equiv (\BR_{\rm MSSM} - \BR_{\rm SM})/\BR_{\rm SM}$.
For the MSSM Higgs boson decay, both $\mhl$ and the corresponding
branching ratios were computed including the radiative corrections due
to the virtual exchange of Standard Model and supersymmetric
particles, as described in \sect{seceb}.  Thus, the $\hl$ branching
ratios depend on $\mha$ and $\tan\beta$ (which fix the tree-level MSSM
Higgs sector properties) and a variety of MSSM parameter that govern
the loop corrections.\iradiativesusy\  
Four scenarios were considered: the minimal and
maximal top-squark mixing cases [see \eq{mhmaxvalue} and surrounding text],
and two additional cases with large $|\mu|=|A_t|$ (for $\mu A_t<0$
and two possible sign choices of $\mu$), where $\mu$ and $A_t$ control
the top-squark mixing.  In the latter two scenarios, significant
renormalization of the CP-even Higgs mixing angle $\alpha$ and 
$\Delta_b$ [see \eq{bbcouplings}] can arise.  

\begin{figure}[t!]
\begin{center}
\includegraphics*[width=\textwidth]{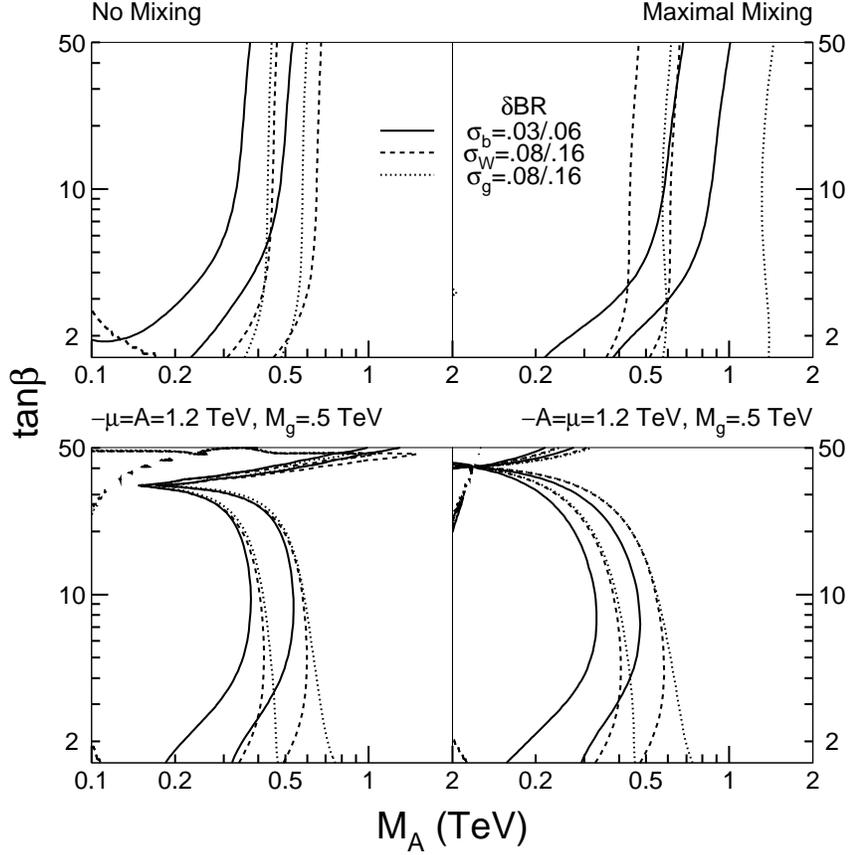}
\end{center}
\caption[0]{\label{mssmdbrs}
Contours of $\delta \BR(b\bar b) = 3$ and 6\% (solid), 
$\delta \BR(WW^*) = 8$ and 16\% 
(dashed) and 
$\delta \BR(gg) = 8$ and 16\% (dotted) [BR deviations are defined in
the text]
in the no ({\it i.e.} minimal) mixing scenario (top left),
the maximal mixing scenario (top right),
and the large $\mu$ and $A_t$ scenario with  
$\mu = -A_t = 1.2$ TeV (bottom left) and 
$\mu = -A_t = -1.2$ TeV (bottom right).  
Taken from \protect\cite{Carena:2001bg}.\ihbrs}  
\end{figure}

In \fig{mssmdbrs}, contours of $\delta \BR$ are plotted for three
$\hl$ decay modes:  $b\bar b$, $WW^*$ and $gg$.  The contours shown
correspond roughly to the $1\sigma$ and $2\sigma$ measurements claimed
by \cite{battaglia}, rescaled for the LC at
$\sqrt{s}=500$~GeV (see also the
$b\bar b$ and $WW^*$ branching ratio precisions given in 
Table~\ref{tab:higgs_br}).\ihmeas\idecoup\ 
In the minimal and maximal scenarios, the
dependence on $\mha$ is nearly independent of $\tan\beta$, and
demonstrates that one can achieve sensitivity to values of $\mha$ that
lie significantly beyond $\sqrt{s}/2$ where direct production at the
LC via $e^+e^-\to \hh\ha$ is kinematically forbidden.  
However in some MSSM parameter regimes [{\it e.g.}, the case of
large $\mu=-A_t$ as shown in \fig{mssmdbrs}(c) and (d)],
a region of $\tan\beta$ may yield almost no constraint on $\mha$.
This is due to the phenomenon of $\mha$-independent (or premature) 
decoupling~\cite{Carena:2001bg}
in which $\cos(\beta-\alpha)$ [which controls the departure from 
the decoupling limit] vanishes at a particular value of 
$\tanb$ independently of the value of $\mha$.\idecouppremature\
Thus, a measured deviation of Higgs branching ratios that
distinguishes $\hl$ from $\hsm$ can place significant constraints on
the heavier non-minimal Higgs states, although the 
significance of the resulting constraints can
depend in a nontrivial way on the value of the MSSM parameters that
control the Higgs radiative corrections.

\subsection{Coupling Determinations -- Intermediate Mass Higgs Boson}
\label{sechc}

For $\mh < 2 m_W$, there are many possible branching ratio
measurements at the LC (see Table~\ref{tab:higgs_br}).  This rich phenomenology
allows for the determination of Higgs
couplings to many of the fermions and bosons.\ihmeas\ihbrs\
For larger masses, decays to $f\bar{f}$ become rarer until the threshold
for decays into top is crossed.  
In this intermediate mass range, the LC can measure
the $W$ and $Z$ couplings more precisely than the LHC
both through Higgs production rates and via branching ratios
for decays into these bosons. 
Whether the observed Higgs boson fully generates the $W$ 
and $Z$ mass can then be checked.

We have noted in Section~\ref{secb} that
precision electroweak measurements in the framework of the 
Standard Model imply that
$\mhsm\lsim 193$~GeV at 95\% CL~\cite{precision}.\ipew\  
Thus, any Higgs boson observed
with a mass much greater than this would imply new physics. At this point,
measurements from a Giga-$Z$ dataset would be particularly useful to
probe this new sector.\igigaz\

\subsubsection{Cross Sections}
\label{sechca}

Techniques described earlier~\cite{haijun_mass,rvk_separate} for 
cross section measurements of both the Higgsstrahlung and $W$-fusion
processes, with subsequent Higgs decays into $b\bar{b}$,
can still be performed for the lower portion of the intermediate
mass range below $\mhsm = 160$~GeV.\ihprod\
Even in this intermediate mass
range, it is beneficial to run at the peak of the cross section
at roughly $\mh + m_Z + 50$~GeV. 
Typical precisions of 
$\Delta\sigma(Z\hsm) / \sigma(Z\hsm) \simeq 5$\% and
$\Delta\sigma(\nu \bar{\nu}\hsm) / \sigma(\nu \bar{\nu}\hsm) \simeq 17$\%
for $\mhsm = 160$~GeV
at $\sqrt{s} = 350$~GeV 
with 500~fb$^{-1}$ can be obtained.\ihmeas\icollparam\

For heavier Higgs bosons in this mass range, cross sections for
both Higgs\-strahlung and $W$-fusion will
need to be extracted from these processes followed by
$\h\rightarrow WW^*$ decays for example as described in
\cite{Borisov:1999mu}. Couplings determined from
$t\bar{t}\h$ and $Z\h\h$ production would clearly need higher
center-of-mass energy.
\ihcoups\

\subsubsection{Branching Ratios}  
\label{sechcb}

Using Higgsstrahlung events at an optimal $\sqrt{s}$, the statistical
error on $\BR(\hsm\rightarrow b\bar{b})$ is still only 6.5\% at $\mhsm =
160$~GeV~\cite{battaglia}.\ihmeas\ihbrs\
  At $\rts=500\gev$, with leptonic decays of
the $Z$ only, the statistical error on this branching ratio reaches
25\% at $\mhsm \simeq 165$~GeV with 250~fb$^{-1}$ and remains below 30\%
for $\mhsm < 200$~GeV with 2000~fb$^{-1}$~\cite{fnal_report}.\icollparam\  
However,
in addition to the leptonic decays of the $Z$, hadronic decays can
also be used to effectively tag the associated $Z$.  Extrapolating
from full LCD detector simulations, it is conservatively estimated
that including the hadronic decays of the $Z$ results in an increase
in signal statistics above background by a factor of four.  With these
assumptions and 500~fb$^{-1}$, again with the optimal $\sqrt{s} \simeq
350$~GeV, the error on the $b\anti b$ branching ratio can then be
estimated to reach 25\% at $\mhsm \simeq 200$~GeV.  Measurement of
branching ratios to $c\bar{c}$, $\tau^+\tau^-$, $gg$, and $\gamma
\gamma$ does not seem feasible in this mass range.

Branching ratios into vector bosons can be measured with good
precision in this intermediate mass range.\ihmeas\
For $\mhsm = 160$~GeV and 500~fb$^{-1}$, a predicted excellent
precision of 2.1\% on $\BR(\hsm \rightarrow WW)$, has been 
reported~\cite{Borisov:1999mu} and
extrapolated estimated precisions of better than 7\% over the mass 
range of 150 to 200~GeV~\cite{fnal_report}.  

To measure $\BR(\h \rightarrow ZZ)$, it will be necessary to 
distinguish hadronic $Z$ decays from hadronic $W$ decays, serving 
as an important benchmark for electromagnetic and hadronic calorimetry.
With the same luminosity, and
assuming that this separation allows one to identify
one of the two $Z$'s in the Higgs decays (through leptons or
$b\bar{b}$) 40\% of the time, the statistical uncertainty of this
branching ratio would be approximately 8\% for  
$\mhsm \simeq 210$~GeV~\cite{fnal_report} degrading to
17\% for $\mhsm = 160$~GeV~\cite{tesla_report} where the branching
ratio into $Z$'s is still small.\ihmeas\

\subsection{Coupling Determinations -- Heavy Higgs Boson}
\label{sechd}

If a SM-like Higgs boson is heavy ({\it i.e.}, $\mh > 2 m_t$),
then new physics beyond the Standard Model must exist
(\eg, see \sect{secgb}; other
examples of possible new physics effects are surveyed in 
\cite{peskinwells,Choudhury:2002qb}).
High-statistics measurements at the
$Z$ peak would again be useful to elucidate these effects.

\subsubsection{Cross Sections}
\label{sechda}

As a specific case, for $\mh = 500$~GeV at $\sqrt{s} = 800$~GeV,
a SM-like Higgs boson would have a width of 70~GeV
and dominant decay modes into $W^+W^-$ (55\%), $ZZ$ (25\%), 
and $t\bar{t}$ (20\%). The production cross section for $Z\h$ would
be 6~fb, but production would be dominated by $\nu \bar{\nu}\h$
production at 10~fb.  With 1000~fb$^{-1}$, one expects 
400 $Z\h$ events where the $Z$ decays to $e^+e^-$ or $\mu^+\mu^-$.
With reasonable selection and acceptance cuts, a measurement
of $\sigma(Z\h)$ to better than 7\% should be feasible.\ihmeas\ihbrs\
See \cite{Choudhury:2002qb} for an independent assessment of the LC discovery
reach for a heavy Higgs boson.

\subsubsection{Branching Ratios}
\label{sechdb}

The LHC will have great difficulty distinguishing $\h\rightarrow t\bar{t}$
decays due to huge QCD $t\bar{t}$ backgrounds.
On the other hand,
this mode should be observable at the LC.\ihbrs\ihmeas\ 
In the SM, the important coupling $g^2_{tt\hsm} \simeq 0.5$ can be
compared to $g^2_{bb\hsm} \simeq 4 \times 10^{-4}$.
If the Higgs boson is heavier than 350~GeV, it will be possible
obtain a good determination of the top-Higgs Yukawa coupling.\iyuk\
Full simulations are needed
for heavy Higgs decays into top, but with reasonable assumptions,
one can expect a statistical error of $\delta \BR /\BR \simeq 14$\% with
500~fb$^{-1}$~\cite{fnal_report}.
Simulations using the TESLA detector of the 
$W^+W^- \rightarrow \hsm \rightarrow t\bar{t}$ process with 1000~fb$^{-1}$
and 6-jet final states
show impressive signal significance for $\sqrt{s} = 1000$~GeV and 
reasonably good significance at $\sqrt{s} = 800$ GeV~\cite{Alcaraz:2000xr}.
These studies find that
a relative error in the top quark Yukawa coupling measurement
better than 10\% can be achieved for Higgs masses in the
350--500~GeV and 350--650~GeV ranges at $\sqrt{s} = 800$~GeV
and 1000~GeV, respectively.\ihmeas\ihcoups\icollparam\

Again assuming detector performance allowing for separation of
hadronic $W$ and $Z$ decays, and using production through 
$W$-fusion, similar techniques for extracting $\BR(\hsm \rightarrow t\bar{t})$
can be applied resulting in estimates on $\BR(\hsm \rightarrow W^+W^-)$ and 
$\BR(\hsm \rightarrow ZZ)$ as shown in Table~\ref{tab:summary}.

\subsection{Summary of Couplings}
\label{seche}

The relative measurement errors for a SM Higgs at various masses are
summarized in Table~\ref{tab:summary}.  As much as possible, the
entries in this compilation have been collected from LCD detector
simulations, particularly for lighter masses.  For uniformity, the
entries have been scaled to 500~fb$^{-1}$, except where noted
otherwise.\icollparam\  The significant measurements of many branching ratios,
couplings and the total width
demonstrates the strength of the LC Higgs program.\ihmeas\

\begin{table}[t!]
\caption{Summary of measurement precisions for 
the properties of a SM-like Higgs boson, $\h$, 
for a range of Higgs boson masses.
Unless otherwise noted (see footnotes below the table), 
$\sqrt{s}=500$~GeV and the total integrated luminosity
is taken to be 500~fb$^{-1}$.\ihmeas\ihbrs\ihcoups\ihmass\ihwidths\
}
\label{tab:summary}
\vspace{0.1in}
\begin{tabular}{c||c|c|c|c|c}  \hline

 $\Delta \mh$ &  \multicolumn{5}{l}{$\simeq 120$~MeV (recoil against leptons
                                          from $Z$)} \\ 
      
                 &  \multicolumn{5}{l}{$\simeq 50$~MeV (direct reconstruction)}
                                                       \\ \hline
 $\mh$~(GeV)     & 120 & 140 & 160 & 200 
                 & 400--500  \\ \hline 
 $\sqrt{s}$~(GeV)    &  \multicolumn{4}{c|}{500}
                 &  800 \\ \hline \hline
 $\Delta \sigma(Z\h) / \sigma(Z\h)$
     &  $4.7$\% & $6.5$\% & $6$\% &  $7$\%   & $10$\% \\ \hline
     
 $\Delta \sigma(\nu\bar{\nu}\h)\BR(b\bar{b})/ \sigma \BR$ 
     & $3.5$\%  & $6$\% & $17$\% 
     &  --             & --            \\ \hline \hline
       
 $\delta g_{\h xx} / g_{\h xx}$ (from $\BR$'s)&  &  &  &  &  
\\ \hline \hline

 $t\bar{t} $     & 6--21\% \dag\ & -- & -- & --  & 10\%   \\

 $b\bar{b}$      & $1.5$\% & $2$\% & $3.5$\% 
                 & $12.5$\% & -- \\ 

 $c\bar{c}$      & $20$\%  & $22.5$\%  & -- & -- & -- \\

 $\tau^+ \tau^-$ & $4$\%   & $5$\%     & -- & -- & -- \\ 

 $\mu^+ \mu^-$   & $15$\% \ddag\  & --        & -- & -- & -- \\    \hline
 
 $WW^{(*)}$      & $4.5$\% & $2$\% & $1.5$\% 
                 & $3.5$\% & $8.5$\%    \\ 

 $ZZ^{(*)}$      & --             & --           & $8.5$\% 
                 & $4$\%   & $10$\%    \\
 
 $gg$            & $10$\%  & $12.5$\%  & -- & -- & -- \\ 
 
 $\gamma \gamma $ & $7$\%  & $10$\%    & -- & -- & -- \\ \hline
  
 $g_{\h\h\h}$        & $20$\% \S\ & --  & -- & -- & -- \\ \hline

 $\Gamma_{\rm tot}$ \dag\dag & $10.1\%$ & $8.2$\% & $12.9$\% & $10.6$\% & $22.3$\% \\ \hline                                                     
\end{tabular}\\[0.5ex]
$\dag$ The range of $ht\bar t$ couplings is 
obtained from $e^+e^-\to t\bar t\h$, with
$\sqrt{s}=500$--800~GeV and 1000~fb$^{-1}$ of data.\\
$\ddag$ based on $\sqrt{s}=800$~GeV and 1000~fb$^{-1}$ of data.\\
\S\ based on $\sqrt{s}=500$~GeV and 1000~fb$^{-1}$ of data.\\
\dag\dag\ indirect determination from $\Gamma(VV^*)/\BR(VV^*)$, $V=W,Z$.
\end{table}

Just as the computer program {\tt ZFITTER}~\cite{Bardin:2001yd} is used
with $Z$ mass, widths, asymmetries and branching ratios to make
global fits for $Z$ couplings, a program 
{\tt HFITTER}~\cite{Battaglia:2000jb} is
now available that performs a global fit taking into account correlations
between measurements of Higgs boson properties. 
Individual couplings of the Higgs boson can then be extracted optimally,
for example through the correct combination of cross section and branching
ratio measurements for such couplings as $g_{\h WW}$ and $g_{\h ZZ}$.
Such precision fits can
be used to probe indirectly for higher-mass states.

\subsection{Total Width}
\label{sechf}

For light Higgs bosons, the predicted SM width is far too small to be
measured directly (and any anomalously wide state will indicate new
physics), but the combination of branching ratios and coupling measurements
allows the indirect and {\it model-independent} measurement of the total 
width through $\Gamma_{\rm tot} = 
\Gamma(\h \rightarrow X)/\BR(\h \rightarrow X)$.\ihwidths\
For $\mhsm< 115$~GeV, the total width measurement would very likely require a 
$\gamma \gamma$ collider, an $e^+e^-$ LC, and input from the 
LHC~\cite{Gunion:1996cn}.\igamc\ 
However, limits
from LEP indicating $\mhsm \gsim 115$~GeV would result in a non-negligible
branching ratio to $WW^*$ with the exciting and attractive prospect of 
an indirect model-independent measurement of the total width using LC
measurements alone.\ihmeas\

First, measurements of $\sigma(\h\nu\nu) \times 
\BR(\h\rightarrow b\bar{b})$ and
$\BR(\h\rightarrow b\bar{b})$ independently (through recoil in Higgsstrahlung)
gives $\Gamma(\h \rightarrow WW^*)$.  Using a similar independent measurement
of $\BR(\h \rightarrow WW^*)$ then gives the total width
through the relation
$\Gamma_{\rm tot} = \Gamma(\h \rightarrow WW^*)/\BR(\h \rightarrow WW^*)$.
Similarly, both $\Gamma(\h\to ZZ^*)$ and $\BR(\h\to ZZ^*)$
can be determined from the Higgsstrahlung process and used to compute 
$\Gamma_{\rm tot} = \Gamma(\h \rightarrow ZZ^*)/\BR(\h \rightarrow ZZ^*)$.  
These two independent results for $\Gamma_{\rm tot}$ can
then be combined. The results are
summarized in Table~\ref{tab:summary}. Even with as little as
200~fb$^{-1}$, $\Gamma_{\rm tot}$ can be found to approximately 10\% for
$\mhsm= 120$~GeV, improving to $\sim 7\%$ for $\mhsm \sim 150$~GeV.
Even better precision can be attained with the introduction of some
model assumptions in the value used for $\Gamma(\hsm \rightarrow WW^*)$, 
{\it i.e.}, assuming the SU(2) relation between $W$ and $Z$ couplings
along with $\sigma_{meas}(Z\hsm)$, or else by using its SM value
directly as a consistency check.\ihwidths\ihmeas\

\begin{figure}[t!]
\begin{center}
\includegraphics*[width=0.8\textwidth]{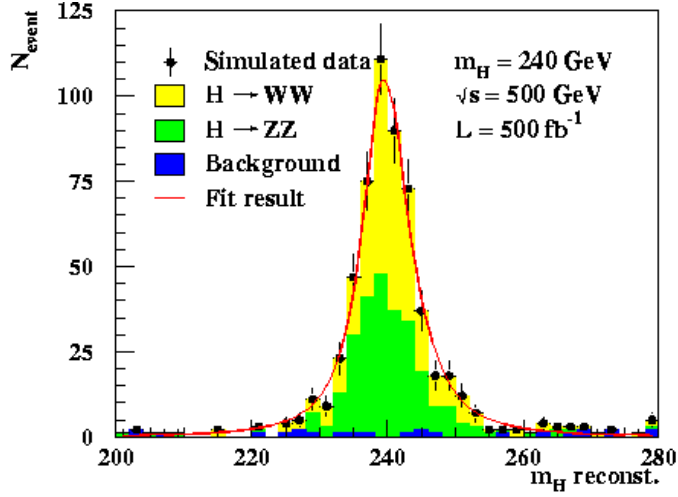}
\end{center}
\caption{\label{fig:widthfit}
Fit to the reconstructed $\hsm$ mass to directly extract 
$\Gamma_{\hsm}$~\protect\cite{Meyers:2001a}.\ihwidths\ihmeas}
\end{figure}

For $\mhsm \gsim 205$~GeV, $\Gamma_{\rm tot}(\hsm) > 2$~GeV and the
total width could be directly resolved with typical LCD detector
resolutions.  
The variations of precision for indirect and direct measurements for
different values of $\mhsm$ and inputs from different 
machines are examined in \cite{Gunion:1996cn,Drollinger:2001bc}.  The
jet-jet mass resolution assumed in \cite{Gunion:1996cn} has been
verified by full simulations~\cite{Ronan} in the LCD Large Detector
resulting in estimated direct measurements of the total width reaching
minimum values of $\simeq 6$\% in the mass range of 240--280~GeV
with 200~fb$^{-1}$.  A more complete analysis 
for $\mhsm = 240$~GeV, where the predicted SM Higgs width is 3.4~GeV,
fits to a convolution of resolution and a Breit-Wigner as
shown in \fig{fig:widthfit} to find 
$\Delta \Gamma_{\hsm} / \Gamma_{\hsm} = 12$\%~\cite{Meyers:2001a}.
The indirect determination described above can
also still be pursued and a combination would allow even better
precisions.

\subsection{Quantum Numbers}
\label{sechg}

The spin, parity, and charge conjugation quantum numbers $J^{PC}$ of
a Higgs boson candidate, generically denoted by $\phi$,
can potentially be determined in a model-independent
way.\ihcp\ihspinparity\  Useful ingredients include the following:

1. If $\phi$ is produced in $\gam\gam$ collisions, then
it cannot have $J=1$ and it must have positive C~\cite{yangsak}.

2. The behavior of the $Z\phi$ Higgsstrahlung cross section at threshold 
constrains the possible values of $J^{\rm PC}$ of the state. If the
spin of the $\phi$ is 2 or less, a cross
section growing as $\beta$ indicates a CP-even object, whereas
a cross section growing as $\beta^3$ signals a CP-odd 
state~\cite{Dova:2003py,Miller:2001bi} with differences as shown in 
\fig{fig:thresh}(a).

\begin{figure}[t!]
\begin{center}
\includegraphics*[width=\textwidth]{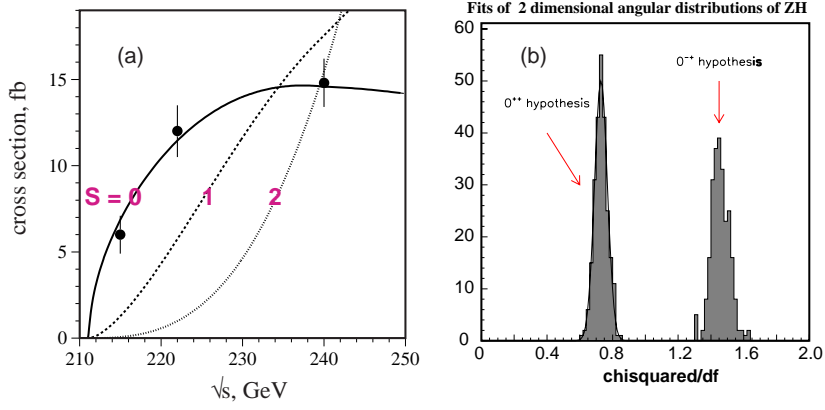}
\end{center}
\caption[0]{\label{fig:thresh}
(a) Behavior of Higgsstrahlung threshold for various spin states
along with typical measurement precisions on the cross 
section.  Taken from~\protect\cite{Dova:2003py}.
(b) Fit to the double-differential angular distribution in
$Z\phi$ events (see text) to distinguish CP-even and 
CP-odd states.  Taken from~\protect\cite{fnal_report}.\ihspinparity}
\vspace{-0.1in}
\end{figure}

3. The angular dependence of the $\epem\to Z\phi$ cross section depends
upon whether the $\phi$ is CP-even, CP-odd, or a
mixture~\cite{Miller:2001bi,Hagiwara:1994sw,Barger:1994wt,Han:2001mi}.
Following \cite{Han:2001mi} we parameterize the $ZZ\phi$ vertex as
\begin{equation}
    \Gamma_{\mu\nu}(k_1,k_2) = 
        a g_{\mu\nu} +
        b \,\frac{k_{1\mu} k_{2\nu} - g_{\mu\nu}k_1\cdot k_2}{\mz^2}  +
   \tilde{b}\,\frac{\eps_{\mu\nu\alpha\beta} k_1^\alpha k_2^\beta}{\mz^2}\,,
    \label{eq:zzh}
\end{equation}
where $k_1$ and $k_2$ are the momenta of the two $Z$s.
The first term arises from a Standard-Model-like $ZZ\phi$ coupling, and the
last two from effective interactions that could be induced by high-mass
virtual particles.
With this vertex the Higgsstrahlung cross section becomes
\beqa
\frac{d\sigma}{d\cos\theta_Z} & \propto 1 &
    + \frac{p_Z^2}{\mz^2} \sin^2\theta_Z
    - 4\, {\rm Im}\left[\tilde{b}\over\tilde{a}\right]
      \frac{v_ea_e}{v_e^2+a_e^2}
      \frac{p_z\sqrt{s}}{\mz^2} \cos\theta_Z \nonumber \\
&&\qquad    +\left|\tilde{b}\over \tilde{a}\right|^2
      \frac{p^2_zs}{2\mz^4} (1+\cos^2\theta_Z)\,,
    \label{eq:zzhX}
\eeqa
where $\theta_Z$, $p_Z$, and $E_Z$
are the scattering angle, momentum, and energy of the final-state $Z$
boson; $v_e$ and $a_e$ are the vector and axial-vector $Ze^+e^-$ couplings;
and $\tilde{a}\equiv a-bE_Z\sqrt{s}/\mz^2$.
The term in \eq{eq:zzhX} proportional to $\cos\theta_Z$ arises
from interference between the CP-even and CP-odd couplings in
\eq{eq:zzh}.\ihcp\ihmeas\
If the CP-odd coupling~$\tilde{b}$ is large enough, it can be extracted
from the forward-backward asymmetry.
Even upper limits on this asymmetry would be interesting.
Note that the CP-even component
of a Higgs boson will typically couple at tree level whereas the
CP-odd component will only couple via one-loop diagrams (typically
dominated by the $t$ quark loop). As a result, one expects the coupling
strength $\tilde b$ to be proportional to $\mz^2/s$ 
times a loop suppression factor.  Thus, an asymmetry measurement may be
able to provide a crude determination of 
the $\tilde b/a$ term.
If $\phi$ is a purely CP-odd state with a one-loop coupling, 
the resulting $Z\ha$ cross section will simply be too small to
provide a useful measurement of the asymmetry.

4. The angular distribution of the fermions in the $Z\to f\anti f$
decays in $Z\phi$ production also reflects the CP nature of
the state $\phi$ \cite{Hagiwara:1994sw,Barger:1994wt}.\ihcp\ihmeas\ 
For the decay $Z \rightarrow e^+ e^-$ or $\mu^+\mu^-$, the following 
angles can be defined: the angle between the initial $e^-$ and the $Z$; 
the angle between the final state $e^-$ or $\mu^-$ and the direction of motion
of the $Z$, in the rest frame of the $Z$; and the angle between the $Z$
production plane and $Z$ decay plane.  Correlations between these angles
can be exploited, {\it e.g.}, 
a fit to the double-differential angular distribution
of the first two of these angles results in a 14$\sigma$ separation between
the $0^{++}$ (CP-even, scalar) and the 
$0^{-+}$ (CP-odd, pseudoscalar)~\cite{fnal_report},
assuming that the $Z\phi$ cross section is independent
of the CP nature of $\phi$ [see \fig{fig:thresh}(b)]. Even more powerful are
fits to the triple-differential angular distribution 
where sufficient luminosity 
can uncover non-standard $ZZ\phi$ couplings.
However, this technique again suffers from the 
difficulty described in the previous item; 
namely, the CP-odd part of the state $\phi$
is typically so weakly coupled to $ZZ$ that there is little
sensitivity to the CP-odd component 
(if there is any significant CP-even component
in $\phi$) or little cross section (if $\phi$ is almost purely CP-odd).

5. If $\phi$ has significant branching ratios to either $\tau^+\tau^-$
or $t\anti t$, channels which are `self-analyzing',
then the decay distribution structure can provide a direct
determination of the ratio $b_f/a_f$ in the
$y_f\anti f (a_f+ib_f\gamma_5)f\phi$ ($f=\tau$ or $t$) Yukawa
coupling structure of 
$\phi$~\cite{Grzadkowski:1995rx,Grzadkowski:1994kv,Kramer:1994jn}. 
\iyuk\ihcp\ihmeas\
Detector simulation studies have been made of the specific decay
$\phi \rightarrow \tau^+ \tau^- \rightarrow \rho^+ \bar{\nu}_{\tau} 
\rho^- \nu_{\tau}$ where likelihood fits were made to
angular correlations 
between decay products of each $\tau$~\cite{Bower:2002zx} resulting 
in distinguishing the CP nature of the $\phi$
to a confidence level greater than 95\% with 500~fb$^{-1}$ at
$\sqrt{s} = 500$~GeV.

6. The angular distributions in the $t\anti t \phi$ final state,
which has adequate cross section for $\rts\gsim 800\gev$
for modest values of $m_\phi\lsim 200\gev$, assuming Yukawa
coupling $y_t\anti t (a_t+ib_t\gamma_5)t\phi$ comparable to SM values, 
appears to provide an excellent
means for determining the CP nature of $\phi$ by allowing one
to probe the ratio $b_t/a_t$ \cite{Gunion:1996bk,Gunion:1996vv}.\ihcp\ihmeas\
An observable
CP-violating asymmetry in $t\anti t \phi$ production (via a T-odd
triple correlation product proposed in \Ref{Bar-Shalom:1995jb})
would provide evidence for a CP-violating Higgs sector.

7. It is likely that the CP properties of the 
$\phi$ can  be well determined using photon polarization asymmetries in
$\gam\gam\to \phi$ 
collisions~\cite{Grzadkowski:1992sa,Gunion:1994wy,Kramer:1994jn}
(see \sect{secj}).\ihcp\ihmeas\

8. If the $\phi$ has substantial $ZZ$ coupling, then
$e^-e^-\to ZZ e^-e^-\to \phi e^-e^-$ 
can be used to probe its CP nature~\cite{Boe:1999kp} 
via the energy distributions of the $\phi$
and the final electrons, which are much harder in the
case of a CP-odd state than for a CP-even state.
Certain correlations are also useful probes of the CP properties of
the $\phi$. However, if the CP-odd portion of $\phi$ couples at one-loop
(as expected for a Higgs boson), there will be either little sensitivity
to this component or little cross section.\ihcp\ihmeas\

\subsection{Precision studies of non-SM-like Higgs bosons}
\label{sechh}

We confine our remarks to the two-doublet Higgs model 
(either the MSSM Higgs sector or a more general 2HDM).\ihmeas\itwohdm\
In the MSSM, we noted in \sect{seced} that for $\mha\lsim\sqrt{s}/2$, 
as long as one is not too close to threshold, it is possible to
observe all Higgs scalars of the non-minimal Higgs sector.  In
particular, in parameter regions away from the decoupling limit, none
of the CP-even Higgs scalars may resemble the SM Higgs boson.\idecoup\ 
Precision studies of all the Higgs bosons will provide a detailed
profile of the non-minimal Higgs sector.
Once $\mha\gsim\sqrt{s}/2$, only
the $\hl$ will be visible at the LC in an approximate decoupling regime
(although there may be some possibilities for observing the heavier
Higgs states produced singly either in association with a $b\bar b$
pair at large $\tan\beta$ where the coupling to $b\bar b$ is enhanced,
or by $s$-channel resonance production at a $\gamma\gamma$ collider\igamc).
\ihddiff\idecoup\
In the more general 2HDM, the simplest
mechanisms for producing the heavier neutral Higgs bosons
are $\epem\to \hh\ha$, $\epem\to\hh b\anti b,\ha
b\anti b$ and $\epem\to \hh t\anti t,\ha t\anti t$.\ihprod\itwohdm\
Charged Higgs bosons can be produced via $\epem\to\hp\hm$
and $\epem\to \hp \anti t b+\hm t\anti b$.
We have seen that none of these processes are guaranteed
to be either kinematically accessible or have a useful rate.\ihddiff\

Values of $\mha$ and $\mhh$  
in excess of 500 GeV to 1 TeV are certainly possible.  In such cases,
a substantial increase of energy for the LC will be required to observe these
states directly, either in association with $b\bar b$ (at large
$\tan\beta$) or via $\hh\ha$ production.\ihddiff\
Measuring the former will provide a 
crucial determination of the $b\anti b$ couplings,
which in the given model context will provide a determination
of $\tanb$, with accuracy determined by the production rates.\ihcoups\
Moreover, if $\hh$ and $\ha$
can be produced at a high rate (by whatever process),
a detailed study of their branching ratios has the potential
for providing vital information regarding the model parameters.\ihbrs\

\begin{figure}[t!]
\begin{center}
\includegraphics*[width=.9\textwidth]{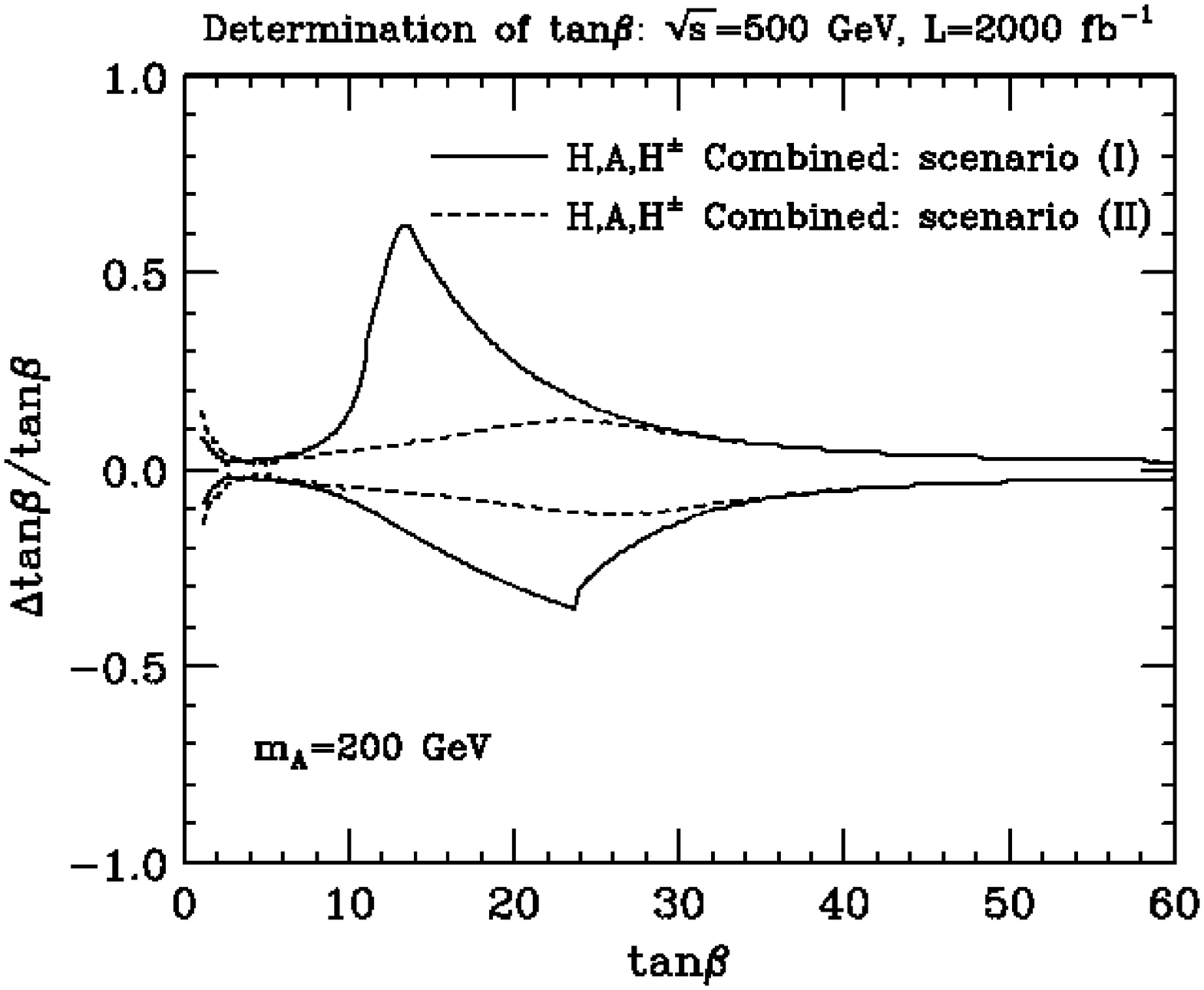}
\end{center}
\caption[0]{\label{fig:tanbdetermination}
The accuracy with which $\tanb$ can be measured, assuming $\mha=200\gev$,
$\sqrt s=500$~GeV and integrated luminosity of $L=2000\fbi$,
after combining in quadrature the errors on $\tanb$ from: 
a) the rate for $\bb\ha,\ \bb\hh \to \bbbb$; 
b) the $\hh\ha\to \bbbb$ rate; 
c) a measurement of the average $\hh,\ha$ total width in $\hh\ha$ production;
d) the $\hp\hm\to \tbtb$ rate; and 
e) the total $\hpm$ width measured in $\hp\hm\to\tbtb$ production.
Results are shown for two supersymmetric scenarios: in (I), the $\hh$ and $\ha$
do not decay to supersymmetric particles; in (II), there are substantial
decays of the $\hh$ and $\ha$ to a pair of the lightest supersymmetric
particles ($\cnone\cnone$). Taken from \protect\cite{Gunion:2001qy}.
\ihmeas\ihcoups}
\end{figure}

In low-energy supersymmetric models, 
the heavy $\hh$, $\ha$ and $\hpm$ would generally
decay to various pairs of supersymmetric particles 
as well as to $b$'s and $t$'s.\ihbrssusymod\
A study of the relative branching ratios and the rates
for specific processes would provide powerful
determinations of $\tanb$ and many of the soft-supersymmetry-breaking
parameters~\cite{Gunion:1996qd,Gunion:1997cc,Feng:1997xv,Barger:2000fi,Gunion:2001qy}.\ihmeas\ihcoups\
In addition, at large $\tanb$
the absolute widths of $\hh,\ha$ ($\hp$) become large enough to
be directly measured in the $b\anti b$ ($t\bar b$) final state and provide
an excellent determination of 
$\tanb$ \cite{Gunion:2001qy}.\footnote{More precisely, what
is being measured via the above techniques is the relevant Yukawa coupling,
which is determined by $\tanb$ at tree level. 
At one-loop, radiative corrections
must be included in order to convert from the Yukawa coupling to
a uniform definition of $\tanb$~\cite{tanbloopdef}.}\ihwidths\iyuk\
The accuracy in the $\tanb$ measurement that can be
achieved after combining all these $\tanb$ determinations in quadrature
are illustrated in fig.~\ref{fig:tanbdetermination} assuming
$\rts=500\gev$, $L=2000\fbi$, and $\mha=200\gev$. Excellent precision
is achieved at low and high $\tanb$ regardless of the specific 
supersymmetric scenario. However, in the range $10<\tanb<30$,  
good accuracy in the $\tanb$ measurement
is only achieved if there are some supersymmetric decays of $\hh$ and $\ha$
against which the corresponding $b\anti b$ decay modes have to compete. 
\ihmeas\

\section{The Giga-$Z$ option---implications for Higgs physics}
\label{seci}

Measurements of the effective leptonic mixing angle 
and the $W$ boson mass to
precisions of $\delta\sweff \simeq 10^{-5}$ and $\delta\mw \approx 6 \mev$ 
at Giga-$Z$ can be exploited in many ways.\igigaz\ipew\
The size of the resulting
$90\%$ CL ellipses were shown in \fig{dsdt}.
These measurements imply that the SM
Higgs boson mass can be determined indirectly to a
precision of about 7\%.  A deviation between the directly observed
value and the value implied by a global SM fit to 
Giga-$Z$ data would require new physics beyond the SM.

If new physics is present then it will be possible
to obtain information about new high mass scales beyond the
direct reach of the collider.  For example, in the MSSM
such information would be of particular importance if 
the heavier scalar top quark, ${\tilde t_2}$, and the heavy Higgs
bosons $\ha$, $\hh$ and $\hpm$ were beyond the kinematical reach of the
LC and background problems precluded their observation at the LHC.
Similar considerations apply to more general extended Higgs sector
models.

\subsection{Giga-$Z$ and the MSSM} 

In the MSSM,
the relation between $\mw$ and $\sweff$ is affected by the parameters
of the supersymmetric sector, especially the
${\tilde t}$ sector. At the 
LC, the mass of the light ${\tilde t}$, $m_{\tilde t_1}$, and the
${\tilde t}$ mixing angle, $\theta_{\tilde t}$, 
should be measurable very well if the process $e^+\,e^- \to {\tilde t}_1 
\bar{{\tilde t}_1}$ is accessible~\cite{lcstop}.\igigaz\igigazhmssm\

\begin{figure}[t!]
\begin{center}
\includegraphics*[width=0.8\textwidth]{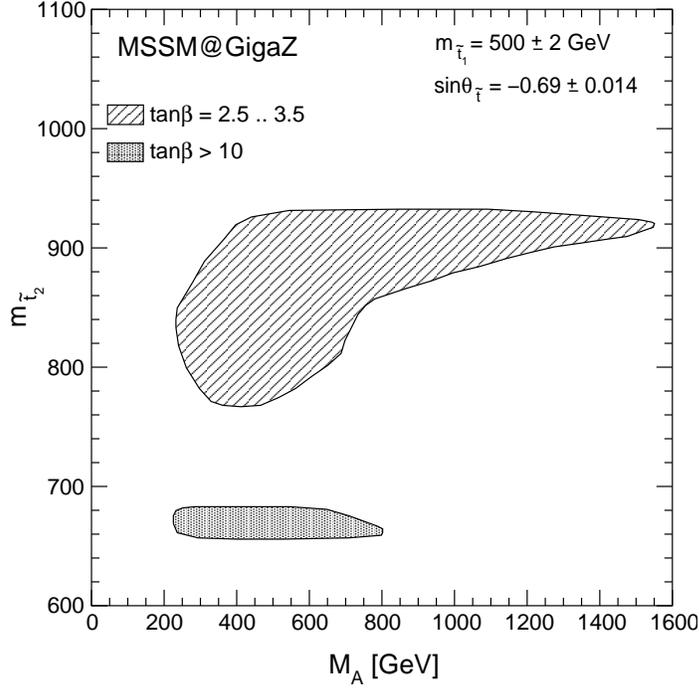}
\end{center}
\caption[0]{\label{fig:MSt2MA}
The region in the [$\mha$, $m_{\tilde t_2}$] parameter space, allowed by
$1\,\sigma$ errors obtained from the Giga-$Z$ measurements of $\mw$ and 
$\sweff$: 
$\mw = 80.400 \pm 0.006 \gev$, 
$\sweff = 0.23140 \pm 0.00001$, 
and from the LC measurement of $\mhl$:
$\mhl = 115 \pm 0.05~({\rm exp.}) \pm 0.5~({\rm theo.}) \gev $. 
$\tanb$ is assumed to be $\tanb = 3 \pm 0.5$ or $\tanb > 10$. 
The other parameters are given by 
$ m_{\tilde t_1} = 500 \pm 2 \gev$,
$\sin\theta_{\tilde t} = -0.69 \pm 0.014$,
$ A_b =  A_t \pm 10\%$,
$\mgl = 500 \pm 10 \gev$,
$\mu = -200 \pm 1 \gev$ and
$M_2 = 400 \pm 2 \gev$.  Taken from \protect~\cite{gigaz}.\igigazhmssm
}
\end{figure}

In \fig{fig:MSt2MA}~\cite{gigaz}, it is demonstrated how 
{\em upper} bounds on $\mha$ and 
$ m_{\tilde t_2}$ can be derived from measurements of $\mhl$, $\mw$
and $\sweff$,
supplemented by precise determinations of 
$ m_{\tilde t_1}$ and $ \theta_{\tilde t}$.
The most restrictive upper bounds are obtained if $\tanb > 10$
(in this case, one assumes that the value of $\tan\beta$ would have
already been determined from 
measurements in the gaugino sector~\cite{tbmeasurement}).
The other parameters values 
are assumed to have the uncertainties as expected from LHC~\cite{lhctdr} and
the LC~\cite{tesla_report}.\igigaz\igigazhmssm\

For low $\tanb$ (where the prediction for $\mhl$ depends sensitively on
$\tanb$) the heavier ${\tilde t}$ mass, $ m_{\tilde t_2}$, can be restricted to
$ 760 \gev \lsim  m_{\tilde t_2} \lsim 930 \gev$ from the $\mhl$, $\mw$ and 
$\sweff$ 
precision measurements. 
The mass $\mha$ varies between $200 \gev$ and $1600 \gev$. 
If $\tanb \ge 10$ (where $\mhl$ has only a mild dependence on $\tanb$),
the allowed region for the ${\tilde t_2}$ is
much smaller, $ 660 \gev \lsim  m_{\tilde t_2} \lsim 680 \gev$, and 
$\mha$ is restricted to $\mha \lsim 800 \gev$.

In deriving the bounds on the heavier ${\tilde t}$~mass, $ m_{\tilde t_2}$, 
the constraints from $\mhl$, $\sweff$ and $\mw$ 
play an important role.\igigazhmssm\ For the bounds
on $\mha$, the main effect comes from $\sweff$.
The assumed value of $\sweff = 0.23140$ differs slightly from the 
corresponding value obtained in the SM limit.
For this value the (logarithmic) dependence on $\mha$ is 
still large enough~\cite{gigaz} so that from 
the high precision in 
$\sweff$ at Giga-$Z$ an {\em upper limit} on $\mha$ can be set.
With an error in $\sweff$ that would be at least ten times larger at 
the LC without the Giga-$Z$ mode,
no bound on $\mha$ could be inferred.

\subsection{Giga-$Z$ and non-exotic extended Higgs sectors} 

Building on the discussion of the general 2HDM given earlier,
one can imagine many situations for which 
the very small Giga-$Z$ $90\%$ CL ellipses illustrated in \fig{dsdt}
would provide crucial (perhaps the only) constraints.\igigaz\ipew\
For example, suppose the LHC observes a $1\tev$ Higgs boson with 
very SM-like properties and no other new
physics below the few-TeV scale.  
We have seen that this is possible in the 2HDM scenarios consistent with
current precision electroweak constraints. Suppose
further that it is not immediately possible to increase
$\sqrt s$ sufficiently so that $\hl \ha$ production is allowed (typically
requiring $\sqrt s> 1.5\tev$ in these models). Giga-$Z$
measurements would provide strong guidance as to the probable masses
of the non-SM-like Higgs bosons of any given non-minimal
Higgs sector. However, it must be accepted that a particular Giga-$Z$
result for $S,T$ might have other non-Higgs interpretations as well.

\section{\boldmath The $\gam\gam$ collider option}
\label{secj}

Higgs production in $\gamma\gamma$ collisions offers a unique 
capability to measure the two-photon width of the Higgs and
to determine its CP composition through control of the photon
polarization.\igamc\ihwidths\ihcp\

The $\gam\gam$ coupling of a SM-like Higgs boson $h_{\rm SM}$ of
relatively light mass receives contributions from loops containing any
particle whose mass arises in whole or part from the vacuum
expectation value of the corresponding neutral Higgs field.  A
measurement of $\Gamma(h_{\rm SM}\to\gam\gam)$ provides the
possibility of revealing the presence of arbitrarily heavy particles
that acquire mass via the Higgs mechanism.\footnote{Loop contributions
from particles that acquire a large mass $M$ from some other mechanism
will decouple as $M^{-2}$ and $\Gamma(h_{\rm
SM}\to\gam\gam)$ will not be sensitive to their presence.}  However,
because such masses are basically proportional to some coupling times
$v$, if the coupling is perturbative the masses of these heavy
particles are unlikely to be much larger than 
$0.5$--$1\tev$.\iperturbativity\   Since
${\rm BR}(h_{\rm SM}\to X)$ is entirely determined by the spectrum of light
particles, and is thus not affected by heavy states, $N(\gam\gam\to
h_{\rm SM}\to X)\propto \Gamma(h_{\rm SM}\to\gam\gam){\rm BR}(h_{\rm SM}\to
X)$ will provide an extraordinary probe for such heavy states.\ihbrs\
Even if there are no new particles that acquire mass via the Higgs
mechanism, a precision measurement of $N(\gam\gam\to\widehat h\to X)$
for specific final states $X$ ($X=b\anti b,WW^*,\ldots$) can allow one
to distinguish between a ${\widehat h}$ that is part of a larger Higgs
sector and the SM $h_{\rm SM}$.\footnote{It may also be possible to
detect some rare Higgs decay modes at the $\gamma\gamma$ collider.\ihextended\
For example, \cite{Asner:2002aa} argues that the 
detection of $\gamma\gamma\to
h_{\rm SM}\to \gamma\gamma$ will be possible.}
The deviations from the SM
predictions typically exceed 5\% if the other heavier Higgs bosons
have masses below about 400 GeV.

The predicted rate for Higgs boson production followed by decay to
final state $X$ can be found in \cite{Gunion:1993ce}.
This rate depends strongly on
$d\mathcal{L}_{\gam\gam}/dy$, the differential $\gam\gam$ collider luminosity,
where $y=m_{\widehat h}/\rts$
and $\rts$ is the $ee$ collider center-of-mass energy.
An important parameter to maximize peak luminosity is 
$\VEV{\lambda\lambda'}$, 
the average value of the product of the helicities of the two colliding
photons after integration over their momentum fractions $z$ and $z'$.
Larger values of this parameter also suppress the dominant $J_z = \pm 2$,
$\gamma\gamma\to b\bar b(g)$ background, which is proportional
to $(1 - \VEV{\lambda\lambda'})$.
The computation of $d\mathcal{L}_{\gam\gam}/dy$ was first considered in 
\cite{Ginzburg:1983vm,Ginzburg:1984yr}. More realistic 
determinations~\cite{cainref}
including beamstrahlung, secondary collisions between scattered electrons
and photons from the laser beam, and other non-linear effects result in
a substantial enhancement of the luminosity in the low-$E_{\gamma\gamma}$
region as shown in \fig{fig:higgsspec}.\icollparam\

\begin{figure}[t!]
\begin{center}
\includegraphics*[width=\textwidth]{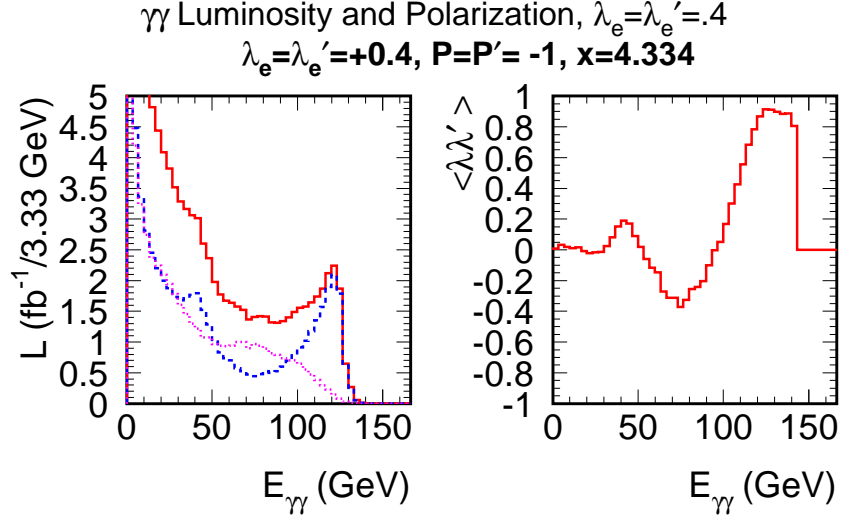}
\end{center}
\caption[0]{\label{fig:higgsspec}
The CAIN \cite{cainref} predictions for the $\gam\gam$ 
luminosity, $L=d\mathcal{L}/dE_{\gam\gam}$, are plotted in units of $\fbi$
per $3.33 \gev$ bin size, for circularly polarized  
photons assuming the NLC-based design for the $\gam\gam$
collider \cite{resourcebook,gunasner}, with $\rts=160\gev$, 
80\% electron beam polarization,
and a 1.054/3 micron laser wavelength.\icollparam\ 
The directions of the photon polarizations and electron helicities 
have been chosen to produce 
a peaked $E_{\gam\gam}$ spectrum. Beamstrahlung
and other effects are included. The dashed (dotted) curve gives the component
of the total luminosity that derives from the $J_z=0$ ($J_z=2$) two-photon
configuration. 
Also plotted is the corresponding value of $\vev{\lam\lam'}$
[given by $\vev{\lam\lam'}=(L_{J_z=0}-L_{J_z=2})/(L_{J_z=0}+L_{J_z=2})$].
Taken from~\protect\cite{gunasner}.
} 
\end{figure}

The choice of parameters that gives a peaked spectrum is well suited
for light Higgs studies. Using the spectrum of~\fig{fig:higgsspec}
as an example, the di-jet invariant
mass distributions for the Higgs signal and
for the $b\anti b(g)$ background for $m_{h_{\rm SM}}=120\gev$ 
are shown in \fig{fig:higgs}~\cite{gunasner}. 
After a nominal year of operation ($10^7$~sec),
$\Gamma(h_{\rm SM}\to\gam\gam){\rm BR}(h_{\rm SM}\to b\anti b)$ 
could be measured with
an accuracy of about $2.9\%$.\ihmeas\igamc\ (The more optimistic error of close
to $2\%$, quoted in \cite{hjikia,Niezurawski:2002aq}
for $m_{h_{\rm SM}}=120\gev$,
is based upon a significantly higher peak luminosity.)
The error for this measurement increases to about $10\%$
for $m_{h_{\rm SM}}= 160\gev$,
primarily due to the decrease of the Higgs di-jet branching fraction
by a factor of 18.

\begin{figure}[t!]
\begin{center}
\includegraphics*[width=0.8\textwidth]{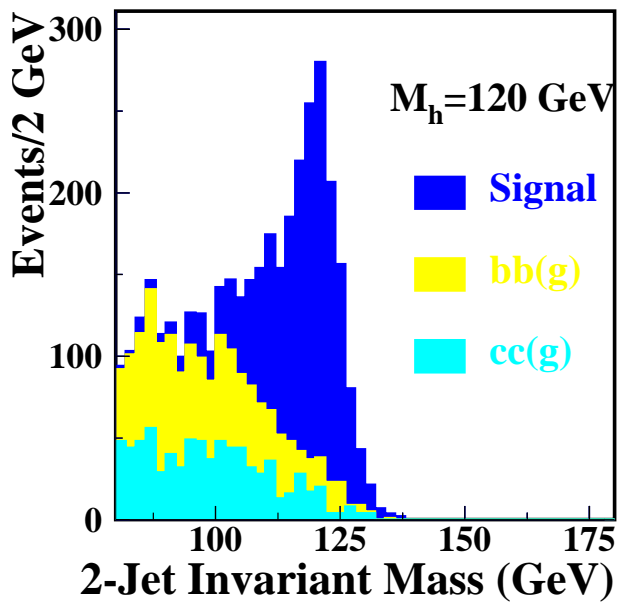}
\end{center}
\caption[0]{\label{fig:higgs}
Higgs signal and heavy quark backgrounds for $\gam\gam\to h$
in units of events per 2 GeV
for a Higgs mass of 120~GeV after one year of running,
with the $\gam\gam$ luminosity 
spectrum of \protect\fig{fig:higgsspec}. Results are
from~\protect\cite{gunasner}.\igamc\ihmeas
}
\end{figure}

In many scenarios, it is possible that by combining this 
result with other types of
precision measurements for the SM-like Higgs boson, 
small deviations can be observed indicating the possible 
presence of heavier Higgs
bosons.  For a 2HDM (either the MSSM or a two-Higgs-doublet model
with partial decoupling),
if $\mhh\sim\mha>\rts/2$ then $e^+e^-\to \hh\ha$  is not possible.\idecoup\
Further, as discussed earlier in association with \fig{wedge}, 
the alternatives of $b\bar{b}\hh$ and $b\bar{b}\ha$ production
will only allow $\hh$ and $\ha$ detection if $\tan\beta$ is 
large \cite{Grzadkowski:2000wj}.\ihddiff\ihnolose\ Thus,
$\gam\gam\to \hh,\ha$ may be the only option
allowing their discovery (without a collider with
higher $\rts$).
The back-scattering kinematics are such that a 
LC for which the maximum energy is
$\rts=630\gev$ can potentially probe Higgs masses as
high as $500\gev$ in $\gam\gam$ collisions.\icollparam\
If $\mhh$ and $\mha$ are known to within roughly $50\gev$
on the basis of precision $\hl$ data, 
then they will be easily detected by choosing the machine $\rts$
and polarization/helicity configurations that yield a $E_{\gam\gam}$
spectrum peaked near the upper edge of the expected mass range
\cite{Muhlleitner:2001kw,gunasner}. But, if it happens that
no constraints have been placed on the $\hh,\ha$ masses,
then it is most appropriate to run entirely at the maximal $\rts$
and to consider how best to cover the portion of the 
LHC wedge~\footnote{The LHC wedge, discussed in Section~\ref{secfc},
corresponds to the moderate $\tanb$ large $\mha$ wedge-like
region in the [$\mha$, $\tanb$] parameter space (see \fig{f:atlasmssm})
where only the $\hl$ of the MSSM Higgs sector can be observed.}
with $\mhh\sim\mha\gsim \rts/2$ (for which $\hh\ha$ pair production
is not possible).\ihnolose\
Assuming a maximum of $\rts=630\gev$, the goal would then be to
roughly uniformly cover the $300\gev\lsim \mha,\mhh\lsim 500\gev$
region of the LHC wedge.\icollparam\  For this purpose, it is best to run 
part of the time
with polarization/helicity directions that yield a peaked (type-II) spectrum 
(with peak at $E_{\gam\gam}\sim 500\gev$ for $\rts=630\gev$) 
and roughly three times as long in a polarization/helicity configuration 
that yields a broad (type-I) $E_{\gam\gam}$ spectrum with good
luminosity and substantial $\vev{\lam\lam'}$ 
in the $350\lsim E_{\gam\gam}\lsim 450\gev$ region.

\begin{figure}[p!]
\begin{center}
\includegraphics[width=0.9\textwidth]{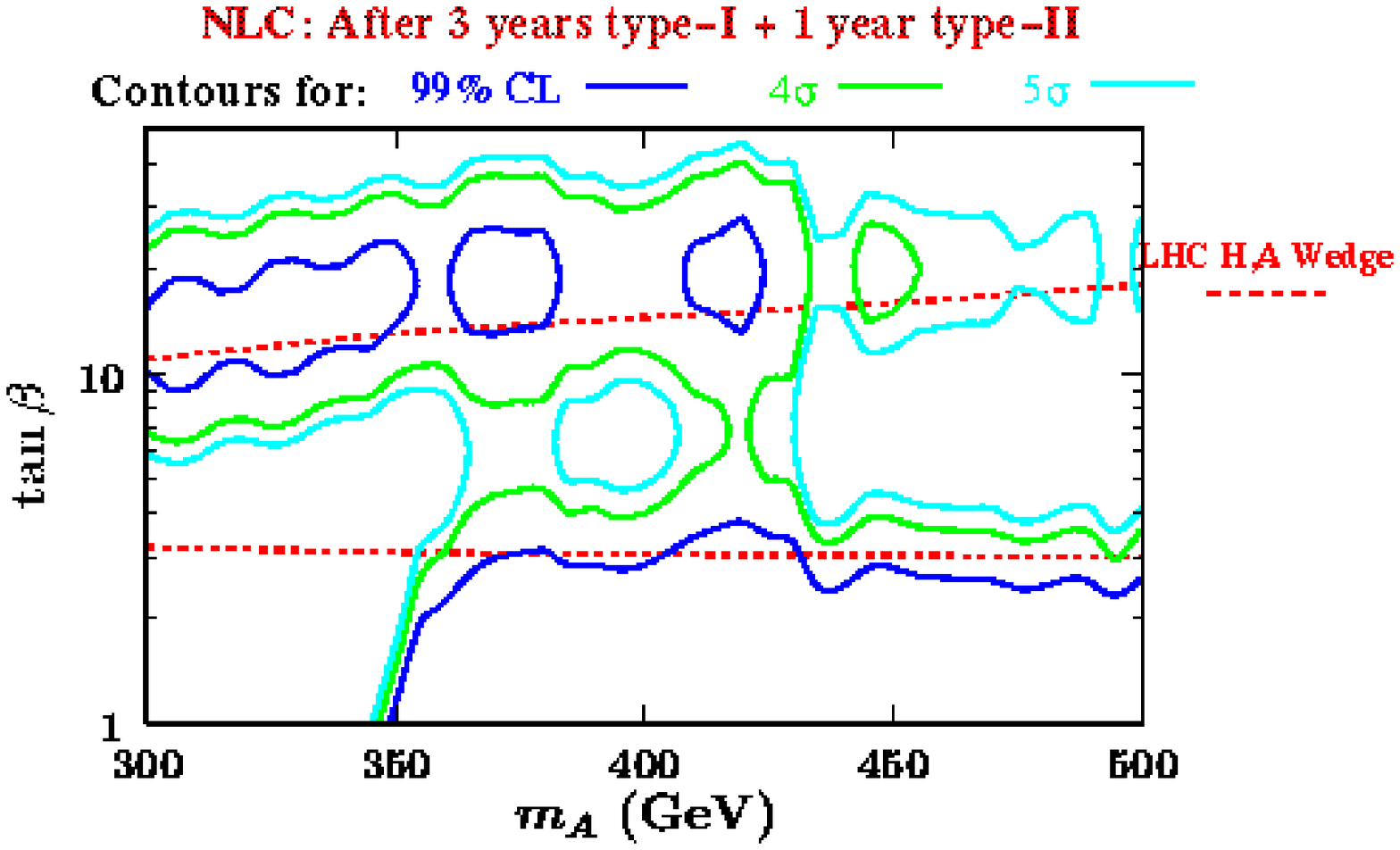}

\includegraphics[width=0.9\textwidth]{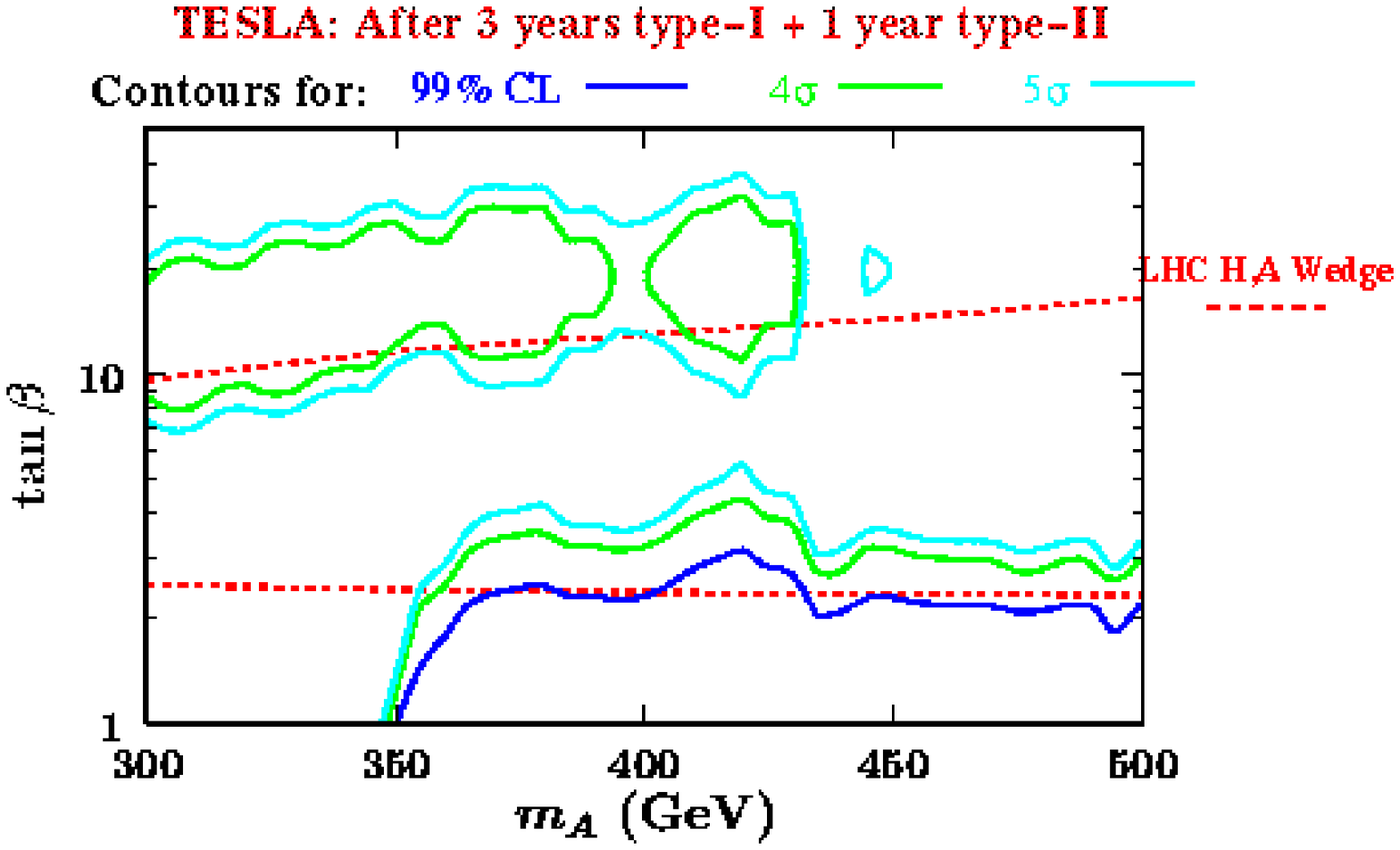}
\end{center}
\caption[0]{\label{wedgeplot}
Assuming a machine energy of $\rts=630\gev$
and employing only $\gamma\gamma\to \hh,\ha\to b\anti b$,
we show the $[\mha,\tanb]$ regions for which three years
of operation 
using the type-I $P\lam_e,P'\lam_e'>0$ polarization configuration
and one year of operation using the type-II $P\lam_e,P'\lam_e'<0$
configuration will yield $S/\sqrt{B}\geq5$, $S/\sqrt{B}\geq 4$
or exclude the $\hh,\ha$ at 99\% CL.\icollparam\
The dashed curves indicate the wedge region from the LHC plot
of \protect\fig{f:atlasmssm}---the lower curve gives 
a rough approximation to the
LEP (maximal-mixing) limits
while the upper curve is that above which $\hh,\ha\to\tau^+\tau^-$
can be directly detected at the LHC.  The upper plot is
for the NLC-based $\gam\gam$ collider design 
and the lower plot is that achieved
with a factor of 2 increase in luminosity, as might be
achieved at TESLA or by using round $e^-e^-$ beams instead of flat beams
at the NLC. 
Results are from \protect\cite{gunasner} as summarized in
\protect\cite{Asner:2002aa}.\ihnolose
}
\end{figure}

For the NLC-based design of the $\gam\gam$ collider
as specified in \cite{resourcebook,gunasner}, after
roughly four years of total running time 
(one year with the peaked
spectrum and three years with the broad spectrum,
assuming that the machine operates $10^7$~sec per year) one can achieve
$>4\sigma$ signals (99\% exclusion) in $\gam\gam\to\hh,\ha\to b\anti b$
for most (all) of the $[\mha,\tanb]$ LHC wedge region.\icollparam\ihnolose\ 
With twice the luminosity of the nominal NLC
design (as might be achieved at TESLA or by using round beams
at the NLC), essentially all of the wedge region is covered
at the $5\sigma$ level except for the lowest $\tanb$ values
in the $\mha>2\mt$ part of the wedge 
(where it is anticipated that $\gam\gam\to\hh,\ha\to t\anti t$ will
yield a viable signal).\ihddiff\  This is summarized in \fig{wedgeplot}.
In fact, for $\hh,\ha$ discovery, the $\gam\gam$ collider is almost
perfectly complementary to the LHC
(and also the LC operation in the $\epem$ collision mode).

The corresponding results for $\gam\gam\to \what h\to b\anti b$
for the wedge region of a 
2HDM model with a light decoupled $\what h$,
and all other Higgs bosons heavier than $\rts$, were given earlier
in \fig{wedge}.\itwohdm\ihddiff\
Again, the $\gam\gam$ collider would allow
$\what h$ discovery in a substantial portion of the wedge region
in which its discovery would not otherwise be possible.\ihddiff\ihnolose\

Once one or more Higgs bosons have been detected, 
precision studies can be
performed including: determination of CP properties; 
a detailed scan to
separate the $\hh$ and $\ha$ in the decoupling limit of a 2HDM; and
branching ratios measurements.\ihcp\ihmeas\  The branching ratios 
to supersymmetric final states
are especially important for determining the basic supersymmetry breaking 
parameters~\cite{Gunion:1995bh,Gunion:1997cc,Feng:1997xv,Muhlleitner:2001kw}.

The CP properties can be determined for 
any spin-0 Higgs $\widehat h$ produced in $\gam\gam$
collisions.  Since $\gam\gam\to {\widehat h}$ is of one-loop order,
whether ${\widehat h}$ is CP-even, CP-odd or a mixture, 
the CP-even and CP-odd parts of ${\widehat h}$ have
$\gam\gam$ couplings of similar size.\igamc\ihcp\ihcpviol\
However, the structure of the couplings is very different:
\begin{equation}
\mathcal{A}_{\,\scriptstyle \textrm{CP-even}}\propto 
\vec \eps_1\cdot\vec \eps_2\,,\quad
\mathcal{A}_{\,\scriptstyle \textrm{CP-odd}}\propto 
(\vec\eps_1\times\vec \eps_2)\cdot \what p_{\rm beam}\,.
\end{equation}
By adjusting the orientation of the initial laser
photon polarization vectors with
respect to one another, it is possible to determine the relative
CP-even and CP-odd content of the resonance ${\widehat h}$
\cite{Grzadkowski:1992sa}. 
If ${\widehat h}$ is a mixture, one can use helicity asymmetries for this 
purpose \cite{Grzadkowski:1992sa,Kramer:1994jn}.
However, if ${\widehat h}$ is either purely CP-even
or purely CP-odd, then one must employ transverse linear polarizations
\cite{Gunion:1994wy,Kramer:1994jn}.\ihcp\
Substantial luminosity with transverse polarization can 
be obtained, although the spectrum is not peaked (see \cite{gunasner}).

One measure of the CP nature of a Higgs boson is the asymmetry
for parallel {\it vs.} perpendicular orientation 
of the linear polarizations of the initial laser beams,
\begin{equation}
\mathcal{A}\equiv{N_{\parallel}-N_{\perp}\over N_{\parallel}+N_{\perp}}\,,
\end{equation}
which is positive (negative) for a CP-even (odd) state. Since 
100\% linear polarization for the laser beams translates into
only partial linear polarization for the colliding photons, 
both $N_{\parallel}$
and $N_{\perp}$ will be non-zero for the signal.  In addition,
the heavy quark background contributes to both. The expected value
of $\mathcal{A}$ must be carefully computed for a given model.  For the
SM Higgs boson with $m_{h_{\rm SM}}=120\gev$, it is found
\cite{gunasner} that $\mathcal{A}$ 
can be measured with an accuracy of about 20\% in one year
of operation using linear polarizations for the two lasers.\ihcp\ihmeas\
This measurement would thus provide a 
moderately strong test of the even-CP nature of $h_{\rm SM}$.
Of course, the linear polarization configuration is not ideal for the most
accurate determination of $\Gamma(\hsm\to\gam\gam)\BR(\hsm\to b\anti b)$,
but would allow measuring this product with an accuracy of about 8\%.
\ihcp\ihmeas\

We end by noting that the $e^-\gam$ and $e^-e^-$ collider
options are most relevant to exotic Higgs scenarios,
as discussed in Section~\ref{seck}.\ihextended\

\section{Concluding Remarks}
\label{secl}

The physical origin of electroweak symmetry breaking is not yet known.
In all theoretical approaches and models, 
the dynamics of electroweak symmetry
breaking must be revealed at the TeV-scale or below.  This energy
scale will be thoroughly explored by hadron colliders, starting with
the Tevatron and followed later in this decade by the LHC.
Even though the various theoretical alternatives can only be confirmed
or ruled out by future collider experiments,
a straightforward interpretation of the electroweak precision data
suggests that electroweak symmetry breaking dynamics is
weakly-coupled, and a Higgs boson with mass between
100 and 200 GeV must exist.\ihmass\  With the
supersymmetric extension of the Standard Model, this interpretation
opens the route to grand unification of all the fundamental forces,
 with the eventual incorporation of gravity in particle physics.\icoupu\

The discovery of the Higgs boson at the Tevatron and/or the LHC
is a crucial first step.  The measurement of Higgs properties at the
LHC will begin to test the dynamics of electroweak symmetry breaking.\ihmeas\
However, a high-luminosity $e^+e^-$
linear collider, now under development, is
needed for a systematic program of precision Higgs measurements.
For example, depending on the value of the Higgs mass, branching
ratios and Higgs couplings can be determined in some cases at the
level of a few percent.\ihmeas\
In this way, one can extract
the properties of the Higgs sector in a comprehensive way,
and establish (or refute) the theory of 
scalar sector dynamics as the mechanism responsible
for generating the fundamental particle masses.\iewsb\

\section*{Acknowledgments}
\addcontentsline{toc}{section}{Acknowledgments}

We are very grateful to Andreas Kronfeld who served with us as a co-convener
of the Higgs Working Group of the American Linear Collider Working
Group.  This chapter is an update of our contribution to the 
American Linear Collider Working Group Resource Book for Snowmass
2001.  We have benefited greatly from Andreas' contributions and counsel.
We would also like to thank David Asner, Marcela Carena, 
Sven Heinemeyer, Heather Logan, Steve Mrenna, 
Michael Peskin, Carlos Wagner, Georg Weiglein,  
Dieter Zeppenfeld and Peter Zerwas for fruitful 
conversations and collaborative efforts that contributed to this work.

J.F.G., H.E.H. and R.V.K. are supported in part by the U.S.
Department of Energy under the respective grants DE-FG03-91ER40674,
DE-FG03-92ER40689 (Task B) and DE-FG02-91ER40661 (Task A).

\clearpage

\printindex             

\end{document}